\documentclass[a4paper, reqno]{amsart}
\usepackage{graphicx}
\usepackage{amsmath}
\usepackage{amsfonts}
\usepackage{amsthm}
\usepackage{amstext}
\usepackage{enumerate}

      \newcommand{\sgn}{{\operatorname{sgn}}}
      \newcommand{\td}{{\operatorname{td}}}
      \newcommand{\crit}{{\operatorname{cr}}}
\renewcommand{\bar}{\overline}
      \newcommand{\Bog}{{\operatorname{bg}}}
      \renewcommand{\i}{{\operatorname{i}}}
     \newcommand{\e}{\operatorname{e}}

     \newcommand{\un}{{\operatorname{un}}}

     \newcommand{\s}{{\operatorname{s}}}
     \renewcommand{\d}{{\operatorname{d}}}
     \newcommand{\dist}{{\operatorname{dist}}}     
     \newcommand{\cl}{{\operatorname{cl}}}

     \newcommand{\an}{{\operatorname{an}}}

     \newcommand{\R}{{\mathbb{R}}}
     \newcommand{\Z}{{\mathbb{Z}}}
     \newcommand{\C}{{\mathbb{C}}}
     
\renewcommand{\sp}{{\operatorname{sp}}}

\renewcommand{\Im}{{\operatorname{Im}}}
\newcommand{\zz}{{\mathbb{Z}}}
\newcommand{\cc}{{\mathbb{C}}}
\newcommand{\rr}{{\mathbb{R}}}
\newcommand{\x}{\mathbf{x}}
\newcommand{\w}{\mathbf{w}}
\newcommand{\y}{\mathbf{y}}
\newcommand{\0}{{\bf 0}}
\newcommand{\kk}{{\bf k}}
\newcommand{\q}{{\bf q}}
\newcommand{\pp}{{\bf p}}
\newcommand{\p}{\mathbf{p}}
\newcommand{\f}{{\rm f}}

\newcommand{\ph}{{\rm ph}}

\def\bbbone{{\mathchoice {\rm 1\mskip-4mu l} {\rm 1\mskip-4mu l}
{\rm 1\mskip-4.5mu l} {\rm 1\mskip-5mu l}}}
\def\one{\bbbone}

     \theoremstyle{plain}
     \newtheorem{thm}{Theorem}[section]
     \newtheorem{prop}[thm]{Proposition}

 \newtheorem{conjecture}[thm]{Conjecture}
     \theoremstyle{definition}

     \numberwithin{equation}{section}
\numberwithin{figure}{section}
\title[Zero]{On the infimum of the 
energy-momentum spectrum of a homogeneous Bose gas}

\author{H.D. Cornean}
\address[H.D. Cornean]{Dept. of Math., Aalborg University\\Fredrik Bajers Vej 7G, 9220 Aalborg, Denmark}
\email{cornean@math.aau.dk}

\author{J. Derezi\'{n}ski}
\address[J. Derezi\'{n}ski]
{Dept. of Math. Methods in Phys., University of Warsaw\\ 
Hoza 74, 00-682 Warszawa, Poland} 
\email{Jan.Derezinski@fuw.edu.pl}

\author{P. Zi\'{n}}
\address[P. Zi\'{n}]{Institute of Theoretical Physics, Uniwersity of Warsaw\\
Hoza 69, 00-681 Warszawa, Poland}
\email{Pawel.Zin@fuw.edu.pl}

\date{\today}

\begin{document}

\begin{abstract} We consider  second quantized homogeneous 
Bose gas in a large cubic box with periodic boundary conditions, 
at zero temperature.
We discuss the energy-momentum spectrum of the Bose gas and its physical
significance. We review various rigorous and heuristic
 results
 as well as open conjectures  about its properties.
Our main aim is to convince the readers, including those with mainly
mathematical background,  that this subject has many interesting
problems for rigorous research. 

In particular, 
we investigate  the upper bound on the infimum of
the  energy for a fixed total momentum $\kk$ 
 given by the expectation value of one-particle excitations over a
squeezed  states. This bound can be viewed as  a rigorous version of the
famous Bogoliubov method.
We show that this approach seems to lead to a (non-physical)
energy gap.

The  variational problem involving squeezed states
can serve as the preparatory step in a
 perturbative approach that should be useful in computing excitation
 spectrum. This version of a perturbative approach to the Bose gas seems
(at least in principle) superior
to the commonly used  approach based on the $c$-number
 substitution.
\end{abstract}
\maketitle
\tableofcontents

\section{Introduction}
\label{s1}

In this paper we would like to review one of
outstanding open problems of quantum physics -- rigorous understanding 
of the 
energy--momentum spectrum of homogeneous Bose gas at 
zero temperature. 
We  
describe various rigorous and heuristic arguments about its shape.
In particular, we discuss
a number of versions of the so-called Bogoliubov approach. We use 
the main idea
of this approach to give rigorous
 upper bounds on the  energy-momentum spectrum of the Bose gas.

There exists  little rigorous work on this
subject.
 We think that
mathematicians avoid this topic not only because of its
difficulty. Unfortunately, it is
 not easy to formulate   questions  in this domain
that  are, on one hand, physically
relevant, and on the other hand, mathematically clean and precise. We
try to ask a number of such questions, some of them rather
ambitious, but some, perhaps,   within the reach of present methods. 
We think that rigorous
  methods of  spectral analysis and  operator theory could be very
  helpful in 
  clarifying this subject.

\subsection{Bose gas in canonical approach}

One can distinguish two possible approaches to the Bose gas at positive
density:
 ``canonical'' -- fixing the density  $\rho$ -- and
``grand-canonical'' -- fixing the chemical potential $\mu$. In most of 
our paper we will concentrate on the
latter setting. Nevertheless, in the introduction, as well as in Section
\ref{Fixed number of particles}
we will stick to the canonical approach.

We suppose that the 2-body potential of an interacting Bose gas 
is described by a real function $v$ defined on $\rr^d$, satisfying
$v(\x)=v(-\x)$. We  assume that $v(\x)$ decays at infinity sufficiently fast.

A typical  assumption on the potentials that we have in mind in our paper is
 \begin{equation}
\hat v(\kk)>0, \ \ \kk\in\rr^d,\label{posi}\end{equation}
where
the Fourier transform of $v$ is given by 
\begin{equation}
\hat v(\kk):=\int_{\rr^d} v(\x)\e^{-\i \kk\x}\d \x.\end{equation} 
Potentials satisfying (\ref{posi}) 
will be called repulsive.
 Note, however, that a large part of our paper does not use
directly   any specific assumption on the potentials.

$n$-body Schr\"odinger Hamiltonian of the homogeneous Bose gas 
acts on  the Hilbert space $L_\s^2((\rr^d)^n)$ (symmetric square integrable
functions on $(\rr^d)^n)$ and is described by the Hamiltonian 
\begin{equation}
H^{n}=
-\sum_{i=1}\frac12\Delta_i+
\sum_{1\leq i<j\leq n}v(\x_i-\x_j)
\label{sch4}\end{equation}
and the momentum operator
\[P^{n}:=\sum_{i=1}^n-\i\nabla_{\x_i}.\]
$(H,P)$ is a collection of $1+d$ commuting self-adjoint operators,
hence we can 
ask about the properties of their joint spectrum, called
the energy--momentum spectrum. 

(\ref{sch4}) describes however only a finite number of particles in an
infinite space. We would like to
investigate homogeneous Bose gas at positive density. It is a little
problematic how to model such a system. A natural solution would be 
restricting (\ref{sch4}) to e.g.
$\Lambda=[-L/2,L/2]^d$, the $d$-dimensional
 cubic box of side length $L$, with Dirichlet boundary conditions. This 
 will, however,  destroy
its translational invariance. Therefore,
following the accepted although somewhat unphysical tradition,
we consider the Bose gas on a torus.
This means in particular that
the potential $v$ is replaced  by
\begin{equation}\label{interakt}
v^L(\x)=\frac{1}{V}\sum_{\kk\in \frac{2\pi}{L}\mathbb{Z}^d}\e^{\i\kk \cdot \x}\hat{v}(\kk),
\end{equation}
where $\kk\in
 \frac{2\pi }{L}\zz^d$ is the discrete momentum variable and $V=L^d$
 is the volume of the box.
Note that $v^L$ is periodic with respect to the domain
$\Lambda$, and $v^L(\x)\to v(\x)$ as $L\to\infty$. 
 The system  on a torus  is described by the Hamiltonian
\begin{equation}
H^{L,n}=
-\sum_{i=1}\frac12\Delta_i^L+
\sum_{1\leq i<j\leq n}v^L(\x_i-\x_j)
\label{sch}\end{equation}
acting on the space $L^2_\s(\Lambda^n)$ 
(symmetric square integrable functions on
$\Lambda^n$). The Laplacian is assumed to have  periodic boundary
conditions.

 Let us denote by $E^{L,n}$ the {\em
  ground state energy in the box}:
 \[E^{L,n}:=\inf \sp H^{L,n},\]
where $\sp K$ denotes the spectrum of an operator $K$.

The total momentum is given by the vector of operators
\[P^{L,n}:=\sum_{i=1}^n-\i\nabla_{\x_i}^L.\]
Its joint spectrum equals $ \frac{2\pi }{L}\zz^d$.

Clearly, $H^{L,n}$ and $P^{L,n}$ commute with each other. Therefore we can
define 
their joint spectrum 
\[\sp(H^{L,n},P^{L,n})\subset \rr\times  \frac{2\pi }{L}\zz^d,\]
which will be called the {\em energy-momentum spectrum in the box}.
By the {\em excitation spectrum in  the box} we will mean
$\sp(H^{L,n}-E^{L,n},P^{L,n})$.

 For $\kk\in  \frac{2\pi }{L}\zz^d$, we can define the Hamiltonian
 $H^{L,n}(\kk)$ to be the restriction of $H^{L,n}$ to the supspace of
 $P^{L,n}=\kk$ and
{\em infimum of the excitation spectrum (IES) in the box} as
\begin{eqnarray}\label{defexcit2}
\epsilon^{L,n}(\kk):=\inf\sp (H^{L,n}(\kk)-E^{L,n})
\end{eqnarray}
By the
{\em infimum of the energy-momentum spectrum in the box} we will mean
$E^{L,n}+\epsilon^{L,n}(\kk)$.

It is believed that the properties of the Bose gas simplify in the
thermodynamic limit. It means that one should fix $\rho>0$,
 take
 the number of particles equal to
 $n=\rho V$, and then 
pass to  the limit
 $L\to\infty$. 
Unfortunately, as far as we know,
the Hamiltonians 
$H^{L,n}-E^{L,n}$ do not have a limit as self-adjoint operators. One can
hope, however, that the IES has some kind of a limit.

Mathematically it is not obvious how to define this limit, since for finite $L$
the IES is defined on the 
 lattice $\frac{2\pi}{L}\zz^d$ and in thermodynamic limit it should be defined
 on $\rr^d$. Below we propose one of possible definitions of the IES in
 thermodynamic limit.

For $\kk\in\rr^d$ and $\rho>0$,  we take $\delta>0$ and set
\begin{eqnarray}\label{ies}
&&\epsilon^\rho
(\kk,\delta)\\
&:=&\liminf_{n\to\infty}
\left(\inf\left\{
\epsilon^{L,n}(\kk')\ :\ \kk'\in
 \frac{2\pi}{L}\zz^d,\; |\kk-\kk'|<\delta,\
 \rho=\frac{n}{L^d}\right\}\right).  \nonumber
\end{eqnarray}
This gives a lower bound 
on the IES
 for the momenta $\kk'$ in the window
 in the momentum space around $\kk$ of diameter 
$2\delta$. The quantity $\epsilon^\rho
(\kk,\delta)$
increases as 
$\delta$ becomes smaller. 
The {\em IES in the thermodynamic
  limit} is defined as
 its supremum (or, equivalently,
its  limit) 
as $\delta\searrow 0$:
\begin{equation}
\epsilon^\rho(\kk):=\sup_{\delta >0}
\epsilon^\rho
(\kk,\delta).\label{begin}\end{equation}

 Under Assumption (\ref{posi})
it is easy to prove that $E^{L,n}$ is finite and
 $\epsilon^{L,n}(0)=\epsilon^\rho(0)=0$ (see
Theorem \ref{thm} and Proposition \ref{prop1}).

\begin{conjecture}\label{conj1a}
We expect that for a large class of repulsive potentials
the following statements hold true: 
\begin{enumerate} 
\item The function
 $\rr^d\ni \kk\mapsto \epsilon^\rho(\kk)\in \rr_+$ is continuous.
\item Let $\kk\in{\mathbb R}^d$. If $L\to\infty$,  $n_L\to\infty$,
 $\frac{n_L}{L^d}\to\rho$,
 $\kk_L\in\frac{2\pi}{L}\zz^d$,
and  $\kk_L\to\kk$ ,
we have that $\epsilon^{L,n_L}(\kk_L)\to \epsilon^\rho(\kk)$.
\item If $d\geq2$, then
$\inf\limits_{\kk\neq0}\frac{\epsilon^\rho(\kk)}{|\kk|}=:c_\crit>0$.
\item There exists the limit
  $\lim\limits_{\kk\to0}\frac{\epsilon^\rho(\kk)}{|\kk|}= : c_\ph>0$.
\item The function ${\mathbb R}^d\ni \kk\mapsto\epsilon^\rho(\kk)$ is
  subadditive, that is, $\epsilon^\rho(\kk_1+\kk_2)\leq\epsilon^\rho(\kk_1)+
\epsilon^\rho(\kk_2)$.
\end{enumerate}
\end{conjecture}

Statements (1) and (2) can be interpreted as some kind of a ``spectral thermodynamic
limit in the canonical approach''.
 Note that if (1) and (2) are true around $\kk=0$, then we can say that there
is ``no gap in the excitation 
spectrum''.

The properties (3) and (4)
of the Bose gas were predicted by Landau in the 40's. 
Shortly thereafter, they were derived by a somewhat heuristic argument
by Bogoliubov \cite{Bog}.

(3) is commonly believed to be responsible for the superfluidity of the Bose
 gas. More precisely, it is argued that because of (3) a drop of Bose gas
 travelling at speed less than $c_\crit$ will experience no friction. This
 argument is described e.g. in the course by Landau-Lifshitz \cite{LaL} and in
 \cite{WdS}, see
 also Section \ref{superfluidity}.

Note that in dimension $d=1$ the statement (3) should be replaced by

\medskip

\noindent\ \ \ \ (3)' If $d=1$, then 
\begin{equation}\epsilon^\rho(\kk
+2\pi\rho)=\epsilon^\rho(\kk).\label{3'}\end{equation}

\medskip

The statement (3)' has a simple rigorous proof, which we will describe 
 in our paper.
 It implies that in dimension $d=1$ the excitation  spectrum
is periodic with the period $2\pi\rho$.
It follows by the well known argument that involves   boosting all
particles simultaneously by the velocity $\frac{2\pi}{L}$
 (see e.g. \cite{L2}).

 Excitation spectrum with the property
 described by (4) is often described as {\em phononic} and the excitations
 with such a spectrum are called {\em phonons}. One also expects that the
 parameter $c_\ph$ coincides with the  speed of sound -- a 
 parameter in principle 
 macroscopically measurable.

Let us describe a heuristic argument for (5). Suppose that
 excitations
can be described by certain elementary
quasiparticles with a dispersion relation $\kk\mapsto\omega(\kk)$. We
assume that  the state consisting of quasiparticles with momenta
$\kk_1,\dots,\kk_n$ has the excitation energy
$\omega(\kk_1)+\cdots+\omega(\kk_n)$. Then it is easy to see that the
IES is the subadditive hull of $\omega(\kk)$, that is
\begin{equation}
\epsilon
(\kk)=\inf\{\omega(\kk_1)+\cdots+
\omega(\kk_n)\ :\ 
\kk_1+\cdots+\kk_n=\kk,\ \ n=1,2,\dots\},\label{epsilon}\end{equation}
which is the largest subadditive function less than $\epsilon$.
 In Appendix \ref{a00} we describe a somewhat more elaborate, but
 still  
heuristic,
    argument that seems to indicate that, in thermodynamic limit,
the Bose gas has a
    subadditive excitation spectrum.

\begin{figure}[h]
\includegraphics[width=0.8\textwidth]{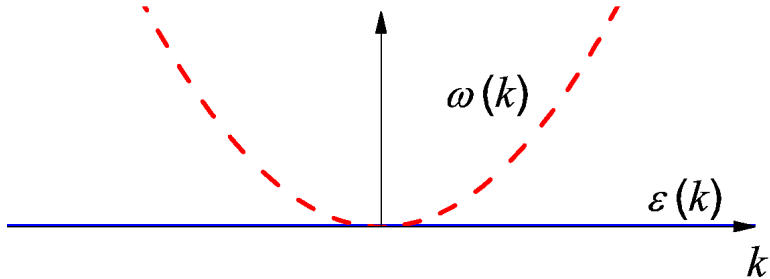}
\caption{Excitation spectrum of the free Bose gas}
\label{fig1.1}
\end{figure}

Note that
 free Bose gas does not satisfy Conjecture \ref{conj1a}. In this case
 $\omega(\kk)=\kk^2$ and  $\epsilon(\kk)=0$, see Fig. \ref{fig1.1}

There are not so many subadditive
 functions.  There exist, however, subadditive functions, which satisfy the
properties described in our conjecture in (3) or (3)', and in (4). We discuss
basic properties of subadditive functions
 in Appendix \ref{a0}.
 We could not find these facts in the
literature, although they probably belong to the folk knowledge.

\subsection{Experimental evidence}

To our experience, most physicists  interested in this subject (but not all)
 would agree that one should
expect Conjecture \ref{conj1a} (as well as the analogous 
Conjecture \ref{conj1} formulated in the grand-canonical setting)
to be true.
 Let us start with a brief account of experimental
evidence for these conjectures.

\begin{figure}[h]
\includegraphics[width=0.8\textwidth]{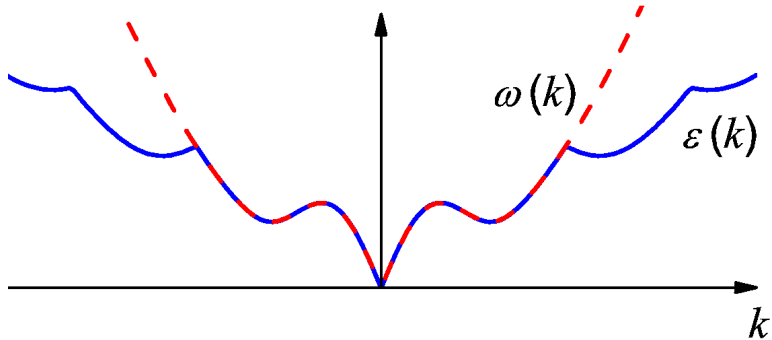}
\caption{Excitation spectrum of Helium IV}
\label{fig1.2}
\end{figure}

\begin{figure}[h]
\includegraphics[width=0.8\textwidth]{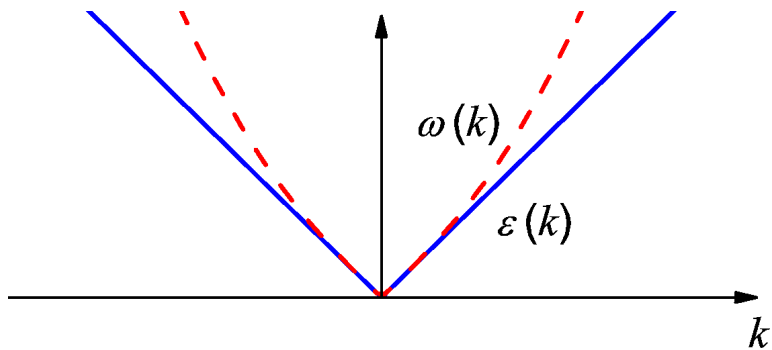}
\caption{Excitation spectrum typical for BE
 condensates of  alcalic metals}
\label{fig1.3}
\end{figure}

Theoretically,
 the cross-section for neutron scattering against a droplet of 
Helium IV at zero temperature
is
approximately proportional to
 the so-called van Hove formfactor
$S(\omega,\kk)$
(\cite{vH}, see also (\ref{formfactor})). 
$S(\omega,\kk)$ is a measure of
the density of excitations of
the Bose gas at energy $\omega$ and momentum $\kk$ at zero
temperature. Therefore, $S(\omega,\kk)$ is zero
below the curve $\kk\mapsto\epsilon^\rho(\kk)$.
It seems reasonable to suppose that more is true: $S(\omega,\kk)$
should be
nonzero everywhere above the curve  $\kk\mapsto\epsilon^\rho(\kk)$.
If in  addition
 Conjecture \ref{conj1a} is true, then the lower boundary
of the support of 
 $S(\omega,\kk)$  should satisfy
 (3), (4) and (5) of this conjecture.

To our understanding, within experimental accuracy,
 experiments on Helium IV 
at low temperatures seem to confirm
the above theoretical expectations.

Actually, experiments seem to say more than this.
At least for low momenta, one
 observes a sharp peak of $S(\omega,\kk)$
along a curve similar to $\kk\mapsto\omega(\kk)$ at Fig.
\ref{fig1.2}, see \cite{WC} and Fig 1 of \cite{M}.
 This curve is interpreted as the 
 dispersion relation of a quasiparticle 
 (elementary excitation spectrum). These quasiparticles
 are called phonons for small momenta and
rotons around the local minimum of the dispersion relation.
To our understanding, experiments indicate
that below  the subadditive hull of the 
elementary excitation
 spectrum the value of $S(\omega,\kk)$ drops down
substantially.  (In the case of
Fig. \ref{fig1.2}, this subadditive hull equals
 $\kk\mapsto\epsilon(\kk)$).

 Experiments involving excited phonon states
are usually successfully interpreted
 in terms of  multi-quasiparticle states whose
 momenta and energies  are additive \cite{M}.
This also implies that the energy of
 multi-quasiparticle states lies above
 $\kk\mapsto\epsilon(\kk)$.

A similar picture arises in the case of Bose-Einstein (BE)
 condensates of
alcalic metals. For example, the reader can consult 
Fig.  2 of \cite{SOKD}, which shows the 
quasiparticle spectrum of the BE condensate of
  ${}^{\rm 87}$Rb around zero
temperature.    Compared to Helium IV,
the main difference is the absence of the rotonic
part of the elementary excitation spectrum,  see Fig. \ref{fig1.3}.

Of course, it is difficult to interpret real experiments in terms of
rigorous statements.
 The setup that we describe in this paper does not
apply in all its details to realistic BE condensates.
First of all, both in Helium IV and alcalic metals, the potential has
typically an attractive part and a hard core. Therefore,
 strictly speaking
it does not belong 
to the class that we would like to consider in this paper. 

In the case of Helium IV, the situation is further complicated by 
the fact that the Schr\"odinger operator of the form (\ref{sch4}) is
not believed
 to describe it  adequately
 -- $3$-body interactions are
probably relevant. 
This problem does not appear in BE condensates of alcalic metals,
where apparently 
one can assume that 
 only 2-body
interactions  play the role.

BE condensates of alcalic metals
 have  a different
 conceptual problem absent in the case of Helium IV:
they do not represent
 the true ground state but only a metastable state.

\subsection{Bogoliubov approximation}

There are many theoretical physics papers devoted to the Bose gas.
 To our surprise, 
their authors usually avoid making
precise statements or conjectures about the excitation
spectrum of the Bose gas. (A notable exception is \cite{WdS}, where a
definition of the IES similar to (\ref{ies}) can be found).

 In \cite{Bog} Bogoliubov proposed
an approximation, which implies that 
the Bose gas should be described by
elementary excitations with the spectrum
\begin{equation}
\omega_\Bog^\rho(\kk)=\sqrt{\frac12\kk^2(\frac12\kk^2+2\hat
  v(\kk)\rho)}.\label{omega}\end{equation} 
Within this approximation,
 the IES equals $\epsilon_\Bog^\rho$, the subadditive hull of
$\omega_\Bog^\rho$ (see  (\ref{epsilon})), which has the properties
 described in Conjecture  \ref{conj1a}.

Note that  if we replace the potential $v$ with $\lambda v$ and the density
$\rho$ with $\rho/\lambda$ (with a positive $\lambda$), then 
neither  $\omega_\Bog^\rho$ nor $\epsilon_\Bog^\rho$  depend on
$\lambda$.
 In fact, it is natural to conjecture that  $\epsilon_\Bog^\rho$
describes the true IES in the 
weak coupling/large density limit.

More precisely, let
 $\epsilon^{\rho,\lambda}(\kk)$ be the IES for the potential
$\lambda v$. 
\begin{conjecture}
Let $d\geq2$. Then for a large class of repulsive potentials we have
\[\lim_{\lambda\searrow0}
\epsilon^{\frac\rho\lambda,\lambda} (\kk)
=\epsilon_\Bog^\rho(\kk).\]
\label{conj-1}\end{conjecture}

Note that Conjecture \ref{conj-1} is certainly wrong in dimension $d=1$,
because of (\ref{3'}).

We do not know 
 complete proofs of a statement similar to
 Conjectures \ref{conj1a}, \ref{conj-1}, as well as 
 their grand-canonical analogs
 described later on in our paper.
 We believe that to prove or
 disprove
 them would be an interesting subject for research in mathematical
 physics.

Many
 theoretical works on the energy-momentum spectrum
of the Bose gas instead of the correct
Hamiltonian $H^{L,n}$ (or its second-quantized version $H^{L}$ and the
grand-canonical version $H_\mu^{L}$) consider
 its modifications. They either replace the zero mode by a $c$-number
or drop some of the terms, or do both modifications 
\cite{Be,HP,GN,Gira,Ta,ZB}. 
 These Hamiltonians have no independent 
justification apart from being approximations to the correct Hamiltonian
 in some
uncontrolled way.  
Let us stress that in our paper we are mostly interested in the correct
 Hamiltonian and not its modifications:
 all our statements will be related either to $H^{L,n}$ or  to the
grand-canonical Hamiltonian  $H_\mu^L$ (the  second
quantization of $H^{L,n}-\mu n$).

\subsection{Organization of the paper}
\label{Organization of the paper}

The paper
 is divided into several sections and appendices. The individual
sections use sometimes slightly different notation and are devoted to
different apsects of the Bose gas.

Section \ref{Fixed number of particles} is devoted to some facts about
Bose 
gas that are naturally formulated in the canonical approach, where we start
with a definite number of
particles and go to thermodynamic limit keeping the density fixed.
The later part of the paper, where we use the grand-canonical
approach, is independent of this section.

We start with a discussion of the Galileian covariance  in a finite
box with periodic boundary consitions. We believe that this is
relevant if one wants to understand the physics of the Bose gas. Even
though what we present is elementary, we did not find most of it in
the literature.

In Subsection \ref{s2a} we discuss the case of dimension $d=1$. We prove that the
excitation
 spectrum is periodic in momentum. This fact is known to some
experts and we do not claim its discovery -- nevertheless, 
we have never seen
it explicitly stated in the literature.

In Subsection \ref{superfluidity} we describe an argument that links the
properties of the excitation spectrum to superfluid behavior. The argument
that we describe is slightly different from
 the one usually stated in the literature \cite{LL,WdS} 
-- in particular it applies to 
 systems  confined to a finite volume.

One of the most imporant quantities in 
in superfluidity is the so-called
critical velocity. There are several non-equivalent
definitions of this concept in  thermodynamic limit. We argue, that
to obtain the definition that is  relevant for
superfluidity one should first take thermodynamic limit for the
excitation spectrum, and only then  compute the critical velocity.

In Subsection \ref{Feynmann's ansatz} we present the variational ansatz due to
 Bijls \cite{Bi} and Feynmann \cite{F}  for the excitation spectrum. 

In 
Subsection \ref{Dependence of the energy on external potential}
we 
describe the (non-rigorous but interesting)
 argument due to Onsager \cite{Pr}
that indicates the phononic character of the
 excitation spectrum obtained by this ansatz.
Our presentation follows that of a recent paper \cite{WdS}.

 Section \ref{Variational approach based on squeezed states} is the central
 part of our paper. Starting with this section,
we switch to the grand-canonical setting. This means that we 
allow the number of particles to vary and we fix the chemical
potential $\mu$. 
We also use the formalism of second quantization.

In Subsection   \ref{s2b} we describe the formalism and formulate Conjecture
\ref{conj1},
the grand-canonical analog of Conjecture  \ref{conj1a}.

In Subsection
\ref{number substitution}, we discuss the Hamiltonian obtained by a $c$-number
substitution of the zero mode. We describe the theorem of Lieb, Seiringer and
Yngvason \cite{LSSY1} 
saying that this approximate Hamiltonian gives the correct energy
density in thermodynamic limit. Note that the result of \cite{LSSY1} is more
general, it concerns an arbitrary temperature. Our presentation sticks to the
zero temperature, which allows for some minor simplifications.

In Subsection \ref{s3} we describe the Bogoliubov approximation.
Its original form was formulated in the canonical setting of fixed density
in the second-quantized formalism. We follow its grand-canonical
version, which can be traced back to
 Beliaev \cite{Be} and Hugenholz - Pines
\cite{HP}, see also a recent review paper by Zagrebnov
and Bru \cite{ZB}. Even though this is a classic reasoning, our presentation
seems to be somewhat different from and 
more satisfactory than what we have seen in the
literature. Its first step is a variational problem involving coherent
states. The second step is the Bogoliubov translation and rotation adapted to
the resulting approximate ground state.
 Our reasoning does not involve the $c$-number substitution: we
treat the zero mode quantum mechanically.

One can try to improve on Bogoliubov's approximation by looking for the
minimum of the energy among translation invariant squeezed states.
To our knowledge, in the context of the Bose gas 
this idea  first appeared in the paper of 
Robinson \cite{Ro}. Robinson
considered a slightly more general class of states -- quasi-free states. He
noticed, however, that in the case  he looked at it is sufficient to
restrict to pure quasi-free states, 
which  coincide with squeezed states.
One should mention also  \cite{CS}, where 
 a variational bound on
the pressure of
 Bose gas in a positive temperature is derived by using quasi-free
 states. In the literature this approach often goes under the name of
 the Hartree-Fock-Bogoliubov method.

Only an upper bound to the ground state energy is considered in 
 \cite{Ro}.
 We go one step further: we show how this method
 can be extended to obtain upper bounds on the infimum of the energy-momentum
spectrum
 by using
 one-particle excitations over squeezed states.

 After finding the minimizing squeezed vector $\Psi$,
it is natural
 to express the Hamiltonian in the new creation/annihilation operators
 $b_\kk^*/b_\kk$, for
 which the new approximate ground state is a Fock vacuum. 
We show that
 the resulting quadratic Hamiltonian has no terms involving $b$, $b^*$, $bb$
 and $b^*b^*$. 
The Hamiltonian 
becomes
\begin{equation}
H=C+\sum_\kk D(\kk) b_\kk^*b_\kk+\hbox{terms of order 3 and 4}.\label{equ1}
\end{equation}

This new form of the Hamiltonian yields immediately an interesting estimate on
the energy-momentum spectrum. In fact, the vectors $b_\kk^*\Psi$ have
precisely the momentum $\kk$ and the energy equal to $C+D(\kk)$.
 One could ask
how good is this estimate. It turns out that it has a serious drawback. 
As we show in Subsection \ref{Thermodynamic limit of the fixed point
  equation},
 under quite general circumstances we have
$D(0)>0$.

Perhaps, this is the most important (even if negative) finding of our paper.
It implies that
at the bottom of its spectrum
1-particle excitations over a squeezed state are poor test functions for the
excitation spectrum.

 In the literature,
the existence of a gap in various approximation schemes that
 try to improve
on the original Bogoliubov's one
 has been noticed by a number of authors \cite{Gira,Ta}. However, to our
knowledge those authors did not consider the correct
Hamiltonian, but always used one of its distorted versions.

The Hilbert space of the homogeneous Bose gas can be naturally factorized into
the tensor product of an infinite family of Hilbert spaces for various values
of the momenta. 
Variational ansatzes involving translation invariant 
squeezed states, as well as particle
excitations over the squeezed states have a common feature --  they are
factorized with respect to this tensor product. We call such states {\em
  uncorrelated}. One can pose a question: how good are uncorrelated states as
variational test functions in many body problems? It seems  to us that they
have serious drawbacks -- in particular, we conjecture that they typically
give spectrum with an energy gap.

In Section \ref{Perturbative approach} we discuss approaches to the Bose gas
based on perturbation theory. We would like to treat
the coupling constant $\lambda$ as a small parameter, keeping the chemical
potential $\mu$ fixed.

In Subsection \ref{Perturbative approach based on the Bogoliubov method} we
describe a naive splitting of the Hamiltonian
into a main part and a perturbation based on the usual
Bogoliubov approach. Unfortunately, this approach seems to  fail
 because of a serious
infrared problem. 
 We also formulate Conjecture
\ref{conj-1a}, 
 which is the grand-canonical analog of Conjecture  \ref{conj-1} and is
 suggested by Bogoliubov's approximation.

In Subsection \ref{Perturbative approach based on improved Bogoliubov method} we
propose a certain systematic procedure for perturbation expansion,
which avoids the infrared problem.  This procedure uses (\ref{equ1}) as the
starting point for the expansion. 
The 3rd and 4th order term are treated as
perturbations. The advantage of this procedure is that it does
not   drop
 any terms from the Hamiltonian.
All the  works on the Bose gas based on perturbation
theory \cite{Be,HP,GN} that we know
involve the $c$-number substitution. This substitution, even if
justified 
for the energy
 density \cite{LSSY1}, is unfounded for finer quantities such  as
the infimum of the  excitation spectrum. Our perturbative procedure
does not 
involve  distorting the Hamiltonian.  Therefore, in our opinion, it is superior
from the physical point of view.

In Section \ref{A priori estimates} we describe various inequalities
 on the Bose gas that can be proved rigorously. 
 These results are consistent with the absence of the
 energy gap and the phononic form of the excitation spectrum. They are
 obtained by relatively simple methods, involving especially the so-called
 $f$-sum rule.
Our presentation is based  on
 the work of Bogoliubov \cite{Bog1}
and on results presented by Stringari
 \cite{St1,St2}.

In appendices we present some background material, to make our paper
accessible to a larger audience. In Appendix \ref{a2} we describe 
technical computations.

\subsection{Additional remarks about the  literature}

Let us make some additional remarks about the literature of the subject.
The case of dimension $d=1$ and repulsive delta interactions has been studied
in detail. Girardeau \cite{Gi} studied the case of ``infinite'' coupling (which
amounts to the Dirichlet boundary conditions). The case of an arbitrary positive
coupling constant was studied 
in \cite{LL,L2}, where arguments for
 the absence of a gap and the phononic shape of  the excitation spectrum
 in the thermodynamic limit are given. \cite{WdS} gives a full rigorous proof
 for the linearity of the excitation spectrum in thermodynamic limit for
 Girardeau's model. Note, however, that the 1-dimensional case is believed to
 be quite different from the case  $d\geq2$.

There exists a large 
literature on the energy density
of the Bose gas. The  energy density can be defined as 
\begin{equation}
e^\rho:=\lim_{L\to\infty}\frac{E^{L,n}}{V},
\label{densi}\end{equation} 
where $\rho=\frac{n}{V}$ is kept fixed. There exist derivations 
of the asymptotics of $e^\rho$ in dimension $d=3$ for small $\rho$
 going back to
\cite{BS,HP,LHY,L1}.  This asymptotics is not restricted to small potentials
-- it covers also the case of hard-core potentials.
One can show rigorously (see \cite{LSSY} and references therein)
that the leading term of this asymptotics correctly
describes the energy density. Similar results can be shown
 in dimension $d=2$.

Note that the energy density is easier to study than the infimum of the
excitation spectrum. Besides, it does not
capture some interesting physical phenomena  the
 excitation spectrum is
responsible for. Note also that  the above mentioned results  involve 
the following limit:
 the scattering length is kept fixed (this can be achieved e.g. by fixing the
 potential) 
and the density goes to zero.
In our paper we usually 
consider a
different limit: the chemical potential is kept fixed and the coupling
constant in front of the potential goes to zero. One can have 
various opinions  comparing the physical relevance
 of the two limits. In any case, the latter limit seems
 more appropriate if one wants to capture the phononic
character of the excitation spectrum.

Another direction of  rigorous 
research involves studying the so-called Gross-Pitaevski limit. Again it concerns
mostly the dimension $d=3$. The quantity that is kept fixed is $\frac{an}{L}$,
where $a$ is the scattering length and $n$ goes to infinity. The
Gross-Pitaevski limit is usually presented with a fixed $L$ and the
scattering length $a$ going to zero, which is achieved by an
appropriate scaling of the potential. Equivalently,
one can fix the potential, consider $L\to\infty$.
and scale the density  as $\rho\sim
L^{1-d}$ as $L\to\infty$. In this limit,  Lieb,
Seiringer and Yngvason have obtained a number of interesting and precise
results \cite{LSSY}.
 In particular, they are able to approximate the behavior of the
$n$-body system by a non-linear effective equation -- the Gross-Pitaevski
equation. Note, however, that in this limit it is difficult
 to say something
interesting about the excitation spectrum, because 
the density,  and hence the
speed of sound and the critical velocity,   go to zero.

The result of  \cite{LSY} (see also
Theorem 5.3 of \cite{LSSY}) can be interpreted as the positivity of
the critical velocity in finite volume. The critical velocity, as
understood in this result, 
 goes to zero in
thermodynamic limit. It is
 presumably  far too low to fully
 account for the
physical phenomenon of superfluidity, see Subsection  \ref{sub-alpha}
for a discussion.



Finally, let us note  that there exists a number of interesting rigorous
 results about Bose gas in
positive temperatures \cite{Bog1,CS,Ho,St1,St2,PSt,LSSY1}.

\vspace{0.5cm} 
\noindent{\bf Acknowledgments.} 
The research of J.D.
  is  supported in part by the grant N~N201~270135.
and was partly done during his visit to the Erwin
 Schr\"odinger Institute as a Senior Research Fellow.
A part of the work of H.C.  was also done during his
 visit to the  Erwin
 Schr\"odinger Institute.
H.C. and J.D. also acknowledge support from the Danish F.N.U. grant 
{\it Mathematical Physics and Partial Differential Equations}. 
 The research of P.Z. was also supported by the 
grant N202 022 32/0701.

The authors acknowledge interesting and useful
 discussions with J.~P.~Solovej 
J.~Yngvason and V.~Zagrebnov. 
They are also grateful to the referees of the previous versions of
the manuscript for their remarks.

\section{Canonical approach}
\label{Fixed number of particles}

In this section we will always work on a Hilbert space of fixed number of
particles. We will use the Hamiltonian
$H^{L,n}$  introduced in the introduction. To simplify the notation, 
we usually 
drop the superscripts $n,L$,  writing e.g. $H$ instead of
$H^{L,n}$. 

\subsection{Free Bose gas}

\begin{figure}[h]
\includegraphics[width=0.8\textwidth]{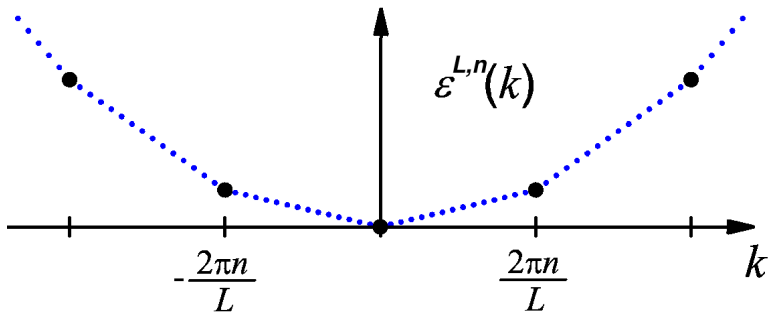}
\caption{Infimum of excitation spectrum of free Bose gas in finite volume}
\label{fig2.1}
\end{figure}

In finite volume, the momentum is restricted to 
\[\kk=
\frac{2\pi}{L}\tilde\kk,\ \ \ \tilde\kk\in\Z^d.\]
It is easy to compute exactly  
the excitation spectrum of the free Bose gas
in a finite volume, see Fig \ref{fig2.1}.
 In particular, in dimension $d=1$,
its infimum is given by
the broken line with vertices at \[
\kk=\frac{2\pi n}{L}\tilde \kk,\ \ \ 
\epsilon^{L,n}(\kk) =\frac{1}{2}\kk^2,\ \ \ \ 
\ \ \ \tilde\kk\in\Z.\]
In an arbitrary dimension, we just add the contributions from each
dimension:
\[\epsilon(\kk_1,\dots,\kk_d)=\epsilon(\kk_1)+\cdots+\epsilon(\kk_d).\]

\subsection{Galileian covariance on a box with periodic boundary
  conditions} 

\begin{figure}[h]
\includegraphics[width=0.8\textwidth]{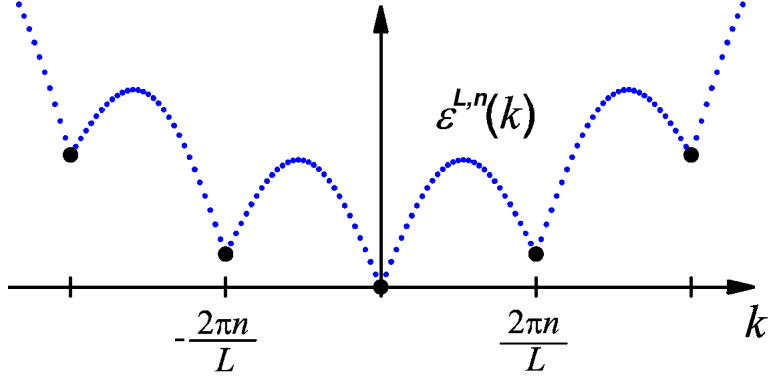}
\caption{Typical 
infimum of excitation spectrum of interacting
 Bose gas in finite volume}
\label{fig2.2}
\end{figure}

In infinite volume the Galileian covariance involves an arbitrary
value of velocity. This is not the case on a box with periodic
boundary conditions (torus), where 
 the
Galileian covariance is only rudimentary. To describe it we will
restrict ourselves to
boosts in the first coordinate.
The following operator, which we will call the {\em boost operator}
adds simultaneously velocity $\frac{2\pi}{L}$
to all particles in the direction of the first
coordinate: 
\[U_1:=\exp\left(\frac{\i  2\pi}{L}\sum_{i=1}^n \x_{i1}\right).\]
($\x_{i1}$ denotes the 1st coordinate of the $i$th particle).
Clearly, $U_1$ preserves  the domain of $H$ and
is a unitary operator on $L_\s(\Lambda)$ satisfying
\begin{eqnarray}
U_1P_1U_1^*&=&P_1- \frac{2\pi n}{L},\label{pop1a}\\
U_1HU_1^*&=&H-\frac{2\pi}{L}P_1+\frac{(2\pi)^2n}{2L^2}.
\label{pop2a}
\end{eqnarray}
($P_1$ denotes the first component of the total momentum).
Hence\begin{equation}
U_1\left(H-\frac{1}{2n}P^2\right)U_1^*=
H-\frac{1}{2n }P^2.\label{pop2a1}
\end{equation}
(\ref{pop2a1}) and (\ref{pop2a})
 imposes a severe 
restriction on the shape of the
excitation spectrum:
\begin{equation}
\sp\left(H-\frac{1}{2n}P^2\right)\label{pio}
\end{equation}
has to be invariant with respect to translations by
$\frac{2\pi n}{L}$, see Fig. \ref{fig2.2}.

\subsection{Critical velocity}

The 
critical velocity in finite volume is defined by 
\[c_{\crit}^{L,n}:=\inf_{\kk\neq0}\frac{\epsilon^{L,n}
(\kk)}{|\kk|}.\]
Recall that we are interested in thermodynamic limit, which involves
$L,n\to\infty$ with $\frac{n}{V}=\rho>0$.

 Note that for the free Bose gas we have
\[c_{\crit}^{L,n}=\frac{\pi}{L}.\]
Thus the  critical velocity is positive, but goes to zero in
thermodynamic limit.

In the interacting case we have a sequence of low energy states with
the momentum and the excitation  energy obtained by boosting the
ground state:
\[\kk=\frac{2\pi n}{L}\tilde\kk,\ \ \ \epsilon=
\frac{(2\pi)^2}{2L^2}n \tilde\kk^2.\]
where $\tilde\kk\in\Z^d$. Expressed in terms of density this gives
\[\kk=2\pi \rho L^{d-1}\tilde\kk,\ \ \ \epsilon=
\frac{(2\pi)^2}{2}\rho L^{d-2}\tilde\kk^2,\]
Therefore, in the general case the critical velocity is not greater than in the case of the free gas:
\[c_{\crit}^{n,L}
\leq\frac{\pi}{L}.\]

 In dimension $d\geq2$, the momentum of these
states escapes to infinity, so in a sense they are not visible in
thermodynamic limit.

\subsection{Bose gas in dimension $d=1$}
\label{s2a}

Bosonic gas in dimension $d=1$ 
seems to have different properties than in
higher dimensions. In particular, statement (3) of Conjecture
\ref{conj1a} should be replaced by (3)', that is

\begin{thm} $\rr\ni \kk\mapsto \epsilon^\rho(\kk)$
 is periodic with the period $2\pi\rho$. \label{pio1} \end{thm}

\begin{figure}[h]
\includegraphics[width=0.8\textwidth,clip]{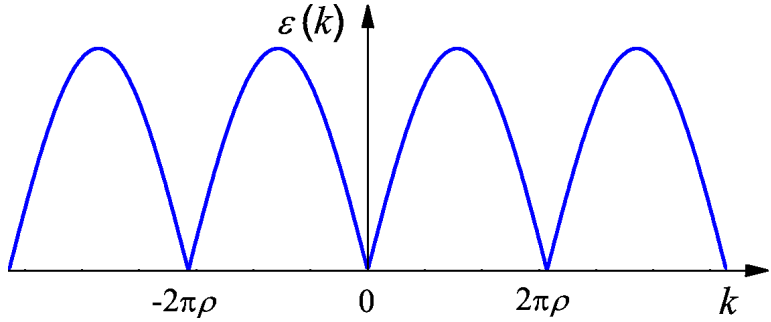}
\caption{Typical infimum of excitation spectrum of  Bose gas in
dimension $d=1$}
\label{fig2.3}
\end{figure}

Theorem  \ref{pio1}  
easily follows from the invariance of the spectrum
(\ref{pio}).

\subsection{Twisted boundary conditions}
\label{sub-alpha}

One can also consider the boost operator for an arbitrary velocity. In
fact, for $\alpha\in\R$ define the unitary operator
\[U_1(\alpha)
:=\exp\left(\frac{\i  \alpha}{L}\sum_{i=1}^n \x_{i1}\right).\]
Let $[\alpha]$ denote $\alpha\mod{}(2\pi)$. We will write
 $P_{1,[\alpha]}$,
$H_{[\alpha]}$ for the momentum and the Hamiltonian with the
boundary condition in the $1$st coordinate twisted by $[\alpha]$.
(It means that the elements of the domain of these operators 
 for $j=1,\dots,n$ satisfy
 \[\Phi\left(\x_1,\dots,\x_j-\frac12 L\e_1,\dots,\x_n\right)
=\e^{\i\alpha}\Phi
\left(\x_1,\dots,\x_j+\frac12 L\e_1,\dots,\x_n\right),\]
where $\e_1$ is the unit vector
 in the direction of the 1s coordinate).

For $\alpha$ not equal to a multiple of $2\pi$ the operator
$U_1(\alpha)$ does not preserve the domain of $H$ and $P$.
 Instead of
identities (\ref{pop1a}) and (\ref{pop2a}) we have
\begin{eqnarray}
U_1(\alpha)P_1U_1(-\alpha)
&=&P_{1,[\alpha]}- \frac{\alpha n}{L},\label{pop1a-}\\
U_1(\alpha)HU_1(-\alpha)
&=&H_{[\alpha]}-\frac{\alpha}{L}P_{1,[\alpha]}+\frac{\alpha^2n}{2L^2},
\label{pop2a-}
\end{eqnarray}

(\ref{pop1a-}) and (\ref{pop2a-}) imply
\begin{equation}
H_{[\alpha]}=U_{1}(\alpha)\left(H+\frac{\alpha}{L}P_1\right) 
U_1(-\alpha)+\frac{\alpha^2n}{2L^2}.\end{equation}
In particular, if $0\leq \alpha_0\leq\pi$, then
\[c_{\crit}\geq\frac{\alpha_0}{L}\]
if and only if
\begin{equation}
H_{[\alpha]}-E\geq\frac{\alpha^2
  n}{2L^2},\ \ \ |\alpha|\leq\alpha_0.\label{glob}
\end{equation}
(\ref{glob}) was used by Lieb, Seiringer and Yngvason as a criterion
for the positivity of critical velocity in \cite{LSY}, see also
Theorem 5.3 of \cite{LSSY}.

\subsection{Superfluidity}
\label{superfluidity}
Let us describe one
 of experiments that show superfluid properties of the  Bose
gas.

A laser beam playing the role of an optical spoon \cite{RKODKHK} 
is directed into a sample of a Bose gas at a sufficiently low
 temperature. The beam makes a rotating motion. If the velocity of
 this motion is lower than a certain critical value, then the sample heats up
 very slowly. For velocities above this critical
 value, the sample heats up much faster.

Let us try to describe an idealized mathematical model of this phenomenon,
 which
 is a version of the well known argument due to Landau described e.g. in
\cite{LaL,WdS}. 


 Since our Bose gas has periodic boundary conditions, we can make
an idealized assumption that the ``laser beam'' travels forever with a
constant velocity $\w=(w,0,\dots,0)\in\rr^d$. We will model it with a weak
travelling potential
$u(\x-t\w)$ interacting with all particles.
 Thus the system is described by the
 Schr\"odinger equation with a time-dependent Hamiltonian
\begin{equation}
\i \partial_t\Psi_t=
\left(H+\sum_{i=1}^nu(\x_i-t\w)\right)\Psi_t.\label{travo}
\end{equation}
Let us replace $\Psi_t$ with 
\[\tilde\Psi_t(\x_1,\dots,\x_n):=
\Psi_t(\x_1+t\w,\dots,\x_n+t\w)
.\]
We obtain a  Schr\"odinger equation with a time-independent Hamiltonian:
\begin{equation*}
\i \partial_t\tilde\Psi_t=
\left(H-\w P
+\sum_{i=1}^nu(\x_i)\right)\tilde\Psi_t.
\end{equation*}

\begin{figure}[h]
\includegraphics[width=0.8\textwidth]{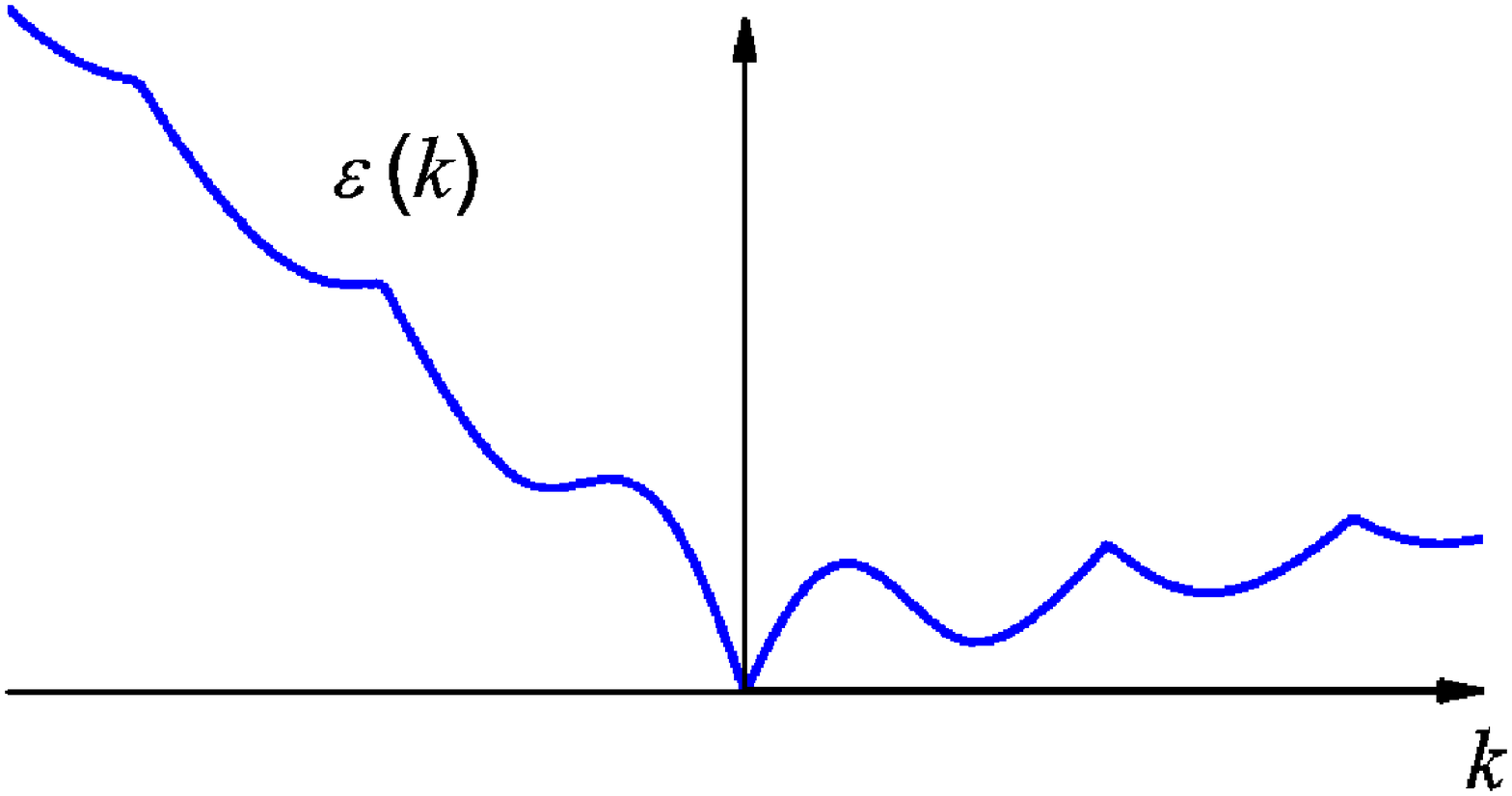}
\caption{Infimum of excitation spectrum of tilted Hamiltonian 
for velocity below $c_\crit$}
\label{fig2.4}
\end{figure}

\begin{figure}[h]
\includegraphics[width=0.8\textwidth]{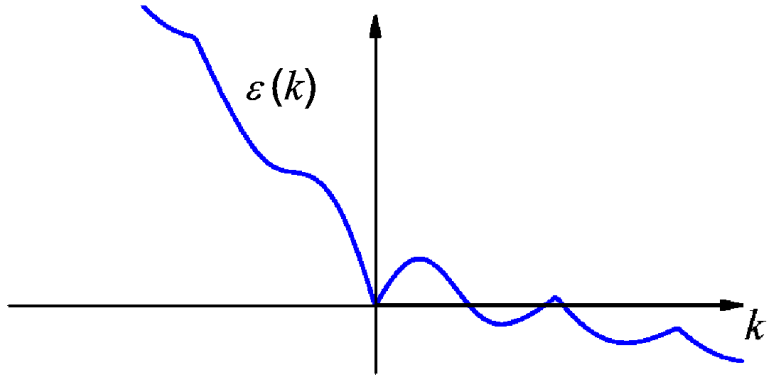}
\caption{Infimum of excitation spectrum of tilted Hamiltonian
for velocity above $c_\crit$}
\label{fig2.5}
\end{figure}

We know that the operator $H$ has a nondegenerate ground state energy
    $E$. The corresponding ground state $\Psi$ is therefore stable with
respect to small  time-independent perturbations.
We ask the question whether
 the state $\Psi$ is stable against a small
 travelling perturbation of the form (\ref{travo}).

Let us first describe a slightly dishonest version 
 of the argument for superfluidity (which actually is  usually 
found in the literature). Let $c_\crit$ denote the critical velocity. 
For $|\w|< c_\crit$,
 the excitation spectrum of the ``tilted
Hamiltonian'' $H-\w P$ looks as at Fig. \ref{fig2.4}, so
 $\Psi$ is its
ground 
state. Hence $\Psi$ hence is stable. 
For $|\w|>c_\crit$    the excitation
spectrum looks as at Fig. \ref{fig2.5}. Therefore
 $\Psi$ is not a ground
state of $H-\w P$
and its energy is close to energies of many other states.
 Hence $\Psi$ is unstable.

\begin{figure}[h]
\includegraphics[width=0.8\textwidth]{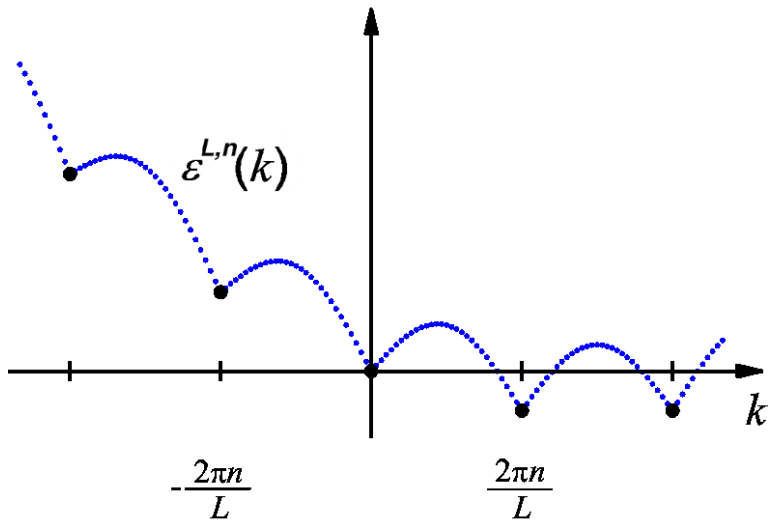}
\caption{Infimum of excitation spectrum of ``tilted Hamiltonian''
in finite volume}
\label{fig2.6}
\end{figure}

Note that we cheated a little in the above argument. The
situation that we described involved a finite volume, 
but the pictures were (as we believe)
typical for the
thermodynamic limit. 
 The actual plot 
in finite volume resembles  Figure \ref{fig2.6}.
In particular
the critical velocity in finite volume
 $c_\crit^{n,L}$ is
small and goes to zero in thermodynamic limit.
 Physical evidence seems to
indicate that the observed phenomenon of superfluidity involves much
larger velocities, which are positive in thermodynamic limit.
 Note that the Figs \ref{fig2.4}
and \ref{fig2.5} represent the excitation spectrum in thermodynamic
limit and do not show the low lying states present in finite volume,
which have a very low critical velocity.

Let us try to present a more physical argument for superfluidity.
 Suppose that 
 the Fourier transform of $u$
 is supported in the ball $|\kk|\leq R$.
Define the {\em critical velocity below the momentum $R$} as
\begin{equation}
c_{\crit,R}:=\inf\left\{\frac{\epsilon(\kk)}{|\kk|}
\:\ \ \kk\neq0,\ \ |\kk|<R\right\}.
\label{spect}\end{equation}   
If the ``tilted Hamiltonian'' $H-\w P$ has no
other eigenstates of energy close to
 $E$ and momentum less that $R$, then the state $\Psi$ will be stable
against the  perturbation (\ref{travo}), at least in the 1st order.
This is the case for $|\w|<c_{\crit,R}$.
On the other hand, for $|\w|\geq c_{\crit,R}$, we can expect many states
with energy close to $E$ and momentum less than $R$, so the stability
will be lost.

Heuristic arguments (e.g. the Bogoliubov method described later on)
suggest that in dimension $d\geq2$
 the critical velocity $c_{\crit,R}^\rho$ is
bounded away from zero even in thermodynamic limit and for an 
arbitrary 
$R$. This can be formulated in the following conjecture:

\begin{conjecture}\label{connj1}
Fix density $\rho$.
Then  \[c_{\crit,R}^\rho:=\lim\limits_{L\to\infty}c_{\crit,R}^{L,n},\ \ \frac{n}{V}=\rho,\]
exists, where $c_{\crit,R}^{L,n}$
is defined as in  (\ref{spect}). Moreover, in dimension $d\geq2$, 
\begin{equation}
\inf_{R\to\infty} c_{\crit,R}^\rho>0.\label{coonj2}
\end{equation}
\end{conjecture}

Note in parenthesis, that Conjecture \ref{connj1} is stronger than
Conjecture \ref{conj1a} (3). In fact, the left hand side of
(\ref{coonj2}) is less than or equal to
$c_\crit^\rho$
introduced in Conjecture \ref{conj1a} (3).




\subsection{Bijls--Feynmann's ansatz}

\label{Feynmann's ansatz}

For $\kk\in \frac{2\pi}{L}\zz$  set 
\begin{equation}
 N_\kk:=\sum_{i=1}^n\e^{\i\kk\x_i},\label{fyny}\end{equation}
acting on $L_\s^2(\Lambda^n)$.

It is well known that  the ground state of $H$
 is nondegenerate. Denote it by  $\Psi$.
We will write (see Appendix \ref{sec-ine})
\[\langle A\rangle:=(\Psi|A\Psi),\ \ \ \ 
\langle\langle 
A,B\rangle\rangle:=\langle A(H-E)^{-1}B\rangle
+\langle
B(H-E)^{-1}A\rangle.\]

 Note
the following identity:
\begin{equation}
\frac12[ N_\kk^{*},[H, N_\kk]]=\frac{\kk^2}{2}n.\
\label{f-sum}\end{equation}
It implies the so
 called  {\em f-sum rule}:
\begin{equation}
\frac12\langle  N_\kk^{*}(H-E) N_\kk\rangle+
\frac12\langle  N_\kk(H-E) N_\kk^*\rangle
=\frac{\kk^2}{2}n.\
\label{fsum}\end{equation}
By the reality of $\Psi$ and $H$, (\ref{fsum}) also equals $\langle N_\kk^*(H-E)N_\kk\rangle$.
Define
\begin{eqnarray}
s_\kk&:=& n^{-1}\langle N_\kk^* N_\kk\rangle;\label{chi1}\\
\chi_\kk&:=& n^{-1}\langle\langle N_\kk^*,N_\kk\rangle\rangle.\label{chi}
\end{eqnarray}
By the reality of $\Psi$ and $H$, (\ref{chi}) also equals $2n^{-1}\langle N_\kk^*(H-E)^{-1}N_\kk\rangle$.

By the Schwarz inequality, we obtain
\begin{equation}\label{sqth}
s_\kk\leq \frac12 |\kk|\sqrt{\chi_\kk}.\end{equation}

Bijls \cite{Bi} and Feynmann \cite{F} 
proposed to consider the following variational ansatz:
\[\Psi_\kk:= N_\kk\Psi/\| N_\kk\Psi\|,\]
to obtain the excitation spectrum of the Bose gas. 

\begin{thm}
We have 
\begin{eqnarray}
P\Psi_\kk&=&\kk\Psi_\kk,\\
\Psi_0&=&\Psi,\\
(\Psi_\kk|(H-E)\Psi_\kk)&=&\frac{|\kk|^2}{2s_\kk}\geq\frac{|\kk|}
{\sqrt{\chi_\kk}}
.\label{bijls}\end{eqnarray}
\end{thm}

\proof To see (\ref{bijls}) we note that
\begin{eqnarray*}
(\Psi_\kk|(H-E)\Psi_\kk)&=&
\frac{\langle N_\kk^*(H-E) N_\kk\rangle+
\langle N_\kk(H-E) N_\kk^*\rangle}
{2\langle N_\kk^* N_\kk\rangle}
\\
&=& \frac{\kk^2}{2 s_\kk} \geq \frac{|\kk|}
{\sqrt{\chi_\kk}}
\end{eqnarray*}
In the above we apply the $f$-sum rule to the numerator and used (\ref{sqth}).
\qed


It is believed that the Bijls-Feynman ansatz gives 
the correct behavior
of the excitation spectrum for low momenta, and in particular it gives
the value
of $c_\ph$, which was defined in  Conjecture \ref{connj1}
(4) \cite{St2}. Let us formulate this as 
a conjecture:

\begin{conjecture} Fix density $\rho$. Let $ \rr^d\ni\kk\mapsto
\epsilon_{\rm BF}^\rho(\kk)$ be
  defined as in  (\ref{ies}) and (\ref{begin})
with $(\Psi_\kk|(H^{L,n}-E^{L,n})\Psi_\kk)$ 
replacing $\epsilon^{L,n}(\kk)$. (In other words, 
$\epsilon_{\rm BF}^\rho(\kk)$ is a thermodynamic limit of          the
left hand side of (\ref{bijls})).
 Then 
\begin{equation}
\lim_{\kk\to0}\frac{\epsilon^\rho(\kk)}{|\kk|}=
\lim_{\kk\to0}\frac{\epsilon_{\rm BF}^\rho(\kk)}{|\kk|}.
\label{eqi2}\end{equation}
\end{conjecture}

\subsection{Speed of sound}

Recall that under broad conditions we are able to prove the existence
of the energy density (\ref{densi}), which 
for typographical reasons in this subsection we will be written
 $e(\rho)$ instead of $e^\rho$.

Another important macroscopic parameter,
which in
principle can be measured experimentally,
is the speed of sound, denoted $c_\s$. It is related to the energy
density by
\begin{equation}
c_\s=\sqrt{\rho e''(\rho)}\label{sppe}\end{equation}
 (see e.g. Appendix \ref{Speed of sound}).

It is believed that for low momenta 
\begin{equation}
\lim_{\kk\to0}\frac{s_\kk}{|\kk|}=\frac{1}{2c_\s},
\label{eqi1}\end{equation} and hence
$(\Psi_\kk|(H-E)\Psi_\kk)\sim c_\s|\kk|$.
 If this is 
the case and if the speed of sound is nonzero, then
 the excitation spectrum given by the Bijls-Feynman ansatz is
 phononic, that is the limit on the right hand side of (\ref{eqi2})
 exists and equals $c_\s$.

\subsection{Relation between the speed of sound and $\chi_\kk$}
\label{Dependence of the energy on external potential}

Instead of arguing for (\ref{eqi1}),
 it seems easier to justify the relation
 \begin{equation}
\lim_{\kk\to0}\chi_\kk=\frac1{c_\s^2}\label{spee}\end{equation}
In this subsection we describe a series of 
heuristic arguments indicating that
 (\ref{spee}) holds in thermodynamic limit.
Given (\ref{spee}) and if the speed of sound is nonzero,  (\ref{bijls})
gives a phononic lower bound on 
 $(\Psi_\kk|(H-E)\Psi_\kk)$.
 This inequality is attributed to Onsager
 \cite{Pr}.
Our
presentation follows that of  \cite{WdS}.

For a given vector $\Phi$ of norm one its 1-particle density is defined as
\[\rho_\Phi(\x):=n\int\cdots\int\bar{\Phi
(\x,\x_2,\dots\x_n)}\Phi(\x,\x_2,\dots,\x_n)
\d\x_2,\dots\d\x_n.\] 
If $\Phi=\Psi$, the ground state of $H$, then $\rho_\Psi$ is a constant equal
to $\rho$. 

The following claim seems to be widely accepted:
\begin{conjecture}
If $L$ is
large and $\Phi$ is a
slowly varying vector close in some sense to the ground state $\Psi$, then
\begin{equation}
(\Phi|H\Phi)\approx\int e(\rho_\Phi(\x))\d
  \x.\label{eqna1}\end{equation} 
\label{coni1}\end{conjecture}

The next claim follows from (\ref{eqna1})
by an application of a perturbation argument:

\begin{conjecture}
Let $u$ be a periodic function 
on $\rr^d$ (or its restriction to $\Lambda$, where we assume that $L$ is a
multiple of the period of $u$). Assume that the mean value of $u$ is
zero and that $u$ varies slowly. Then for large $L$ we have
\begin{eqnarray}
\frac{1}{Ve''(\rho)}\int u(\x)^2\d \x
&=&\frac{2}{V}\left\langle \sum_i u(\x_i)(H-E)^{-1}\sum_i u(\x_i)\right\rangle.
\label{perti3}\end{eqnarray}
\label{coni}\end{conjecture}

Let us give an argument in favor of Conjecture \ref{coni}. 
Consider the perturbed Hamiltonian 
\begin{equation}
H_\tau:=H+\tau
\sum_i u(\x_i).\end{equation}
Let $E_\tau$ be the ground state energy of $H_\tau$. By the usual perturbation
theory, 
\begin{eqnarray}
\frac{\d}{\d \tau}(E_\tau-E)&=&0,\label{perti1}\\
\frac{\d^2}{\d \tau^2}(E_\tau-E)
&=&-2\left\langle \sum_i u(\x_i)(H-E)^{-1}\sum_i
u(\x_i)\right\rangle.\label{perti2} 
\end{eqnarray}

Applying the principle (\ref{eqna1}), we obtain
\begin{eqnarray}
E_\tau
&=&\inf\left\{(\Phi|H\Phi)+\tau\int\rho_\Phi(\x) u(\x)\d\x\ :\
\|\Phi\|=1\right\}\nonumber
\\
&\approx&\inf\left\{\int e(\rho_1(\x))\d\x+\tau\int\rho_1(\x)u(\x)\d\x\right\}
\label{asf1}\\
&\approx
&\nonumber
 Ve(\rho)\\\label{asf2}
&&+\inf\left\{\int\frac{ e''(\rho)}{2}(\rho_1(\x)-\rho)^2\d\x
+\tau\int(\rho_1(\x)-\rho)u(\x)\d\x\right\},
\end{eqnarray}
where in (\ref{asf1}) and (\ref{asf2}) we minimize over positive functions $\rho_1$ such
that
$\int\rho_1(\x)\d\x=n$.
The minimum of  (\ref{asf2}) is attained at
\[\rho_1(\x):=\rho-\frac{\tau}{e''(\rho)} u(\x).\]
Therefore,
\begin{eqnarray*}
E_\tau&\approx
&
Ve(\rho)-\int\frac{\tau^2}{2 e''(\rho)}u(\x)^2\d\x.
\end{eqnarray*}
Invoking (\ref{perti2}) we obtain (\ref{perti3}).

Now  (\ref{spee}) will follow from (\ref{sppe}) and
the following claim:
\begin{conjecture}
$\chi_\kk$ is well defined in thermodynamic
limit and satisfies
\begin{eqnarray} \lim_{\kk\to0}\chi_\kk=\frac{1}{e''(\rho)\rho}.
\label{clai}\end{eqnarray}
\end{conjecture}

Let us justify (\ref{clai}). We assume that $L$ is large. Clearly,
\begin{eqnarray}
\langle N_\kk^*(H-E)^{-1}N_\kk\rangle&=&
\left\langle\sum_i\cos(\kk\x_i)(H-E)^{-1}\sum_j\cos(\kk\x_j)\right\rangle
\nonumber\\
&&\!\!\!+\left\langle\sum_i\sin(\kk\x_i)(H-E)^{-1}\sum_j\sin(\kk\x_j)\right\rangle
\nonumber\\
&&\!\!\!
+2\Im \left\langle\sum_i\sin(\kk\x_i)(H-E)^{-1}\sum_j
\cos(\kk\x_j)\right\rangle
\label{eqna2}.\end{eqnarray}
The ground state 
$\Psi$ is a real vector.
 The Hamiltonian $H$, and hence also $(H-E)^{-1}$, is a real operator.
Therefore, the last term in
(\ref{eqna2}) is zero. 

 We note that 
$\cos\kk \x$ and $\sin\kk\x$ are slowly varying for small $\kk$. 
Therefore,
we can apply (\ref{perti3}) and obtain
\begin{eqnarray*}
\frac{2}{V}\langle N_\kk^*(H-E)^{-1}N_\kk\rangle&\approx&
\frac{1}{Ve''(\rho)}\int(\cos^2(\kk\x)+\sin^2(\kk\x))\d \x\\
&=&\frac{1}{e''(\rho)}.
\end{eqnarray*}

\section{Bogoliubov approach in the grand-canonical setting}
\label{Variational approach based on squeezed states}

\subsection{Grand-canonical approach to the Bose gas}
\label{s2b}

As realized by Bogoliubov \cite{Bog}, even if one is interested in properties of the Bose
 gas with a fixed but large number of particles, it is convenient
 to use the second quantized 
description of the system, 
 allowing for an arbitrary number of particles.

An additional
 reformulation of the problem 
was noted already by Beliaev \cite{Be}, Hugenholz - Pines
\cite{HP} and others. 
Instead of studying the Bose gas in the canonical formalism, fixing the 
density, it is  mathematically  more
convenient to use the grand-canonical formalism and fix 
the chemical potential. Then one can pass from the chemical potential to
the density by the Legendre transformation. 

More precisely, for 
 a given chemical potential $\mu\geq0$
on the symmetric Fock space
 \[\Gamma_\s(L^2(\Lambda)):=\mathop{\oplus}\limits_{n=0}^\infty 
L_\s^2(\Lambda^n),\]
we define
 the grand-canonical
Hamiltonian 
\begin{eqnarray*}
H_\mu^L&:=&\mathop{\oplus}\limits_{n=0}^\infty\left( H^{L,n}-\mu n\right)\\
&=&\int a_\x^* \bigl(-\frac12\Delta_\x-\mu\bigr) a_\x\d \x\nonumber\\
&&+\frac12\int\int a_\x^*a_\y^* v^L(\x-\y)a_\y a_\x\d
\x\d \y.
\end{eqnarray*}
The second quantized
momentum and number operators are defined as
\begin{eqnarray*}N^L&:=&\mathop{\oplus}\limits_{n=0}^\infty n=\int a_\x^*a_\x\d\x,\\
P^L&:=&
\mathop{\oplus}\limits_{n=0}^\infty P^{n,L}=-\i\int a_\x^*\nabla_\x^La_\x\d\x
.\end{eqnarray*}

It is convenient to pass to the momentum representation:
\begin{eqnarray}\label{2B}
H_\mu^L 
&=&\sum_{\kk}(\frac{1}{2}\kk^2 -\mu)a^*_{\kk}
a_{\kk}\nonumber \\
&&+\frac{1}{2V}\sum_{\kk_1,\kk_2,\kk_3,\kk_4}\delta(\kk_1+\kk_2-\kk_3-\kk_4){\hat v}(\kk_2-\kk_3)
a^*_{\kk_1}a^*_{\kk_2}a_{\kk_3}a_{\kk_4},\nonumber 
\\
N^L&=&\sum_{\kk}a^*_{\kk}a_\kk,\nonumber\\
P^L&=&\sum_{\kk}\kk a^*_{\kk}a_\kk.\nonumber
\end{eqnarray}
where we used (\ref{interakt}) to replace $v^L(\x)$ with
 the Fourier coefficients  $\hat v(\kk)$.
 Note that ${\hat v}(\kk)=\hat
 v(-\kk)$,
and $a_\x=V^{-1/2}\sum_\kk \e^{\i \kk\x}a_\kk$.

The ground state energy in the grand-canonical approach is defined as
\begin{equation}\label{lagrtra}
E_\mu^L={\rm inf\; sp}H_\mu^L=\inf_{n\geq 0}(E^{L,n}-\mu n).
\end{equation} 
$E_\mu^L$ is a decreasing concave function of $\mu$. To go back to the
canonical condition (fixed number of particles) one uses
\begin{eqnarray}
n&=&-\partial_\mu E_\mu^L.\label{rho}\end{eqnarray}

Both $E^{L,n} $ and $E_\mu^L$ are finite for a large class of potentials, which
follows from a simple and well-known rigorous result, which we state below.

\begin{thm}
Suppose that $\hat v(\kk)\geq0$, $\hat v(0)>0$ and $v(0)<\infty$. Then
$H^{L,n}$ and $H_\mu^L$ are bounded from below and
\begin{eqnarray}E^{L,n}&
\geq &\frac{\hat{v}(0)}{2V} n^2-\frac{v(0)}{2}n,\label{result1}\\[3mm]
E_\mu^L&\geq&- V\frac{(\frac12v(0)+\mu)^2}{2\hat v(0)}
.\label{result}\end{eqnarray}\label{thm}
\end{thm}

\proof Let us drop the subscript $L$. Set
\begin{eqnarray}\label{densop}
 N_\q &:=&\sum_\kk
 a_{\kk+\q}^*a_\kk=\int\e^{\i\q\x}a_\x^*a_\x\d\x,\end{eqnarray}
(which is the second quantized version of (\ref{fyny})).
Then, by a simple commutation, using that
 $N_0=N$ and that $\hat{v}(\kk)$ is non-negative, we obtain
\begin{eqnarray}
\nonumber
H_\mu &\geq &-\mu N
+\frac{1}{2V}\sum_{\kk_1,\kk_2,\kk_3,\kk_4}\delta(\kk_1+\kk_2-\kk_3-\kk_4){\hat
  v}(\kk_2-\kk_3) 
a^*_{\kk_1}a^*_{\kk_2}a_{\kk_3}a_{\kk_4}\\
&=&\frac{1}{2V}\sum_{\kk}\hat{v}(\kk)
N_\kk^{*}N_\kk-(\mu+\frac{v(0)}{2})N \nonumber\\
&
\geq& \frac{\hat v(0)}{2V}N^2-(\mu+\frac{v(0)}{2})N.\label{densop2}
\end{eqnarray}
Setting $\mu=0$ and $N=n$ in (\ref{densop2}) we obtain
 (\ref{result1}). Minimizing  (\ref{densop2}) over $N$
 proves (\ref{result}). \qed

 For $\kk\in  \frac{2\pi }{L}\zz^d$ 
let $H_\mu^L(\kk)$ denote the Hamiltonian $H_\mu^L$ restricted
 to the space $P=\kk$. The {\em IES in the box} is efined as
\begin{align}\label{defexcit}
\epsilon_\mu^L(\kk):=\inf\sp H_\mu^L(\kk).
\end{align}


For $\kk\in\rr^d$ we define the {\em IES at the thermodynamic
  limit} 
\begin{align}\label{defexclim}
\epsilon_\mu(\kk):=\sup_{\delta >0}\left(\liminf_{L\to\infty}\left(\inf_{\kk'\in
      \frac{2\pi}{L}\zz^d,\; |\kk-\kk'|<\delta}
\epsilon_\mu^L(\kk')\right)\right ). 
\end{align}



Throughout most of 
our paper, the chemical potential $\mu$ is considered to be the natural
parameter of our problem. 
It often can be proven 
that the energy density exists in thermodynamic
limit
for a fixed $\mu\geq0$
\begin{equation}e_\mu:=\lim_{L\to\infty}\frac{E_\mu^L}{V}.\end{equation}
and is related to $e^\rho$ of (\ref{densi}) by the  Legendre transformation
\[e_\mu=\inf_\rho\{e^\rho-\mu\rho\}.\]
By definition,
 $e_\mu$ is decreasing and concave. Hence it is differentiable almost
everywhere. At the points of differentiability, we can pass from
the  grandcanonical  to the canonical approach by
\begin{equation}
-\partial_\mu
e_{\mu}=\rho\label{begino}\end{equation}
   At the points where (\ref{begino})
has a 
 unique solution $\mu(\rho)$, we should be able to relate
the canonical and the grandcanonical IES:
\begin{conjecture}
 $\epsilon^\rho(\kk)=\epsilon_{\mu(\rho)}(\kk)$.
\end{conjecture}

The following proposition is one of few rigorous facts that can be
easily shown about the IES

\begin{prop}\label{prop1} At zero total momentum, the excitation
  spectrum has a global minimum where it equals zero:
$\epsilon^{L,n}(0)=\epsilon^\rho(0)=0 $ and
$\epsilon_\mu^L(0)=\epsilon_\mu(0)=0$. 
\end{prop} 
\noindent{\bf Proof.}
Each $E^{L,n}$ is a non-degenerate eigenvalue of $H^{L,n}$,
 and $H^{L,n}$ commutes with the
total momentum and space inversion. Thus each $E^{L,n}$ 
corresponds to zero total 
momentum, and hence by
\eqref{lagrtra} so does $E_\mu^L$. Hence $\epsilon^{L,n}(0)
=\epsilon_\mu^L(0)=0$.
\qed 

Let us now formulate the conjectures about $\epsilon_\mu(\kk)$ (analogous to
the  Conjecture \ref{conj1a} about $\epsilon^\rho(\kk)$):

\begin{conjecture}\label{conj1}
We expect the following statements to hold true: 
\begin{enumerate} 
\item The map $\rr^d\ni \kk\mapsto \epsilon_\mu(\kk)\in \rr_+$ is continuous.
\item Let $\kk\in{\mathbb R}^d$. If $L\to\infty$,
  $\kk_L\in\frac{2\pi}{L}\zz^d$, and 
 $\kk_L\to \kk$, we have that $\epsilon_\mu^{L}(\kk_L)\to
  \epsilon_\mu (\kk)$.
\item If $d\geq2$, then 
$\inf_{\kk\neq0}\frac{\epsilon_\mu(\kk)}{|\kk|}=:c_\crit>0$.
\item The limit
  $\lim_{\kk\to0}\frac{\epsilon_\mu(\kk)}{|\kk|}=:c_\ph>0$ exists.
\item $\kk\mapsto\epsilon_\mu(\kk)$ is subadditive.
\end{enumerate}
\end{conjecture}

\subsection{$c$-number substitution}
\label{number substitution}
One of the steps of the
  Bogoliubov method consists in  replacing the operators 
$a_\0^*$, $a_\0$  with $c$-numbers:
\begin{equation}
a_\0^*a_\0\approx |\alpha|^2,\ \ \ a_\0\approx\alpha,\ \ \  a_\0^*\approx \bar\alpha.
\label{alpha}\end{equation}
This means replacing $H_\mu^L$  by 
\begin{eqnarray}\label{c-number}
H_\mu^L(\alpha)
&=& \frac{\hat v(0)}{2V}|\alpha|^4-\mu|\alpha|^2+
{\sum_\kk}'\left(\frac12\kk^2-\mu
+\frac{\hat v(0)+\hat v(\kk)}{V}|\alpha|^2\right)a_\kk^*a_\kk
\\\nonumber&&+{\sum_\kk}'\left(\frac{\hat v(\kk)\alpha^2}{2V}a_\kk^*a_{-\kk}^*
+\frac{\hat v(\kk)\bar\alpha^2}{2V}a_\kk a_{-\kk}\right)\\&&\nonumber
+{\sum_\kk}'\bar\alpha\frac{\hat v(\kk_1)+\hat
  v(\kk_1+\kk_2)}{2V}a_{\kk_1+\kk_2}^*a_{\kk_1}a_{\kk_2}
\\\nonumber&&+
{\sum_\kk}'\alpha\frac{\hat v(\kk_1)+\hat
  v(\kk_1+\kk_2)}{2V}a_{\kk_1}^*a_{\kk_2}^* a_{\kk_1+\kk_2}\\&&
+\frac{1}{2V}\mathop{{\sum}'}\limits_{\kk_1,\kk_2,\kk_3,\kk_4}
\delta(\kk_1+\kk_2-\kk_3-\kk_4){\hat v}(\kk_2-\kk_3)
a^*_{\kk_1}a^*_{\kk_2}a_{\kk_3}a_{\kk_4} 
 . \nonumber
\end{eqnarray}
Here ${\sum_\kk}'$ denotes the sum over all
$\kk\in\frac{2\pi}{L}\backslash\{0\}$. 
Note that $H_\mu^L(\alpha)$ is the Wick symbol of the operator $H^L$ with respect
to the zero mode. It is easy to compute also its anti-Wick symbol
\[\tilde H_\mu^L(\alpha)=H_\mu^L(\alpha)-\frac{2\hat v(0)}{V}|\alpha|^2
+\frac{\hat v(0)}{V}+\mu
-
{\sum_\kk}'\frac{\hat v(0)+\hat v(\kk)}{V}a_\kk^*a_\kk.\]
(See e.g. Appendix \ref{Wick and anti-Wick symbol} for the definitions and
basic properties of Wick and anti-Wick symbols).

The following theorem is  due to Lieb, Seiringer and Yngvason
\cite{LSSY1}. 

\begin{thm}
Assume that the energy density $e_\mu$ exists.
 Assume also that $\hat v_\kk$ is bounded.
Then
\begin{equation}
e_\mu=\lim_{L\to\infty}V^{-1}
\inf\{\inf\sp H_\mu^L(\alpha)\ :\ \alpha\in\cc\}.\label{asso}\end{equation}
Thus we can replace $H_\mu^L$ with 
the Hamiltonian $H_\mu^L(\alpha)$ when computing the energy
density.
\end{thm}

\proof The anti-Wick symbol of the number operator $N^L$ with respect to the
zero mode is
\[\tilde N^L(\alpha)=|\alpha|^2-1+{\sum_\kk}'a_\kk^*a_\kk.\]
Note that for $\varphi:=\sup_\kk\hat
v(\kk)$,
\[H_\mu^L(\alpha)-\tilde H_\mu^L(\alpha)\leq
\frac{2 \varphi}{V}\tilde N^L+\frac{\hat v(0)}{V}-\mu.\]
Now
\begin{eqnarray*}
\inf\sp H_\mu^L
&\leq&\inf\left\{\inf\sp H_\mu^L(\alpha)\ :\ \alpha\in\cc\right\}\\
\\&\leq&
\inf\left\{\inf\sp \tilde H_\mu^L(\alpha)+\frac{2\varphi}{V}\tilde N^L
(\alpha)
\ :\ \alpha\in\cc\right\}+\frac{\hat v(0)}{V}-\mu
\\
&\leq&\inf\sp H_{\mu-\frac{2\varphi}{V}}^L+\frac{\hat v(0)}{V}-\mu\\
&\leq&\inf\sp H_{\mu_1}^L+\frac{\hat v(0)}{V}-\mu,
\end{eqnarray*}
where in the last inequality $\mu_1<\mu$ and $V$ is large enough.
Dividing both sides by $V$ and letting $L\to\infty$ we obtain
\begin{equation}
e_\mu\leq \lim_{L\to\infty}V^{-1}
\inf\{\inf\sp H_\mu^L(\alpha)
\ :\ \alpha\in\cc\}\leq e_{\mu_1}
.\label{asso1}\end{equation}
Now $[0,\mu]\ni\mu_1\mapsto e_{\mu_1}$ is  a finite concave
    function, hence it  is continuous, which
    implies 
(\ref{asso}). \qed

\subsection{Bogoliubov method}
\label{s3}

Let us describe a version of the Bogoliubov approximation adapted
to the grand-canonical approach. A similar discussion can be found  
 e.g.  in the review of Zagrebnov-Bru \cite{ZB}. 

In what follows we will always use the grand-canonical approach. We will drop
$\mu$ from $H_\mu^L$,  $\epsilon_\mu^L(\kk)$, etc.

For $\alpha\in{\mathbb C}$, we define  the displacement or Weyl 
operator of the zeroth mode:
\begin{equation}
W_\alpha:=\e^{-\alpha a_\0^*+\bar\alpha a_\0},\label{weyl}\end{equation}
and the corresponding coherent vector
$\Omega_\alpha:=W_\alpha^*\Omega$. Note that $W_\alpha$ is the only
Weyl operator commuting with the momentum, and hence $\Omega_\alpha$
is the only coherent vector of momentum zero.


Let us
apply the ``Bogoliubov translation'' to the zero mode of 
$H^L$. This means making the substitution
\begin{eqnarray}a_\0=\tilde a_\0+\alpha,& a_\0^*=\tilde
  a_\0^*+\bar\alpha,\nonumber
\\
a_\kk=\tilde a_\kk,& a_\kk^*=\tilde a_\kk^*,&\kk\neq0.\label{trans}
\end{eqnarray}
Note that
\[\tilde a_\kk=W_\alpha^* a_\kk W_\alpha,\ \ \ \tilde a_\kk^*=W_\alpha^* a_\kk^* W_\alpha,\]
and thus the operators with and without tildes
 satisfy the same
commutation relations.  In
addition, the  annihilation operators with tildes kill
 the ``new vacuum'' $\Omega_\alpha$.


 For notational simplicity, in what follows we drop the
 tildes and we obtain
\begin{eqnarray}\label{3B2}
 H^L  &=&- \mu |\alpha|^2 +\frac{{\hat v}(0)}{2V} |\alpha|^4
\\ \nonumber
&+ &  \left(\frac{{\hat v}(0)}{V}|\alpha|^2 -  \mu \right) (\bar\alpha a_\0 + \alpha a_\0^* )
\\ \nonumber
&+& \sum_\kk \left(\frac{1}{2}\kk^2-\mu + \frac{\left({\hat v}(0)+{\hat
    v}(\kk)\right)}{V} |\alpha|^2\right)a^*_{\kk}
a_{\kk} 
\\ \nonumber
&+&  \sum_{\kk} \frac{{\hat v}(\kk)}{2V}
\left(\bar\alpha^2a_\kk a_{-\kk} +\alpha^2a_\kk^*a_{-\kk}^* \right)
\\ \nonumber
&+&  \sum_{\kk,\kk'}
 \frac{{\hat v}(\kk)}{V}
(\bar\alpha a_{\kk+\kk'}^*a_\kk a_{\kk'} + \alpha a_\kk^*a_{\kk'}^*a_{\kk+\kk'})
\\ \nonumber
&+&\sum_{\kk_1,\kk_2,\kk_3,\kk_4}\delta(\kk_1+\kk_2-\kk_3-\kk_4)\frac{{\hat v}(\kk_2-\kk_3)}{2V}
a^*_{\kk_1}a^*_{\kk_2}a_{\kk_3}a_{\kk_4}.\nonumber 
\end{eqnarray}

The expectation value of the state given by $\Omega_\alpha$ 
equals the
constant term of  (\ref{3B2}), that is
\begin{equation}(\Omega_\alpha|H^L\Omega_\alpha)=- \mu |\alpha|^2 +\frac{{\hat
    v}(0)}{2V} |\alpha|^4. \label{expo}\end{equation}
(\ref{expo}) is minimized for $|\alpha|^2 = \mu \frac{V}{{\hat
    v}(0)}$. 
 This choice kills also
the linear term on the second line of  (\ref{3B2}).

Let us choose $\alpha$ so as to minimize (\ref{expo}). This means,
 we choose  $\e^{\i\tau}$ and set
 $\alpha=\e^{\i\tau}\frac{\sqrt{V\mu}}{\sqrt{\hat v(0)}}$. Then the
Hamiltonian becomes
\begin{eqnarray} H^L
 &:=&-  V\frac{ \mu^2}{2{\hat v}(0)} 
\label{nonum}\\ \nonumber 
&&+{\sum_\kk} \left(\frac{1}{2}\kk^2 +{\hat
    v}(\kk)\frac{\mu}{{\hat v}(0)} \right)a^*_{\kk} 
a_{\kk} \\\nonumber&&+ {\sum_\kk}  {\hat v}(\kk)\frac{\mu}{2 {\hat v}(0)}
\left(\e^{-\i2\tau}a_\kk a_{-\kk} +
\e^{\i2\tau}a_\kk^*a_{-\kk}^* \right)\\\nonumber
&&+
 \sum_{\kk,\kk'}
 \frac{{\hat v}(\kk)\sqrt{\mu}}{\sqrt{\hat v(0)V}}
(\bar\e^{\i\tau} a_{\kk+\kk'}^*a_\kk a_{\kk'} + \e^{\i\tau}
 a_\kk^*a_{\kk'}^*a_{\kk+\kk'})\\\nonumber
&&+
\sum_{\kk_1,\kk_2,\kk_3,\kk_4}\delta(\kk_1+\kk_2-\kk_3-\kk_4)\frac{{\hat v}(\kk_2-\kk_3)}{2V}
a^*_{\kk_1}a^*_{\kk_2}a_{\kk_3}a_{\kk_4}.
\end{eqnarray}

(Note that we have  made no approximation yet).
The first 3 lines of (\ref{nonum}) form a quadratic Hamiltonian,
which will be denoted by
 $H_{\Bog}^L$.
Now let us make the assumption that
 $H_{\Bog}^L$ can be treated as an approximation to $H^L$. For
a
possible justification for this approximation see Subsection
\ref{Perturbative approach based on the Bogoliubov method}.

It is easy to find the excitation spectrum of  $H_{\Bog}^L$. To
this end,
for $\kk\neq0$ we
 make the substitution
  \begin{eqnarray}
&a_\kk^*=c_\kk b_\kk^*-\bar s_\kk b_{-\kk},&
a_\kk=c_\kk b_\kk-s_\kk b_{-\kk}^*,\label{rota}\end{eqnarray}
Let $\theta=(\theta_\kk)$ be a sequence such that $\theta_\0=0$ and
\[c_\kk:=\cosh|\theta_\kk|,\ \  s_\kk:=-\frac{\theta_\kk}{|\theta_\kk|}
\sinh|\theta_\kk|.\]
Introduce the unitary operator
\begin{equation}
U_{\theta}:=\prod_\kk\e^{-\frac12
\theta_\kk a_\kk^*a_{-\kk}^*+\frac12\bar
    \theta_\kk a_\kk a_{-\kk}}.\label{uthe}\end{equation}
Note that
\[U_\theta^*
a_\kk
U_\theta=b_\kk,\ \
\ \ 
U_\theta^*
a_\kk^*U_\theta
=b_\kk^*,\]
and hence $a_\kk,a_\kk^*$ satisfy the same commutation relations as
$b_\kk,b_\kk^*$. Note also that we have
$s_\kk=s_{-\kk}$ and $c_\kk =c_{-\kk}=\sqrt{1+\bar s_\kk s_\kk}$.

The zero mode has to be treated separately.
Let us introduce the operators $p_0$ and $x_0$ which are defined as 
\begin{eqnarray}
p_0 = \frac{1}{\sqrt{2}} (\e^{\i\tau} a^*_\0 + \e^{-\i\tau} a_\0  ),  \ \ \ \ \ x_0 =
\frac{\i}{\sqrt{2}} (- \e^{\i\tau} a^*_\0+ \e^{-\i\tau} a_\0  ).
\end{eqnarray}
They
 are self-adjoint
 operators and satisfy the commutation relation $[x_0,p_0] = \i$. 
As we can see they are the "momentum" and "position" of the zero'th mode.

We choose the Bogoliubov rotation that  kills double creators and
annihilators, which   amounts to
\begin{eqnarray}
s_\kk&=&\frac{\alpha}{\sqrt2|\alpha|}\left(\left(1-\left(\frac{\hat
  v(\kk)\frac{\mu}{\hat
  v(0)}}{\frac12\kk^2+\hat
  v(\kk)\frac{\mu}{\hat
  v(0)}}\right)^2\right)^{-1/2}-1\right)^{1/2}. \label{sk4}\end{eqnarray} 
and $c_\kk=\sqrt{1+|s_\kk|^2}$. We obtain
\begin{eqnarray}
H_{\Bog}^{L}&=&E_{\Bog}^{L}+\mu  p_0^2
+{\sum_\kk}' \omega_{\Bog}(\kk) b_\kk^*b_\kk,\label{elem}
\end{eqnarray}
where
 the elementary excitation spectrum is
\begin{equation}
\omega_{\Bog}(\kk)=
\sqrt{\frac12 \kk^2(\frac12 \kk^2+2\hat v(\kk)\frac{\mu}{\hat v(0)})}.
\label{omega1}\end{equation}
and the energy is
\begin{eqnarray}
E_{\Bog}^{L}&=
& - V\frac{ \mu^2}{2{\hat v}(0)} -{\sum_\kk}\frac12
\left(\bigl(
\frac12\kk^2+\hat v(\kk)\frac{\mu}{\hat v(0)}\bigr)-\omega_{\Bog}(\kk)\right),\end{eqnarray}
(where the sum above includes the zero mode again).

The IES of $H_{\Bog}^L$ is given by
\begin{equation}
\epsilon_{\Bog}
(\kk)=\inf\{\omega_{\Bog}(\kk_1)+\cdots+\omega_{\Bog}(\kk_n)\ :\ 
\kk_1+\cdots+\kk_n=\kk,\ \ n=1,2,\dots\}.\label{epsilon1xs}\end{equation}

Note that $\omega_{\Bog}(\kk)$ and  $\epsilon_{\Bog}(\kk)$ are
so far restricted to $\frac{2\pi}{L}\Z^d$. But these functions are
well defined also for all values $\kk\in\R^d$. In fact, the
thermodynamic limit of the IES of $H_{\Bog}^{L}$ is simply
$\R^d\ni\kk \mapsto \epsilon_{\Bog}(\kk)$.

We have  (in any dimension)
\begin{enumerate}\item
  $\inf_{\kk\neq0}\frac{\omega_{\Bog}(\kk)}{|\kk|}=\inf\sqrt{\frac12(\frac12\kk^2+2\frac{\hat
      v(\kk)\mu}{\hat v(0)})}
= :c_{\crit,\Bog}>0$;
\item
  $\lim_{\kk\to0}\frac{\omega_{\Bog}(\kk)}{|\kk|}=\sqrt{\mu}=:c_{\ph,\Bog}>0$. 
\end{enumerate}
Therefore, by Theorem \ref{subba3} (1) and (2) we have
\begin{enumerate}\item $\inf_{\kk\neq0}\frac{\epsilon_{\Bog}(\kk)}{|\kk|}
=c_{\crit,\Bog}$;
\item $\lim_{\kk\to0}\frac{\epsilon_{\Bog}(\kk)}{|\kk|}=c_{\ph,\Bog}$.
\end{enumerate}
Thus,   $\epsilon_{\Bog}(\kk)$ 
 has all the properties 
described in Conjecture \ref{conj1}.

We can also compute that for small $|\kk|$
\begin{equation}
s_\kk\approx \frac{\e^{\i\tau}}{\sqrt{2}}\mu
^{1/4}|\kk|^{-1/2}.\label{sk}\end{equation}

 $(\ref{elem})$ has no ground state because of the zero mode. Let $\Psi$ be any
vector that minimizes all nonzero modes.
Clearly, for $\kk\neq0$
\[(\Psi|a_\kk^*a_\kk\Psi)=|s_\kk|^2\approx\frac{\mu^{1/2}}{2|\kk|} .\]

Let
\[N':={\sum_\kk}'a_\kk^*a_\kk\]
be the number of particles away from the  zero mode. 
Clearly
 the density of
particles away from the zero mode in the state $\Psi$ equals
\begin{equation}
\frac{1}{V}(\Psi|N'\Psi)=
\frac{1}{V}{\sum_\kk}'|s_\kk|^2.\label{denso}\end{equation}
We expect that for large $L$, (\ref{denso}) converges
 to \begin{equation}
\frac{1}{(2\pi)^d}\int |s_\kk|^2\d\kk.\label{d1}\end{equation}

Note that in dimension $d=1$ there is a problem with the formula
(\ref{d1}), since 
 $|\kk|^{-1}$ is  integrable only in dimension $d>1$.
Therefore, for $d=1$  (\ref{d1}) diverges.
Thus, the Bogoliubov approximation is problematic for $d=1$ if  we
 keep the density of particles  $\rho$ fixed as $L\to\infty$.
 To our knowledge,
(\ref{sk}) and the above described problem of the Bogoliubov approximation in
 $d=1$ was first noticed in \cite{GN}.

Nevertheless, in spite of the breakdown of the Bogoliubov approximation,
 many authors believe that also in $d=1$ the IES 
exhibits the behavior $\epsilon_\mu(\kk)\approx c_\ph|\kk|$ with $c_\ph>0$
 for low momenta,
 see e.g.
\cite{Pop}, Chapter 6, \cite{LL,L2}.

\subsection{Improving the Bogoliubov method}
\label{s5}

For $\frac{2\pi}{L}{\mathbb Z}^d\ni \kk\mapsto \theta_\kk\in\cc$,  a square
summable sequence with $\theta_\kk=\theta_{-\kk}$, 
let $U_\theta$ be defined as in (\ref{uthe}). (This time we allow $\theta_\0$
to be nonzero).
For $\alpha\in\cc$, let
 $W_\alpha$ be defined as in (\ref{weyl}).

 $U_{\alpha,\theta}:=U_\theta W_\alpha $
is the general form of a Bogoliubov transformation commuting with $P^L$.
Let $\Omega$ denote the  vacuum vector.  Note that 
\[\Omega_{\alpha,\theta}:=U_{\alpha,\theta}^*\Omega\]
is the general form of a squeezed  vector of zero momentum.

One of our next objectives is to
 look for the squeezed  vector that minimizes the expectation value of
$H^L$. As in Section \ref{s3}, we will also compute the Hamiltonian $H^L$
 expressed in new creation and annihilation operators adapted to the new
 vacuum $\Omega_{\alpha,\theta}$. We do this in two steps. First we perform
 the Bogoliubov translation (\ref{trans}), which results in the expression
 (\ref{3B2}). Then we perform the Bogoliubov rotation (\ref{rota}). This time,
 however, we apply it to all the modes, including $0$.

The Hamiltonian after these substitutions in the Wick ordered form equals
\begin{align}
   H ^L
&=
B^L+C^Lb_0^*+\bar C^L b_0\nonumber
\\&+\frac12\sum_\kk O^L(\kk)b_\kk^*b_{-\kk}^*+
\frac12\sum_\kk\bar O^L(\kk)b_\kk b_{-\kk} +\sum_\kk
D^L(\kk)b_\kk^*b_\kk\nonumber
\\&
+\hbox{terms  higher order in {\it b}'s}\label{non}.
\end{align}
Clearly,
\[(\Omega_{\alpha,\theta}|H^L \Omega_{\alpha,\theta})=B^L,
\qquad (b_\kk^*\Omega_{\alpha,\theta}|H^L 
b_\kk^*\Omega_{\alpha,\theta})
=B^L+D^L(\kk).\]
Therefore, we obtain rigorous bounds
\[E^L\leq B^L,\ \ \ E^L+\epsilon^L(\kk)\leq B^L+D^L(\kk).\]

If
 we require that $B^L$ attains its minimum, then we will later on show
 that  $C^L$  and
 $O^L(\kk)$
vanish for all $\kk$. Henceforth we drop the superscript $L$.

\begin{eqnarray}
B&=&-\mu|\alpha|^2+\frac{{\hat v}(0)}{2V}|\alpha|^4
\nonumber\\
&&+\sum_\kk\left(\frac{\kk^2}{2}-\mu
+\frac{({\hat v}(\kk)+{\hat v}(0))}{V}|\alpha|^2\right)|s_\kk|^2\nonumber\\&&
-\sum_\kk\frac{{\hat v}(\kk)}{2V}
(\bar\alpha^2s_\kk c_\kk+\alpha^2\bar  s_\kk c_\kk)\nonumber\\
&&+\sum_{\kk,\kk'}\frac{{\hat v}(\kk-\kk')}{2V}c_\kk s_\kk c_{\kk'}\bar s_{\kk'}\nonumber\\
&&+\sum_{\kk,\kk'}\frac{{\hat v}(0)+{\hat
    v}(\kk-\kk')}{2V}|s_\kk|^2|s_{\kk'}|^2;\nonumber\\ 
C&=&\left(\frac{{\hat v}(0)}{V}|\alpha|^2-\mu + \sum_\kk \frac{({\hat v}(0)+{\hat v}(\kk))}{V}|s_\kk|^2 \right)(\alpha c_0 - \bar \alpha s_0)
\nonumber\\
&&+
\sum_\kk\frac{{\hat v}(\kk)}{V} \left( \alpha s_0 c_\kk \bar s_\kk - \bar\alpha c_0 c_\kk s_\kk \right);
\nonumber
\end{eqnarray}
In order to express $D(\kk)$ and $O(\kk)$, it is convenient to
introduce
\begin{eqnarray}
f_\kk:&=&\frac{\kk^2}{2}-\mu\nonumber\\
&&+|\alpha|^2\frac{{\hat v}(0)+{\hat v}(\kk)}{V}
+\sum_{\kk'}\frac{{\hat v}(\kk'-\kk)+{\hat v}(0)}{V} |s_{\kk'}|^2,
\label{gk11}\\
g_\kk:&=&\alpha^2\frac{{\hat v}(\kk)}{V}
-\sum_{\kk'}\frac{{\hat v}(\kk'-\kk)}{V}s_{\kk'}c_{\kk'}.\label{gk}
\end{eqnarray}
(Note that $f_\kk$ is real).
\begin{eqnarray}
D(\kk) &=& f_\kk (c_\kk^2 + |s_\kk|^2) - c_\kk (s_\kk \bar g_\kk + \bar s_\kk
g_\kk), \label{di1}
\\ 
O(\kk) &=& -2c_\kk s_\kk f_\kk + s_\kk^2 \bar g_\kk + c_\kk^2 g_\kk.\label{di2}
\end{eqnarray}
The main intermediate step of the calculations leading to the above
result is 
described in Appendix \ref{a2}.

\subsection{Conditions arising from minimization of the energy over
  $\alpha$} 
\label{s9}
We demand that
$B$ attains a minimum. To this end we first compute the derivatives
with 
respect to $\alpha$ and $\bar\alpha$:
\begin{eqnarray}
\partial_\alpha B&=&
\left(-\mu+\frac{{\hat v}(0)}{V}|\alpha|^2+\sum_\kk\frac{({\hat v}(0)+{\hat
    v}(\kk))}{V}|s_\kk|^2 \right)\bar\alpha \nonumber\\
&&
-\sum_\kk\frac{{\hat v}(\kk)}{V}\bar s_\kk c_\kk\alpha,\nonumber\\
\partial_{\bar\alpha }B&=&
\left(-\mu+\frac{{\hat v}(0)}{V}|\alpha|^2+\sum_\kk\frac{({\hat v}(0)+{\hat
    v}(\kk))}{V}|s_\kk|^2 \right)\alpha\nonumber\\
&&
-\sum_\kk\frac{{\hat v}(\kk)}{V}s_\kk c_\kk\bar\alpha . \nonumber
\end{eqnarray}

Note that
\begin{eqnarray} \nonumber
C = c_0 \partial_{\bar\alpha }B - s_0\partial_\alpha B,
\end{eqnarray}
so that the condition
\begin{eqnarray} 
\partial_{\bar \alpha} B = \partial_{\alpha}B = 0\label{row1}
\end{eqnarray}
entails $C=0$.
The condition (\ref{row1}) yields
\begin{equation}
\mu=\frac{\hat v(0)}{V}|\alpha|^2+\sum_{\kk'}\frac{\hat v(0)+\hat
  v(\kk')}{V}|s_{\kk'}|^2 - \frac{\alpha^2}{|\alpha|^2} \sum_{\kk'}\frac{\hat
  v(\kk')}{V} \bar s_{\kk'}c_{\kk'}. \label{mu}\end{equation}
This allows to eliminate $\mu$ from the expression for $f_\kk$:
\begin{eqnarray}
f_\kk:&=&\frac{\kk^2}{2}+|\alpha|^2\frac{{\hat v}(\kk)}{V}\nonumber\\
&&+\sum_{\kk'}\frac{{\hat v}(\kk'-\kk)-\hat v(\kk')}{V}|s_{\kk'}|^2
+\frac{\alpha^2}{|\alpha|^2}\sum_{\kk'}\frac{\hat v(\kk')}{V} \bar
s_{\kk'}c_{\kk'}. 
\label{fk}
\end{eqnarray}

\subsection{Conditions arising from minimization of the energy over
  $s_\kk$} 
\label{s9a}

 Computing the derivative with respect to $s_\kk$, $\bar s_\kk$ we can
  use 
\[\partial_{s_\kk }c_\kk=\frac{\bar s_\kk}{2c_\kk},\quad
\partial_{\bar s_\kk }c_\kk=\frac{ s_\kk}{2c_\kk}.\]

\begin{eqnarray}
\partial_{s_\kk }B&=& f_\kk\bar s_\kk - \frac{\bar g_\kk}{2}
 \left(c_{\kk}+\frac{|s_\kk|^2}{2c_\kk}\right) - g_\kk\frac{\bar
 s_{\kk}^2}{4c_\kk},\label{gig1}
\\
\partial_{\bar s_\kk }B&=& f_\kk s_\kk - \frac{g_\kk}{2} \left(c_{\kk}+
\frac{|s_\kk|^2}{2c_\kk}\right) - \bar g_\kk\frac{ s_{\kk}^2}{4c_\kk}.
\label{gig2}
\end{eqnarray}

One can calculate that
\begin{equation} \nonumber
O(\kk) =  \left(-2 c_\kk + \frac{|s_\kk|^2}{c_\kk} \right)\partial_{\bar s_\kk}B -
 \frac{s_\kk^2}{c_\kk}\partial_{ s_\kk}B .
\end{equation}
Thus $\partial_{s_\kk}B=\partial_{\bar s_\kk}B
=0$  entails $O(\kk)=0$.

(\ref{gig1}) and (\ref{gig2}) also imply
\begin{eqnarray} \nonumber
s_\kk\partial_{s_\kk}B-\bar s_\kk\partial_{\bar s_\kk}B &=& \frac{c_\kk}{2} (g_\kk \bar s_\kk-\bar g_\kk s_\kk ), 
\end{eqnarray}
and hence
\begin{equation}\label{row4}
g_\kk \bar s_\kk = \bar g_\kk s_\kk .
\end{equation}

It is convenient to introduce the parameters
\begin{eqnarray*}
S_\kk&:=&2s_\kk c_\kk,\\
C_\kk&:=&c_\kk^2+|s_\kk|^2.
\end{eqnarray*}
Now, using (\ref{row4}) we can write
\begin{eqnarray}
D(\kk) &=&  C_\kk f_\kk - S_\kk \bar g_\kk ,
\\ 
O(\kk) &=& -S_\kk f_\kk + C_\kk g_\kk.
\end{eqnarray}
Equating $O(\kk)$ to zero and assuming that $f_\kk\neq0$
we obtain
\begin{eqnarray}
D(\kk)&=&\sgn f_\kk\sqrt{f_\kk^2-|g_\kk|^2},\label{fixo1}\\
S_\kk&=&\frac{g_\kk}{D(\kk)},\label{fixo2}\\
C_\kk&=&\frac{f_\kk}{D(\kk)}\label{fixo3}.
\end{eqnarray}
We will  keep $\alpha^2$ instead of $\mu$ as the
parameter of 
the theory, hoping that one can later on express $\mu$ in terms
of $\alpha^2$. We set
 $\e^{\i\tau}:=\frac{\alpha}{|\alpha|}$. Then we can write
\begin{eqnarray}
f_\kk:&=&\frac{\kk^2}{2}+|\alpha|^2\frac{{\hat v}(\kk)}{V}\nonumber\\
&&+\sum_{\kk'}\frac{{\hat v}(\kk'-\kk)-\hat v(\kk')}{2V}(C_{\kk'}-1)
+\sum_{\kk'}\frac{\hat v(\kk')}{2V} 
\e^{\i2\tau}\bar S_{\kk'},
\label{fixo5}\\
g_\kk:&=&\alpha^2\frac{\hat v(\kk)}{V}
-\sum_{\kk'}\frac{\hat
  v(\kk'-\kk)}{2V}S_{\kk'}.\label{fixo6} 
\end{eqnarray}
Then we can express $\mu$ by
\begin{equation}
\mu=\frac{\hat v(0)}{V}|\alpha|^2+\sum_{\kk'}\frac{\hat v(0)+\hat
  v(\kk')}{2V}(C_{\kk'}-1) - \e^{\i2\tau} \sum_{\kk'}\frac{\hat
  v(\kk')}{2V} \bar S_{\kk'}. \label{mu1}\end{equation}

One can express the minimizing conditions in the following theorem.

\begin{thm}\begin{enumerate}\item
Suppose that $|\alpha|^2>0$ and $\e^{\i\tau}$ are fixed parameters.
 Let the first derivative of $B$
 with respect to $\alpha,\bar\alpha,(s_\kk), (\bar s_\kk)$ vanish.
Let $f_\kk$, $g_\kk$ be given by
(\ref{fixo5}), (\ref{fixo6}). 
 For any $\kk\in\frac{2\pi}{L}\zz^d$ we have then
 $f_\kk^2\geq|g_\kk|^2$, and the equations
 (\ref{fixo1})--(\ref{fixo3})
hold.
\item
We have \begin{eqnarray}
\left[\begin{array}{cc}
\partial_{\bar\alpha}\partial_{\alpha} B&\partial_{\bar\alpha}^2 B\\
\partial_{\alpha}^2 B&\partial_{\alpha}\partial_{\bar\alpha} B
\end{array}\right]
=\left[\begin{array}{cc}
f_0& g_0\\ \bar g_0& f_0
\end{array}\right].\label{matr}\end{eqnarray}
 (\ref{matr}) is positive/negative definite iff $D(0)$ is
positive/negative. (\ref{matr}) is zero iff $D(0)=0$. Besides, 
\begin{eqnarray} 
D(0) =  2\sgn f_0
\sqrt{\frac{\hat v(0)}{V}\alpha^2 \sum_{{\kk}}\frac{\hat v({\kk})}{2V}
  \bar S_{{\kk}} }
\label{gapp}\end{eqnarray}
\end{enumerate}\label{thht}\end{thm}

In the above theorem we included all possibilities that guarantee the
stationarity of $B$. Clearly,  the case of $D(\kk)<0$ seems
physically irrelevant. But this is equivalent to 
$f_\kk<0$. 
Therefore, under the additional condition $D(\kk)\geq0$, we can drop
$\sgn f_\kk$ from  (\ref{fixo1}).

In the case of the  zero momentum we have an additional argument for
the positivity of $D(0)$ given in (2).
 $D(0)\geq0$
is in fact equivalent to the condition (\ref{matr})$\geq0$, which 
 is  necessary for the existence of
minimum of $B$.

Let us compute the ground state energy in the improved Bogoliubov
method. 
Inserting (\ref{mu}) to the expression for $B$ we obtain
\begin{eqnarray}
B&=&-\frac{\hat v(0)}{2V}\left(|\alpha|^2+\sum_{\kk}|s_{\kk}|^2\right)^2
+\sum_{\kk}\frac{{\kk}^2}{2}|s_{\kk}|^2\nonumber\\
&&+\sum_{\kk}\frac{\hat v({\kk}-{\kk}')-\hat v({\kk}')-v({\kk})}{2V}|s_{{\kk}'}|^2|s_{\kk}|^2\nonumber\\
&&+\sum_{\kk}\frac{\hat v({\kk}')}
{4V}(\e^{\i2\tau}\bar S_\kk+\e^{-\i2\tau} S_\kk)
|s_{\kk}|^2\nonumber\\
&&+\sum_{\kk,\kk'}\frac{\hat v({\kk}-{\kk}')}{8V}S_{{\kk}'}\bar
 S_{\kk},\nonumber
\end{eqnarray}
where recall that $|s_\kk|^2=\frac12(C_\kk-1)$.
Using  (\ref{mu})  again to eliminate $|\alpha|^2$ in favor of $\mu$, and then
computing the derivative with respect to $\mu$ we obtain
\[-\partial_\mu B=|\alpha|^2+\sum_{\kk}|s_{\kk}|^2.\]
Therefore, the grand-canonical density is given by
\begin{equation}\label{densy1}
\rho=\frac{|\alpha|^2+\sum_{\kk}|s_{\kk}|^2}{V}.
\end{equation}

\subsection{Thermodynamic limit of the fixed point equation}
\label{Thermodynamic limit of the fixed point equation}

One can ask whether the method described in the previous two
 sections
has a well defined limit as $L\to\infty$.
 A natural way to take this limit, at least formally, involves the
 following steps. We put
$\alpha=\sqrt{V\kappa }\e^{\i\tau}$,
 for some fixed parameter $\kappa>0 $ having the
interpretation of the density of the condensate.
We expect $s_{\kk}$ (and hence $S_\kk$, etc.)
to converge to a function depending on ${\kk}\in\rr^d$
in a reasonable class. Finally, we replace
$\frac{1}{V}\sum\limits_{\kk}$ by $\frac{1}{(2\pi)^d}\int\d {\kk}$.
Thus  equations  (\ref{fixo5}), (\ref{fixo6}) and (\ref{mu1}) 
 are replaced with
\begin{eqnarray}
f_\kk&=&\frac{\kk^2}{2}+\kappa \hat v(\kk)+\frac{1}{2(2\pi)^d}\int(\hat
v(\kk'-\kk)-\hat v(\kk'))(C_{\kk'}-1)\d\kk'\nonumber\\
&&+\frac{\e^{\i2\tau}}{2(2\pi)^d}\int\hat v(\kk')\bar
S_{\kk'}\d\kk',\label{fk1} \\
g_\kk&=&\kappa \e^{\i2\tau}\hat v(\kk)-\frac{1}{2(2\pi)^d}\int\hat v(\kk'-\kk)
S_{\kk'}\d\kk'\label{gk1},\\
\mu&=&\hat v(0)\kappa +\frac{1}{2(2\pi)^d}
\int(\hat v(0)+\hat v(\kk'))(C_{\kk'}-1)\d\kk'\nonumber\\&&
-\frac{\e^{\i2\tau}}{2(2\pi)^d}\int\hat v(\kk')\bar S_{\kk'}\d\kk'.
\label{mu1a}
\end{eqnarray}

We also obtain (in the physical case of positive $D$)
\begin{eqnarray} \label{positive}
D(0) = 
2\sqrt{\hat v(0)\kappa  \frac{1}{2(2\pi)^d}\int\d {\kk}
\hat v(\kk) \bar S_{\kk} }.
 \end{eqnarray}
(\ref{positive}) is typically positive -- thus the quadratic part of the
Hamiltonian (\ref{non})  seems to have a gap.

One can try to find
 $\alpha,(S_\kk)$ satisfying the minimization condition
by iterations.  
A natural starting point seems to be
$S_\kk=0$. Then, by (\ref{mu}) or (\ref{mu1a}), 
$\mu=\hat
  v(0)\kappa$.
After one iteration we obtain
\begin{eqnarray*}
f_\kk&=&\frac{\kk^2}{2}+\kappa\hat v(\kk),\\
g_\kk&=&\kappa\hat v(\kk),\\
D(\kk)&=&\sqrt{(\kk^2/2)^2+\kk^2\kappa\hat
    v(\kk)} ,\\
S_\kk&=&\frac{\kappa\hat v(\kk)}{\sqrt{(\kk^2/2)^2+\kk^2\kappa\hat
    v(\kk)} }
\end{eqnarray*}
Thus $D(\kk)= \omega_{\Bog,\mu}(\kk)$ given by
 (\ref{omega1}) -- we obtain
 the grand-canonical   Bogoliubov approximation.

In the case of finite $L$  we cannot continue iterations
 because of $S_0=\infty$. 

In thermodynamic limit, the value at zero may not matter, since $\kk$ is a
 continuous variable.  $S_\kk$ for
 small $\kk$ behaves as $\sim |\kk|^{-1}$ (this was noted already in
  (\ref{sk4})). In dimension $d=1$, if we try
 to do the next 
iteration we obtain divergent integrals. Thus, we cannot
 continue iterations. However in dimensions $d\geq2$ the integrals are
 convergent and we can do the next
 iteration (and presumably we can keep on going).

The energy gap appears already at the second iteration.

\subsection{Uncorrelated states}
\label{Uncorrelated states}

Let ${\mathcal H}_0$ denote the space spanned by $1$ and let ${\mathcal H}_{\{\kk,-\kk\}}$
denote the space spanned by $\e^{\i \kk\x}$ and $\e^{-\i\kk\x}$. Clearly,
\begin{equation} L^2(\Lambda)={\mathcal H}_0\oplus
\left(\mathop{\oplus}\limits_{\{\kk,-\kk\}}{\mathcal
  H}_{\{\kk,-\kk\}}\right).\label{tensor1}\end{equation}
The sum in (\ref{tensor1}) runs over all two-element sets 
 of the form $\{\kk,-\kk\}$
  with $\kk\in\frac{2\pi}{L}\zz^d$.

The exponential property of Fock spaces yields
\begin{equation}
\Gamma_\s(L^2(\Lambda))=\Gamma_\s({\mathcal H}_0)\otimes
\left(\mathop{\otimes}\limits_{\{\kk,-\kk\}}\Gamma_\s({\mathcal
  H}_{\{\kk,-\kk\}})\right).\label{tensor}\end{equation}

(See e.g. \cite{Ta3} for the definition of the tensor product of an infinite family of
 Hilbert spaces used in (\ref{tensor}). Note that
  in each of the factors of the tensor product of (\ref{tensor})
 we distinguish a normalized vector -- the vacuum
 vector). 

We will say that a vector $\Psi\in \Gamma_\s(L^2(\Lambda))$ is uncorrelated with
respect to (\ref{tensor}), or simply {\em uncorrelated}, iff it is of the form
\[\Psi=\Psi_0\otimes\left(\mathop{\otimes}\limits_{\{\kk,-\kk\}}
\Psi_{\{\kk,-\kk\}})\right) 
\]
for some
\[\Psi_0\in\Gamma_\s({\mathcal H}_0),\ \
\Psi_{\{\kk,-\kk\}}\in\Gamma_\s({\mathcal   H}_{\{\kk,-\kk\}}) .
\]

We
define the {\em uncorrelated ground state energy in the box}
\[E_\un^L:= \inf\left\{
(\Psi|H^L
\Psi)\ :\ \Psi\ \hbox{is uncorrelated and of norm 1}\right\}
.\]

 For $\kk\in\frac{2\pi}{L}\zz^d$ we define the {\em
uncorrelated IES in the box}
\[\epsilon_\un^L(\kk):= \inf\left\{
(\Psi|H^L(\kk) \Psi)-E_\un^L\ :\ \Psi\ \hbox{is uncorrelated},\ 
\|\Psi\|= 1\right\},\] 
 and for 
$ \kk\in\rr^d$ we define the  {\em
uncorrelated IES in the thermodynamic limit}
\[
\epsilon_\un(\kk)
:=\sup_{\delta>0}\left(\liminf_{L\to\infty}\left(\inf_{\kk'\in
     \frac{2\pi}{L} \zz^d,\; |\kk-\kk'|
<\delta} 
\epsilon_\un^L(\kk')\right)\right).\]

Clearly, from the mini-max principle we obtain:
\[E^L\leq E_\un^L,\qquad \ E^L+ \epsilon^L(\kk)\leq 
E_\un^L+ \epsilon_\un^L(\kk).\]

\begin{conjecture}
\label{conj2} We believe the following statements to hold true:
\begin{enumerate}
\item The map $\rr^d\ni \kk\mapsto \epsilon_\un(\kk)\in \rr$ is positive and
continuous
  away from $0$.
\item Let $\kk\in{\mathbb R}^d\backslash\{0\}$.
 If $L\to\infty$, $\kk_L\in\frac{2\pi}{L}\zz^d$, and
$\kk_L\to \kk$, then  $\epsilon_\un^{L}(\kk_L)\to \epsilon_\un(\kk)$.
\item $\sup_{\kk\neq0}\epsilon_\un(\kk)>0$.
\end{enumerate}
\end{conjecture}

\vspace{0.5cm}

Thus we conjecture that  using
 only uncorrelated states in variational determination
of excitation spectrum have serious limitations. 
We expect the results to be well
behaved in thermodynamic limit, but they will
probably not capture
the  phononic behavior at the bottom of the IES,
and in particular we will obtain an energy gap.

Note that the squeezed vectors $\Omega_{\alpha,\theta}$ and the particle
excitations over the squuezed vectors  $b_\kk^*\Omega_{\alpha,\theta}$ are
examples of uncorrelated vectors. Therefore, 
the expectation values of $H^L$ in
these vectors give an
upper bound on $E_\un^L$
 and $E_\un^L+\epsilon_\un^L(\kk)$. We showed that for these
expectation values one should expect an energy gap -- we expect this gap to
persist even for more general uncorrelated states.

 Thus, in order to obtain
more satisfactory bounds it seems that one needs to  use {\em correlated}
vectors. Note that the Bijls-Feynman variational vector
 $N_\kk\Psi/\|N_\kk\Psi\|$ is
correlated  for $\kk\neq0$, even if $\Psi$ is uncorrelated.

\section{Perturbative approach}
\label{Perturbative approach}

In this section we will use the grand-canonical formalism.
We replace the potential $ v(\x)$  with $\lambda v(\x)$, where $\lambda$ is a
(small) positive constant. 
 We will drop $\mu$
from most symbols and instead we will make the dependence on $\lambda$
explicit.
 Thus instead  $H_\mu^L$ we will write $H^{\lambda,L}$.

\subsection{Perturbative approach based on the Bogoliubov method}
\label{Perturbative approach based on the Bogoliubov method}
Let us go back to the Bogoliubov method described in Subsection \ref{s3}.
Using the formula (\ref{nonum}) we can split the Hamiltonian as
\[H^{\lambda,L}=  \lambda^{-1}H_{-1}^L+H_{0}^L
+\sqrt\lambda H_{\frac12}^L+\lambda H_{1}^L,\]
where
\begin{eqnarray} H_{-1}^L
 &:=&-  V\frac{ \mu^2}{2{\hat v}(0)} ,
\\ \nonumber 
 H_{0}^L
 &:=&{\sum_\kk} \left(\frac{1}{2}\kk^2 +{\hat
    v}(\kk)\frac{\mu}{{\hat v}(0)} \right)a^*_{\kk} 
a_{\kk} \\\nonumber&
+&  {\sum_\kk}  {\hat v}(\kk)\frac{\mu}{2 {\hat v}(0)}
\left(\e^{-\i2\tau}a_\kk a_{-\kk} +
\e^{\i2\tau}a_\kk^*a_{-\kk}^* \right),\\\nonumber
H_{\frac12}^L&:=&
 \sum_{\kk,\kk'}
 \frac{{\hat v}(\kk)\sqrt{\mu}}{\sqrt{\hat v(0)V}}
(\bar\e^{\i\tau} a_{\kk+\kk'}^*a_\kk a_{\kk'} + \e^{\i\tau}
 a_\kk^*a_{\kk'}^*a_{\kk+\kk'}),
\\H_{1}^L&:=&
\sum_{\kk_1,\kk_2,\kk_3,\kk_4}\delta(\kk_1+\kk_2-\kk_3-\kk_4)\frac{{\hat v}(\kk_2-\kk_3)}{2V}
a^*_{\kk_1}a^*_{\kk_2}a_{\kk_3}a_{\kk_4}.\nonumber
\end{eqnarray}

Note that $H_{n}^L$, $n=-1,0,\frac12,1$, do
 not depend on $\lambda$. This suggests that one can
 try to apply methods of perturbation theory to compute the ground state
 energy of $H^{\lambda,L}$ treating  $\sqrt{\lambda}H_{\frac12}+\lambda H_{1}$
 as a small perturbation of the quadratic Bogoliubov Hamiltonian
\begin{equation}
\lambda^{-1}H_{-1}^{L}+H_{0}^{L}.\label{phf}\end{equation} 

It is also tempting to compute 
 the excitation spectrum, applying perturbation methods
to the same splitting   of  $H^{\lambda,L}$ restricted to the sector of fixed
 momentum $\kk$.
Unfortunately, when one tries to implement this idea  one encounters serious
difficulties due to the infrared problem: the operator
(\ref{phf}) does not have a ground state, neither
 globally, nor in fixed momentum sectors (because of the $\kk=0$
 mode).
Further on we will describe a natural approach that should help solve the
 infrared problem and should give a better starting point for the perturbation methods.

In any case, the splitting suggests the following conjecture (which is
the
 grand-canonical version of Conjecture \ref{conj-1}).
 Let $\epsilon_{\mu}^\lambda(\kk)$ be the grand-canonical
IES for the potential
$\lambda v$ and let $\epsilon_{\Bog,\mu}(\kk)$ be given by
(\ref{epsilon1xs}). 
\begin{conjecture}
Let $d\geq2$. Then for a large class of repulsive potentials we have
\[\lim_{\lambda\searrow0}
\epsilon_{\mu}^\lambda(\kk)
=\epsilon_{\Bog,\mu}(\kk).\]
\label{conj-1a}\end{conjecture}

\subsection{Perturbative approach based on improved Bogoliubov method}
\label{Perturbative approach based on improved Bogoliubov method}

We fix the size of the box,
$\mu$ and $\e^{\i\tau}$ and we assume
that we solved the fixed point equation described in Subsection \ref{s9a} and there is an energy gap. We assume
that the solution is unique.
 The expression for
 $H^{\lambda,L}$  Wick-ordered  with respect to the operators $b_\kk$,
$b_\kk^*$ allows us to write
\begin{equation}
H^{\lambda,L}=
\lambda^{-1}H_{-1}^{\lambda,L}+H_0^{\lambda,L}+\sqrt{\lambda}H_{\frac12}^{\lambda,L}
+\lambda
H_1^{\lambda,L},\label{split}\end{equation}
where $\lambda^{-1}H_{-1}^{\lambda,L}=B$ is the constant term,
$H_1^{\lambda,L}=\sum_\kk D(\kk)b_\kk^*b_\kk$ is the quadratic term,
$H_{\frac12}^{\lambda,L}$ and $H_1^{\lambda,L}$ are respectively the third and fourth order
parts of $H$ in  operators $b_\kk$  and $b_\kk^*$, see (\ref{non}).

The splitting (\ref{split}) can be used to set up a perturbatve approach for
computing the energy density and excitation spectrum. The presence of a gap
will be actually an advantage in this case.

 More
precisely, let us consider the following Hamiltonian
\begin{equation}H^{\lambda,\delta,L}:=
\delta^{-1}H_{-1}^{\lambda,L}+H_0^{\lambda,L}+\sqrt{\delta}
H_{\frac12}^{\lambda,L} 
+\delta
H_1^{\lambda,L},\label{split1}\end{equation}
where $\delta$ is an additional parameter introduced for bookkeeping
reasons. We treat $\delta^{-1}H_{-1}^{\lambda,L}+H_0^{\lambda,L}$ as the unperturbed
operator and the rest as a perturbation depending on the small parameter
$\delta$.  $\delta^{-1}H_{-1}^{\lambda,L}+H_0^{\lambda,L}$ has a ground state
$\Omega_{\alpha,\theta}
$, and even a
mass gap, so the perturbation expansion in terms of $\delta$
 for the ground state vector and
energy is well defined and for small $\delta$
\[\Psi^{\lambda,\delta,L}=
\sum_{n=0}^{\infty}(\delta^n\Psi_n^{\lambda,L}+\delta^{n+\frac12}
\Psi_{n+\frac12}^{\lambda,L}),\
\ \ 
E^{\lambda,\delta,L}=\sum_{n=-1}^{\infty}\delta^nE_n^{\lambda,L},\]
where $\Psi_0^{\lambda,L}=\Omega_{\alpha,\theta}$.
(It is easy to see that  all powers of $\delta$ for the energy are integral).
At the end we will substitute $\lambda$ for
$\delta$:
\begin{equation}
\Psi^{\lambda,L}\sim\sum_{n=0}^{\infty}(\lambda^n\Psi_n^{\lambda,L}+\lambda^{n+\frac12}
\Psi_{n+\frac12}^{\lambda,L}),\
 \ \
E^{\lambda,L}\sim\sum_{n=-1}^{\infty}\lambda^nE_n^{\lambda,L}.\label{asym1}\end{equation}

Let $\epsilon^L(\kk)$ be the subadditive hull of
$D^L(\kk)$. Assume that for some $\kk_1,\dots\kk_n$ with
$\kk=\kk_1+\cdots+\kk_n$ we have
$\epsilon^L(\kk)=D^L(\kk_1)+\cdots+D^L(\kk_n)$.
 This implies that  the vector 
$(n!)^{-1/2}b_{\kk_1}^*\cdots b_{\kk_n}^*\Omega_{\alpha,\theta}$ 
is at the bottom of the spectrum of 
 $\delta^{-1}H_{-1}^{\lambda,L}+H_0^{\lambda,L}$ in the sector of
momentum 
$\kk$. Again we can write down the perturbation expansion in terms of $\delta$
for the excitation spectrum, convergent for small $\delta$:
\begin{equation*}
\Psi^{\lambda,\delta,L}(\kk)
=\sum_{n=0}^{\infty}(\delta^n\Psi_n^{\lambda,L}(\kk)
+\delta^{n+\frac12}\Psi_{n+\frac12}^{\lambda,L}(\kk)) 
,\ \ \ 
\epsilon^{\lambda,\delta,L}(\kk)
=\sum_{n=-1}^{\infty}\delta^n\epsilon_n^{\lambda,L}(\kk)
,\end{equation*}
where $\Psi_0^{L,\lambda}(\kk)=b_\kk^*\Omega_{\alpha,\theta}$.
Then we put $\delta=\lambda$ obtaining
\begin{equation}
\Psi^{\lambda,L}(\kk)
\sim\sum_{n=0}^{\infty}(\lambda^n\Psi_n^{\lambda,L}(\kk)
+\lambda^{n+\frac12}\Psi_{n+\frac12}^{\lambda,L}(\kk)) 
,\ \ \ 
\epsilon^{\lambda,L}(\kk)
\sim\sum_{n=-1}^{\infty}\lambda^n\epsilon_n^{\lambda,L}(\kk)
.\label{asym2}\end{equation}

Of course, we do not claim that the power series (\ref{asym1}) and
(\ref{asym2})
have a nonzero radius of convergence.
We only hope that they 
are in some sense asymptotic to the physical
quantities.

We hope that the perturbation expansions  (\ref{asym1}) and
(\ref{asym2})
 survive thermodynamic limit. We do not expect that the $n$th
 terms of these expasions will be of order $O(\lambda^n)$. However, 
we  hope that each next term will give a better approximation, as
expressed in the following conjecture:
\begin{conjecture}
\begin{enumerate}\item
For any $n$, there exist
\begin{eqnarray*}
e_n^\lambda&:=&\lim_{L\to\infty}\frac{E_n^{\lambda, L}}{V},\\
\epsilon_n^\lambda(\kk)&:=&\lim_{L\to\infty}\epsilon_n^{\lambda, L}(\kk_L),\ \
\kk_L\to \kk.
\end{eqnarray*}
\item 
\begin{eqnarray*}
\lim_{\lambda\searrow0}e_{-1}^\lambda&=&\frac{\mu^2}{2\hat v(0)},\\
\lim_{\lambda\searrow0} e_{0}^\lambda&=&
-\frac{1}{(2\pi)^d}\int\frac12
\left(\bigl(
\frac12\kk^2+\hat v(\kk)\frac{\mu}{\hat
  v(0)}\bigr)-\omega_{\Bog,\mu}(\kk)\right)\d \kk,\\
\lim_{\lambda\searrow0}\epsilon_0^\lambda(\kk)&=&\epsilon_{\Bog}(\kk).\end{eqnarray*}
\item  For some $0<\sigma_1<\sigma_2\cdots$ with
 $\lim\limits_{n\to\infty}\sigma_n=\infty$, 
\begin{eqnarray*}
\lambda^ne_{n}^\lambda&=&O(\lambda^{\sigma_n}),\ \ n=1,2,\dots\\
\lambda^n\epsilon_n^\lambda(\kk)&=&O(\lambda^{\sigma_n}),\ \ \ n=1,2,\dots.
\end{eqnarray*}
\item For $\sigma_n$ as above and all $n$,
\begin{eqnarray*}
\sum_{j=0}^n\lambda^j\epsilon_j^\lambda(0)
&=&O(\lambda^{\sigma_n}).\end{eqnarray*}
\end{enumerate}
\end{conjecture}

(1) is the existence of thermodynamic limit of the
    perturbation expansion. 
(2) tells us that the lowest order terms in this expansion agree with the
    quantities obtained in the Bogoliubov approximation. (3) means that the
later terms in the expansion are in some sense
 lower order than earlier. (4) says that there is 
 no gap in the excitation spectrum 
 at $\kk=0$.

It seems that a result similar to the above 
conjecture could  be easier to prove than a result about the true energy
density and the true infimum of the excitation spectrum.

Let us sum up the procedure that we propose to compute various 
quantities for Bose gas with $\lambda$ small and fixed $\mu$. We will call it
the {\em Improved Bogoliubov Approach}
\begin{enumerate}
\item Find a translation invariant squeezed state $\Omega_{\alpha,\theta}$
  minimizing the expectation value of the Hamiltonian $H^{\lambda,L}$.
\item Split the Hamiltonian as in  (\ref{split}):
\begin{equation*}
H^{\lambda,L}=
\lambda^{-1}H_{-1}^{\lambda,L}+H_0^{\lambda,L}+
\sqrt{\lambda}H_{\frac12}^{\lambda,L}
+\lambda
H_1^{\lambda,L}\end{equation*} according to the power in
creation/annihilation operators adapted to  $\Omega_{\alpha,\theta}$.
\item Introduce the fictitious Hamiltonian with an additional coupling
  constant $\delta$.
\begin{equation*}
H^{\lambda,\delta,L}=
\delta^{-1}H_{-1}^{\lambda,L}+H_0^{\lambda,L}+
\sqrt{\delta}H_{\frac12}^{\lambda,L}
+\delta
H_1^{\lambda,L}.\end{equation*}
\item Compute the desired quantity perturbatively, obtaining a 
  (formal) power series $c^{\lambda,\delta,L}=\sum_n\delta^n c_n^{\lambda,L}$.
\item Go to thermodynamic limit with each term in the series separately, obtaining $c_n^{\lambda}:=
\lim_{L\to\infty}c_n^{\lambda,L}$.
\item Set $\delta=\lambda$, obtaining the power series
 $c^{\lambda}=\sum_n\lambda^n c_n^{\lambda}$, which is the final expression
 for the desired quantity.
\end{enumerate}

\subsection{Approach with Isolated Condensate}

In the literature there are many works that are based on a somewhat different
approach to the Bose gas with small $\lambda$ and fixed $\mu$.
 This approach is sometimes called the {\em Approach with
 Isolated Condensate}. We would like to compare it to the Improved
 Bogoliubov's Approach.

Let us describe the basic steps of this approach:
\begin{enumerate}
\item Make the $c$-number substitution, obtaining the Hamiltonian
 $H^{\lambda,L}(\alpha)$
  as in (\ref{c-number}).
\item Substitute $\alpha=\sqrt{\lambda^{-1}\kappa  V}$ and split the
  Hamiltonian as
\begin{equation*}
H^{\lambda,L}(\alpha)=
\lambda^{-1}H_{-1}^{\kappa ,L}+H_{0}^{\kappa ,L}+
\sqrt{\lambda}H_{\frac12}^{\kappa ,L}
+\lambda
H_1^{\kappa ,L}.\end{equation*}
according to the power of $\lambda$.
\item Compute perturbatively the ground state energy, obtaining a (formal)
  power series
  $E^{\lambda,\kappa ,L}=\sum_n\lambda^n E_n^{\kappa ,L}$.
\item Compute the desired quantity as a (formal) power series
  $c^{\lambda,\kappa ,L}=\sum_n\lambda^n c_n^{\kappa ,L}$.
\item Minimize  (up to the desired order in $\lambda$) 
$E^{\lambda,\kappa ,L}$, obtaining  $\kappa ^{\lambda, L}$  as a function of $\lambda,L$.
\item Substitute  $\kappa ^{\lambda,L}$ 
 in the expression for the desired quantity, obtaining
  $c_n^{\lambda,L}= c_n^{\kappa ^{\lambda,L},L}$.
\item Go to thermodynamic limit with each term of the series separately, obtaining
$c_n^{\lambda}=\lim_{L\to\infty} c_n^{\lambda,L}$.
The  final expression
 for the desired quantity is
\[c^\lambda=\sum_n\lambda^n c_n^{\lambda}.\]
\end{enumerate}

 As
proven by \cite{LSSY1} (see Section \ref{c-number}), the Approach with Isolated
Condensate is exact in thermodynamic limit for the energy
density. In the case of finer quantities,
 such as the infimum of the
 excitation spectrum or Green's functions, we do not see
 why thermodynamic limit should make
 this
approach  exact.

 Improved Bogoliubov Approach and  Approach with Isolated Condensate
seem to have a lot in common. In both of them 
the main step 
involves
 calculations with a quadratic Hamiltonian perturbed by 3rd and 4rth order
perturbations.
 In both approaches the quadratic term does not contain a term linear
 in creation/annihilation operators.
 Note also that in both procedures  the dependence of the final
quantities on the
coupling constant $\lambda$  can be quite complicated, and not given
just by a power series.

The Approach with Isolated Condensate
 may seem simpler technically, since
the perturbation expansion is applied to a simpler splitting, whereas
in the Improved Bogoliubov Approach
the first step involves solving a complicated
fixed point equation.
It is however quite clear, that the Improved Bogoliubov Approach is physically
better justified than the
Approach with Isolated Condensate. In the former no term is dropped. In the
 latter, at the very beginning we drop an important term from the
 Hamiltonian.

\section{Observables}
\label{A priori estimates}

In this section we work in the grand-canonical approach. We drop the subscript
$\mu$ and $L$, so that $H_\mu^L$ is denoted by $H$. (In particular, in order
not to clutter the notation we hide the dependence on $L$, which however plays
an important role in what follows).

\subsection{Spontaneous symmetry breaking}

The Hamiltonian $H$ is invariant with respect to the transformation generated
by the number operator $\e^{\i \tau N}$. Consequently, its ground state can
be chosen to have a definite  number of particles. It is
however believed that in thermodynamical limit this gauge invariance is
spontaneously broken. (In fact, it is broken in the Bogoliubov method).

Let us try to
 describe this symmetry breaking rigorously.
Following Bogoliubov
\cite{Bog1}, we
 perturb the Hamiltonian by a non-physical perturbation
\begin{equation}
H_{\nu}:=H_{}-\nu\sqrt{V}(a_\0^*+a_\0),
\label{nnu}\end{equation}
where $\nu>0$.
$H_{\nu}$ depends on the gauge:
\begin{equation}
\e^{\i \tau N}H_{\nu}\e^{-\i \tau N}=H_{}-\nu\sqrt{V}(\e^{\i\tau}a_\0^*
+\e^{-\i\tau}a_\0).
\end{equation}

Let us assume that $H_\nu$ has a unique ground state given by the 
vector $\Psi_\nu$. 
Note that the Hamiltonian $H_\nu$ is real, therefore we can assume
$\Psi_\nu$ to be real as well.
 $H_\nu$ is translation invariant and the group of translations of the torus
is compact. Hence we can take
$\Psi_\nu$ to be translation invariant. The expectation value with respect
to the 
vector $\Psi_\nu$ will be denoted 
\[\langle\cdot\rangle_\nu:=
(\Psi_\nu|\cdot \Psi_\nu).\]

Because of the translation invariance we have
\begin{eqnarray}
\langle a_\kk\rangle_\nu&=&0,\ \ \kk\neq0;\nonumber\\
\langle a_\kk^*a_{\kk'} \rangle_\nu&=&0,\ \ \kk\neq\kk';\nonumber\\
\langle a_\kk a_{-\kk'} \rangle_\nu&=&0,\ \ \kk\neq\kk'.\label{trans1}
\end{eqnarray}

Thus the nontrivial one- and two-point correlation function are
\begin{eqnarray}
\langle a_\0\rangle_\nu&=&\langle a_\0^*\rangle_\nu;\label{cor1}\\
\langle a_\kk^*a_{\kk} \rangle_\nu&=&\langle a_{-\kk}^*a_{-\kk} \rangle_\nu;
\label{cor2}\\
 \langle a_\kk a_{-\kk} \rangle_\nu
&=& \langle a_\kk^* a_{-\kk}^*\rangle_\nu.
\label{cor3}
\end{eqnarray}
and  the expressions  (\ref{cor1}),  (\ref{cor2}) and  (\ref{cor3}) are all
      real.
Their reality follows from the reality of the Hamiltonian
      (\ref{nnu}) and the reality of $\langle \cdot\rangle_\nu$.

Let us assume that there exists the limit
\begin{equation}
\langle\cdot\rangle:=\lim_{\nu\searrow0}\lim_{L\to\infty}
\langle\cdot\rangle_{\nu},\label{dsa}\end{equation}
 as a state  on a suitable
family ${\mathfrak A}$ of observables.

Clearly, ${\mathfrak A}$ is invariant with respect to the Hermitian
conjugation. Moreover,
 the group of translations $\e^{\i \x P}\cdot\e^{-\i\x P}$ and the
 dynamics  $\e^{\i \x H}\cdot\e^{-\i\x H}$
act on  ${\mathfrak A}$.

Clearly, the ground state of 
$\e^{\i \tau N}H_{\nu}\e^{-\i \tau N}$ (before taking the thermodynamic limit)
is  given by 
 $\e^{\i\tau N}\Psi_{\nu}$. Replacing $\Psi_\nu$
with $\e^{\i\tau N}\Psi_{\nu}$ and performing
the limit (\ref{dsa}), we obtain a new state on  ${\mathfrak A}$.
If (\ref{cor1}) or (\ref{cor3}) are non-zero, then
this new state differs from
$\langle\cdot\rangle_\nu$ :
 (\ref{cor1})
 has to be multiplied with $\e^{\i\tau}$ and (\ref{cor3}) with $\e^{\i2\tau}$.

Clearly, (\ref{trans1}) are true if we replace $\langle\cdot\rangle_\nu$ with
$\langle\cdot\rangle$. It is natural to assume that the following limits
exist:
\begin{eqnarray}
\rho&:=&
\lim_{\nu\searrow0}
\lim_{L\to\infty}\frac{\langle N\rangle_\nu}{V};\\
\sqrt{\kappa}&:=&
\lim_{\nu\searrow0}
\lim_{L\to\infty}\frac{\langle a_\0\rangle_\nu}{\sqrt{V}}
 ;\label{cor1a}\\
\langle a_\kk^*a_{\kk} \rangle&=&\lim_{\nu\searrow0}
\lim_{L\to\infty}
\langle a_{\kk}^*a_{\kk} \rangle_\nu,\
\ \kk\neq0;
\label{cor2a}\\
 \langle a_\kk a_{-\kk} \rangle
&=& \lim_{\nu\searrow0}
\lim_{L\to\infty}\langle a_\kk a_{-\kk}\rangle_\nu,\ \ \ \kk\neq0.
\label{cor3a}
\end{eqnarray}
Clearly,
  the expressions  (\ref{cor1a}),  (\ref{cor2a}) and  (\ref{cor3a}) are again
      real.
$\rho$ is the density and $\kappa$ can be interpreted as the density of the
  condensate. Both depend on $\mu$.

\subsection{A priori estimates}

We will use notation explained in an abstract setting in Appendix 
\ref{Green's  functions}, where the reader will also find some general
remarks about  Green's functions and their motivation.
 In particular, for a pair of operators $A$, $B$ we the {\em static
   Green's function} is defined as
\begin{eqnarray*}
\langle\langle 
A,B\rangle\rangle_\nu&:=&\langle
A(H_\nu-E_\nu)^{-1}B \rangle_\nu
+\langle
B(H_\nu-E_\nu)^{-1}A\rangle_\nu .\end{eqnarray*}
Recall the operator
\begin{eqnarray*} N_\q &:=&\sum_\kk
 a_{\q+\kk}^*a_\kk=\int\e^{\i\kk\x}a_\x^*a_\x\d\x,\end{eqnarray*}
(see (\ref{densop}) and (\ref{fyny})


We will tacitly assume that we can perform thermodynamic limit of various
observables, such as
\begin{eqnarray*}
\langle\langle 
a_\kk^*,a_\kk\rangle\rangle&:=&
\lim_{\nu\searrow0}\lim_{L\to\infty}
\langle\langle 
a_\kk^*,a_\kk\rangle\rangle_\nu
;\\
\langle\langle 
a_\kk,a_{-\kk}\rangle\rangle&:=&
\lim_{\nu\searrow0}\lim_{L\to\infty}
\langle\langle 
a_\kk,a_{-\kk}\rangle\rangle_\nu
;\\
s_\kk&:=&
\lim_{\nu\searrow0}\lim_{L\to\infty}\frac{\langle N_\kk^{*} N_\kk\rangle_\nu}{\langle
  N\rangle_\nu};\\
\chi_\kk&:=&
\lim_{\nu\searrow0}\lim_{L\to\infty}\frac{\langle\langle N_\kk^{*}, N_\kk\rangle\rangle_\nu}{\langle
  N\rangle_\nu}.\end{eqnarray*}
(Compare with the definition of $s_\kk$ and $\chi_\kk$ in (\ref{chi1})
and (\ref{chi})). 

In this setting we have the following analog of (\ref{f-sum}):
\begin{equation}
\frac12[ N_\kk^{*},[H_\nu, N_\kk]]=\frac{\kk^2}{2}N+ \frac{\nu\sqrt{V}}{2}
(a_\0+a_\0^*).
\end{equation}
It implies the so-called
 {\em f-sum rule}:
\begin{equation}
\frac12\langle  N_\kk^{*}(H_\nu-E_\nu) N_\kk\rangle_\nu+
\frac12\langle  N_\kk(H_\nu-E_\nu) N_\kk^{*}\rangle_\nu
=\frac{\kk^2}{2}\langle N\rangle_\nu+\nu\sqrt{V}\langle a_\0\rangle_\nu.
\label{f-sum1}\end{equation}

By the Schwarz inequality and taking thermodynamic limit, we obtain
\begin{equation}\label{sqth1}
s_\kk\leq \frac12 |\kk|\sqrt{\chi_\kk}.\end{equation}

In the theorem below,  (\ref{poq1}) is due to Pitaevski and Stringari
\cite{PSt,St1}, and (\ref{poq2}) is the zero-temperature version of the 
famous $\frac{1}{\kk^2}$ Theorem 
of Bogoliubov \cite{Bog1}.

\begin{thm}
\begin{eqnarray}
\langle a_\kk^*a_\kk\rangle&\geq&\frac{\kappa}
{4s_\kk\rho}-\frac12,\label{poq0}\\
&\geq&\frac{\kappa}
{2|\kk|\sqrt{\chi_\kk}\rho}-\frac12,\label{poq1}\\
\langle\langle a_\kk,a_\kk^*\rangle\rangle&\geq&
\frac{\kappa}{\rho \kk^2}\label{poq2}\\&&+\left|
\langle\langle a_\kk, a_{-\kk}\rangle\rangle+
\frac{\kappa}{ \rho\kk^2}\right|.\nonumber
\end{eqnarray}
\end{thm}

\proof
To simplify the presentation, our proof will be  not
quite rigorous, since we will ignore $\nu$ and
 skip thermodynamical limit involving
$\lim\limits_{\nu\searrow0}\lim\limits_{L\to\infty}$.

We set $A^*=a_\kk$ and $B:=N_\kk$ in the uncertainty relation (\ref{pui})
 and we obtain
\[\left(\langle
a_\kk^*a_\kk\rangle+\frac12\right)
\langle N_\kk ^* N_\kk \rangle\geq\frac14|\langle
a_\0\rangle|^2.\]
This proves (\ref{poq0}).
Now (\ref{sqth1}) implies    (\ref{poq1}).


To prove (\ref{poq2}) introduce the operators
\begin{eqnarray*}
Q_\kk&:=& N_\kk + N_\kk ^*,\\
R_\kk&=&\i [Q_\kk,H].
\end{eqnarray*}
We obtain
\begin{eqnarray*}
[Q_\kk,[H,Q_\kk]]&=&2\kk^2N-\kk^2 Q_{2\kk},\\
{}[Q_\kk,a_\kk]&=&-a_\0-a_{2\kk},\\ 
{}[Q_\kk,a_{-\kk}^*]&=&a_\0^*+a_{-2\kk}^*.
\end{eqnarray*}
Therefore,
\begin{eqnarray}
\frac12\langle\langle R_\kk,R_\kk\rangle\rangle
&=&\frac12\langle[Q_\kk,[H,Q_\kk]\rangle\nonumber
\\
&=&\langle N\rangle\kk^2,\label{bogo-}\\
\langle\langle a_\kk,R_\kk\rangle\rangle&=&\i\langle [Q_\kk,a_\kk]\rangle \nonumber\\
&=&-\i\langle a_\0\rangle,\label{bogo1}\\
\langle\langle a_{-\kk}^*,R_\kk\rangle\rangle&=&\i\langle
	       [Q_\kk,a_{-\kk}^*]\rangle
\nonumber\\
&=&\i\langle a_\0^*\rangle\label{bogo2}.
\end{eqnarray}
(\ref{bogo1}) and (\ref{bogo2}) are sometimes called {\em Bogoliubov sum
  rules} \cite{Bog1,St1}.

For a complex parameter $t$ we have
\begin{eqnarray}\label{eqna}
-\i\langle a_\0\rangle+\i t\langle a_\0^*\rangle&=&
\langle\langle(a_\kk+t a_{-\kk}^*),R_\kk\rangle\rangle.
\end{eqnarray}
We take the square of the absolute value of (\ref{eqna}), apply
(\ref{epsi3a}),   and we obtain
\begin{eqnarray*}&&
|\langle a_\0\rangle|^2-t\langle a_\0^*\rangle^2
-\bar t\langle a_\0\rangle^2+|t|^2|\langle a_\0\rangle|^2\\
&\leq&
\langle\langle(a_\kk+t a_{-\kk}^*),(a_\kk^*+\bar t a_{-\kk})\rangle\rangle
\langle\langle R_\kk ,R_\kk\rangle\rangle.
\end{eqnarray*}
Taking into account (\ref{bogo-}), we obtain
\begin{eqnarray*}
0&\leq&
\langle\langle a_\kk,a_\kk^*\rangle\rangle-\frac{|\langle a_\0\rangle|^2}{\langle
  N\rangle \kk^2}\\
&&+t\left(
\langle\langle 
a_{-\kk}^*,a_\kk^*\rangle\rangle+\frac{\langle a_\0^*\rangle^2}{\langle
  N\rangle \kk^2}\right)\\
&&+\bar t\left(
\langle\langle 
a_{\kk},a_{-\kk}\rangle\rangle+\frac{\langle a_\0\rangle^2}{\langle
  N\rangle \kk^2}\right)\\
&&+| t|^2\left(
\langle\langle 
a_{-\kk}^*,a_{-\kk}\rangle\rangle-\frac{|\langle a_\0\rangle|^2}{\langle
  N\rangle \kk^2}\right)
.\end{eqnarray*}
Using
\begin{eqnarray*}\langle a_\0\rangle=\langle a_\0^*\rangle,&
\langle\langle 
a_{-\kk}^*,a_{-\kk}\rangle\rangle=\langle\langle 
a_{\kk}^*,a_{\kk}\rangle\rangle,&
\langle\langle 
a_{-\kk}^*,a_{\kk}^*\rangle\rangle=
\langle\langle 
a_{\kk},a_{-\kk}\rangle\rangle,\end{eqnarray*}
we obtain
\[\left|\langle\langle a_\kk,a_{-\kk}\rangle\rangle
+\frac{\langle a_\0\rangle^2}{\langle
  N\rangle \kk^2}\right|^2\leq
\langle\langle 
a_{\kk}^*,a_{\kk}\rangle\rangle-\frac{|\langle a_\0\rangle|^2}{\langle
  N\rangle \kk^2},\]
which implies 
 (\ref{poq2}). 
\qed

Note the following consequence of (\ref{poq2}):
\begin{eqnarray}\langle\langle 
a_{\kk}^*,a_{\kk}\rangle\rangle-
\langle\langle 
 a_{\kk}^*,a_{-\kk}^*\rangle\rangle&\geq\frac{
2\kappa}{\rho\kk^2}.\label{poq9}
\end{eqnarray}

\begin{thm}
Let $\epsilon(\kk)$ be the IES at momentum $\kk$. Then
\begin{eqnarray}\epsilon(\kk)^2&\leq&
\frac{\kk^2\rho}{2\kappa}\Big(\frac{\kk^2}{2}-\mu+\rho\hat
  v(0)\label{wg}\\
&&\!\!\!\!\!\!\!\!\!\!+\frac{1}{2(2\pi)^d}
\int\hat v(\kk)\left(2\langle a_{\q+\kk}^* a_{\q+\kk}\rangle+
\langle a_{\q+\kk}^* a_{-\q-\kk}^*\rangle+
\langle a_{\q+\kk} a_{-\q-\kk}\rangle\right)\d\q\Big),\nonumber\\
\epsilon(\kk)^2&\leq&
\left(\frac{\kk^2}{2}\right)^2+2\kk^2\int\frac{|\q|^2}{2}\left\langle
a^*_{\q} a_{\q}
\right\rangle\d\q
\label{wag1}
\\&&+\rho\int\d \x(1-\cos \kk\x)\nabla_{\hat \kk}^{(2)}v(\x)\langle a_\0^*a_\x^*a_\x
a_\0 \rangle,
\nonumber\end{eqnarray}
where $\hat \kk$ denotes $|\kk|^{-1}\kk$ and
$\nabla_{\hat \kk}^{(2)}v(\x)$ denotes the second derivative of $v$ in the
direction of $\hat \kk$.
\label{thml}\end{thm}
\proof To prove (\ref{wg}),
we use (\ref{epsi2}) with $A^*=a_\kk-a_{-\kk}^*$ and $B=N_\kk$ 
and  then
go to thermodynamic limit. 

To prove (\ref{wag1}) we use  (\ref{epsi9}) with $A=N_\kk$
\qed

Both estimates of Theorem (\ref{thml})  indicate the
phononic character of the excitation spectrum.
The estimate (\ref{wg}) is due to Wagner \cite{Wa,St1}) and involves the
symmetry breaking parameter $\kappa$. The estimate (\ref{wag1}) 
involves the kinetic energy and the pair correlation function
$\langle a_\0^*a_\x^*a_\x a_\0\rangle$ (here $0$ refers to the position), but does not
involve $\kappa$, hence it can be applied to situations without symmetry
breaking. This estimate comes from \cite{Pu,PW}, see also \cite{St1}.  

\subsection{Green's functions of the Bose gas}

Let us  consider two-point Green's functions of
the Bose gas.
 We assume that
$\langle\cdot\rangle$ is the state obtained by the limiting procedure
 in thermodynamic limit, and we will ignore the complications
 due to
thermodynamic limit.

We define a $2\times2$ matrix of Green's functions
\[G(z,\kk):=\left[\begin{array}{cc}
G_{11}(z,\kk)&G_{21}(z,\kk)\\
G_{12}(z,\kk)&G_{22}(z,\kk)
\end{array}\right],\]
\begin{eqnarray*}
G_{11}(z,\kk)&=&
\langle a_\kk(H-E-z)^{-1}a_\kk^*\rangle
+\langle a_\kk^*(H-E+z)^{-1}a_\kk\rangle,\\
G_{21}(z,\kk)&=&
\langle a_{-\kk}^*(H-E-z)^{-1}a_\kk^*\rangle
+\langle a_\kk^* (H-E+z)^{-1}a_{-\kk}^*\rangle,\\
G_{12}(z,\kk)&=&
\langle a_\kk (H-E-z)^{-1}a_{-\kk}\rangle
+\langle a_{-\kk}(H-E+z)^{-1}a_\kk \rangle,\\
G_{22}(z,\kk)&=&
\langle a_{-\kk}^* (H-E-z)^{-1}a_{-\kk}\rangle
+\langle a_{-\kk}(H-E+z)^{-1}a_{-\kk}^*\rangle.
\end{eqnarray*}

Note that, using the notation of Appendix \ref{Green's functions},
\[G_{ij}(z,\kk)=G_{A_i,B_j}(z),\]
 where
 $A_1:=a_\kk$, $A_2:=a_{-\kk}^*$ and
$B_1:=a_\kk^*$, $B_2:=a_{-\kk}$. We use the conventions
 described
 in this appendix for the meaning of Green's functions both
away from the
 real line and on the real line.

It is a general fact, which does not depend on the details of the system,  that
\begin{eqnarray*}
G_{11}(z,\kk)&=\overline{G_{11}(\bar z,\kk)}&= G_{22}(-z,-\kk),\\
G_{12}(z,\kk)&=\overline{G_{21}(\bar z,\kk)}&= G_{12}(-z,-\kk).
\end{eqnarray*}
By the reflection invariance of the Bose gas
\begin{eqnarray}
G_{ij}(z,\kk)&=&G_{ij}(z,-\kk).\end{eqnarray}

Obviously, for any observable $A$,
 $\langle A^*\rangle=\overline{\langle A\rangle}$. But the state
  $\langle\cdot\rangle$ is real, hence $\overline{\langle A\rangle}
=\langle\,\bar A\,\rangle$. Note also that
$H=\bar H$, $a_\kk=\bar a_\kk$. Therefore,
\begin{eqnarray*}
G_{12}(z,\kk)=G_{21}(z,\kk).\end{eqnarray*}

Note that $G_{11}(0,\kk)=\langle\langle a_\kk^*,a_\kk\rangle\rangle$
and $G_{12}(0,\kk)=\langle\langle a_\kk,a_{-\kk}\rangle\rangle$, hence by (\ref{poq9})
\begin{eqnarray}
G_{11}(0,\kk)-G_{12}(0,\kk)\geq \frac{c}{\kk^2}.
\label{ine}\end{eqnarray}

Let us introduce the ``full mass operator''
\begin{eqnarray*}
&&\Sigma(z,\kk)\\
&=&\left[\begin{array}{cc}
\Sigma_{11}(z,\kk)&\Sigma_{12}(z,\kk)\\
\Sigma_{21}(z,\kk)&\Sigma_{22}(z,\kk)
\end{array}\right]
:=\frac{1}{2\pi}\left[\begin{array}{cc}
G_{11}(z,\kk)&G_{12}(z,\kk)\\
G_{21}(z,\kk)&G_{22}(z,\kk)
\end{array}\right]^{-1}\\
&=&\frac{1}{2\pi}\left(G_{11}(z,\kk)G_{22}(z,\kk)
-G_{12}(z,\kk)G_{21}(z,\kk)\right)^{-1}
\left[\begin{array}{cc}
G_{22}(z,\kk)&-G_{12}(z,\kk)\\
-G_{21}(z,\kk)&G_{11}(z,\kk)
\end{array}\right]
.\end{eqnarray*}

Consequently,
\begin{eqnarray*}
\Sigma_{11}(0,\kk)-\Sigma_{12}(0,\kk)
&=&\frac{1}{2\pi\left(G_{11}(0,\kk)-G_{12}(0,\kk)\right)}.
\end{eqnarray*}
(\ref{ine}) implies
\begin{eqnarray*}
\frac{\kk^2}{2\pi c}\geq\Sigma_{11}(0,\kk)-\Sigma_{12}(0,\kk)\geq0,
\end{eqnarray*}
and in particular
\begin{eqnarray}
\Sigma_{11}(0,0)-\Sigma_{12}(0,0)&=&0.\label{theor}
\end{eqnarray}
(\ref{theor}) was first proven in the framework of perturbation theory for the
Bose gas with isolated condensate, and is sometimes called the Hugenholz-Pines
Theorem, \cite{HP}, see
also \cite{GN}. The proof that we present is valid for the
correct Hamiltonian of the Bose gas and is due to Bogoliubov \cite{Bog1}.

Note that (\ref{theor}) implies that $G(z,\kk)$ has a singularity at
$(z,\kk)=(0,0)$, which is an argument for the absence of a gap in the
excitation spectrum. Bogoliubov \cite{Bog1} gives also an argument for
the phononic shape of the excitation spectrum. The argument is based
on the assumption that
 $\Sigma(z,\kk)$
is regular in $z$, $\kk$ around $(0,0)$. 
Note that
\begin{eqnarray}\nonumber
\det\Sigma(z,\kk)
&=&\Sigma_{11}(z,\kk)\Sigma_{22}(z,\kk)-\Sigma_{12}(z,\kk)\Sigma_{21}(z,\kk) 
\end{eqnarray}
 is invariant with respect to the transformations
$\kk\mapsto-\kk$ and $z\mapsto-z$. Finally, by (\ref{theor}),
we know that
 $\Sigma(0,0)=0$. Therefore,
\begin{eqnarray}\nonumber
\det\Sigma(z,\kk)
&=&\gamma z^2+\beta \kk^2+O(|z|^4+|\kk|^4)
.\label{deter}
\end{eqnarray}
We have $\bar{\det\Sigma(z,\kk)}=
\det\Sigma(\bar z,\kk)$. Hence $\gamma$ and $\beta$ are
real as well.
For purely imaginary nonzero 
$z$,  $\det\Sigma(z,\kk)$ is nonzero. 
Hence
$\gamma$ and $\beta$ cannot have 
the same  sign.
Therefore, 
$\delta=-\frac{\beta}{\gamma}\geq0$.

Assume now that $\beta,\gamma$ are not zero. Then $0<\delta<\infty$,
and 
\[\det\Sigma(\sqrt{\delta}|\kk|,\kk)=O(|\kk|^4).\] Hence,  for small
$\omega,\kk$, the Green's function
 $G(\omega,\kk)$ has a sharp peak along
$\omega(\kk)=\sqrt{\delta}|\kk|$.

\appendix
\section{Energy-momentum spectrum of quadratic Hamiltonians}
\label{a0}

 Suppose
that we consider a quantum system described by the Hamiltonian
\begin{equation}
H=\int_{\rr^d} \omega(\kk)a_\kk^*a_\kk\d\kk,\label{quad}\end{equation}
with the the total momentum
\[P=\int_{\rr^d} \kk a_\kk^*a_\kk\d\kk,\]
both acting on the Fock space $\Gamma_\s(L^2(\rr^d))$.
We will call the function $\omega$ appearing in $H$
the {\em elementary excitation spectrum} of our
quantum system and we will assume it to be nonnegative.


Clearly, the ground state energy of $H$ is $0$. The excitation spectrum of
 (\ref{quad})  is not arbitrary -- it has to be a subadditive function.
This appendix  describes a number of easy results
about subadditive functions. 
 They are quite straightforward
 and  probably they mostly belong to the folk
wisdom. However, we have never seen them explicitly described 
in the literature, and we believe them to be relevant for physical properties
 of Bose gas.


The Hamiltonian of interacting Bose gas is not 
purely quadratic. Nevertheless, some arguments indicate
that the infimum
of  its excitation spectrum in thermodynamic limit 
is subadditive.
A heuristic  argument  in favor of this conjecture is described in the next
appendix. 

Even if one questions this  argument, there exists
 another motivation for a study
of subadditive functions.
Quadratic Hamiltonians are often used in statistical physics as approximate
effective Hamiltonians. In particular, this is the case of 
the Bogoliubov Hamiltonian $H_{\Bog}$.

We will show  that there exists a large
 class of subadditive functions
with the properties
 properties described by Conjecture \ref{conj1a} or \ref{conj1}, 
(3) or (3)', and  (4) (which correspond to the superfluidity or periodicity,
 and to a finite speed of sound).
 We will also show that if the elementary excitations possess these
 properties, then so does the IES.

Let $\rr^d\ni \kk\mapsto\epsilon(\kk)\in
\rr$ be a nonnegative function. We say that it is
{\em subadditive} iff
\[\epsilon(\kk_1+\kk_2)\leq \epsilon(\kk_1)+\epsilon(\kk_2),\ \ \
\kk_1,\kk_2\in\rr^d.\]

Let $\rr^d\ni \kk\mapsto\omega(\kk)\in
\rr$ be another
 nonnegative function. We define the {\em subbadditive hull of
  $\omega$} to be
\[\epsilon(\kk):=
\inf\{\omega(\kk_1)+\cdots+\omega(\kk_n)\ :\ \kk_1+\cdots+\kk_n=\kk,\
n=1,2,\dots\}.\]
Clearly, $\epsilon(\kk)$ is  subadditive and satisfies
$\epsilon(\kk)\leq\omega(\kk)$. 

Clearly, if $\omega(\kk)$ is the elementary excitation spectrum of a quadratic
Hamiltonian, and $\epsilon(\kk)$ its subadditive hull, then $\epsilon(\kk)$ is
the infimum of its excitation spectrum.

Let us state and prove some facts about subadditive functions and subadditive
hulls, which seem to be relevant for the homogeneous Bose gas.

\begin{thm}
Let $f$ be an increasing concave function on $[0,\infty[$ 
with $f(0)\geq0$. Then $f(|\kk|)$ is
subadditive. 
\label{subba}\end{thm}

\noindent{\bf Proof.}
\begin{eqnarray}\nonumber
f(|\kk_1+\kk_2|)&\leq &f(|\kk_1|+|\kk_2|)\\
\nonumber&\leq&
\frac{|\kk_1|}{|\kk_1|+|\kk_2|}
f(|\kk_1|+|\kk_2|)+
\frac{|\kk_2|}{|\kk_1|+|\kk_2|}
f(0)\\
\nonumber
&&+
\frac{|\kk_2|}{|\kk_1|+|\kk_2|}
f(|\kk_1|+|\kk_2|)+
\frac{|\kk_1|}{|\kk_1|+|\kk_2|}
f(0)
\\
\nonumber
&\leq& f(|\kk_1|)+ f(|\kk_2|).
\end{eqnarray}
\qed

We can generalize Theorem \ref{subba} to periodic functions.

\begin{thm}
Let $f$ be an increasing concave function on $[0,\frac{\sqrt d}{2}]$ 
with $f(0)\geq0$. Define $\epsilon$ to be the function on $\rr^d$ periodic
with respect to the lattice $\zz^d$ such that if $\kk\in[-\frac12,\frac12]^d$,
then $\epsilon(\kk)=f(|\kk|)$ (which defines $\epsilon$ uniquely). Then $\epsilon$
is subadditive.
\label{subba1}\end{thm}

\proof We can extend $f$ to a concave increasing function defined on
  $[0,\infty[$,  e.g.
  by putting $f(t)=f(\frac{\sqrt d}{2})$ for $t\geq\frac{\sqrt d}{2}$.

Let $\kk_1,\kk_2\in\rr^d$. Let
$\pp_1,\pp_2\in[-\frac12,\frac12]^d$ such that $\kk_i-\pp_i\in\zz^d$. Let $\pp\in [-\frac12,\frac12]^d$ such that $\kk_i+\kk_2-\pp
\in\zz^d$. Note that $|\pp|\leq|\pp_1+\pp_2|$. Now
\begin{eqnarray}\nonumber
\epsilon(\kk_1+\kk_2)&=&f(|\pp|)
\leq
f(|\pp_1+\pp_2|)
\\
\nonumber
&\leq&\cdots\\
\nonumber
&\leq& f(|\pp_1|)+ f(|\pp_2|)=\epsilon(\kk_1)+\epsilon(\kk_2),
\end{eqnarray}
where in $\dots$ we repeat the estimate of the proof of Theorem  \ref{subba}.
\qed

Obviously, we have
\begin{thm}
Let $\epsilon_0$ be subadditive and $\epsilon_0\leq\omega$. Let $\epsilon$ be
the subadditive hull of $\omega$. Then $\epsilon_0\leq\epsilon$.
\label{subba2}\end{thm}

In the case of the Bose gas with repulsive interactions
 we expect that 
 the excitation
spectrum may have resemble that of a quadratic Hamiltonian with the properties
 described by the following two theorems, which easily
 follow from
Theorems \ref{subba}, \ref{subba1} and \ref{subba2}:

\begin{thm}
Suppose that $\omega\geq0$ is a spherically symmetric function on $\rr^d$ 
 and $\epsilon$ is its subadditive hull.
\begin{enumerate}
\item $\epsilon$ is spherically symmetric.
\item If $\inf\limits_{\kk\neq0}\frac{\omega(\kk)}{|\kk|}=c$, then
$\inf\limits_{\kk\neq0}\frac{\epsilon(\kk)}{|\kk|}=c$.
\item If $\liminf_{\kk
\to0}\frac{\omega(\kk)}{|\kk|}=c$, then
  $\epsilon(\kk)\leq c|\kk|$. \item
Suppose that for some $c>0$, we have $\omega(\kk)
\geq c\min\left(|\kk|,1\right)$.
 Then
 \[\liminf_{\kk\to0}\frac{\omega(\kk)}{|\kk|}=c_\ph\ \hbox{ implies }\
\lim_{\kk\to0}\frac{\epsilon(\kk)}{|\kk|}=c_\ph.\]
\end{enumerate}\label{subba3}\end{thm}

\begin{thm}
Suppose that $\omega\geq0$ is an even function on $\rr$
periodic with respect to $\zz$. Let 
 $\epsilon$ be its subadditive hull.
\begin{enumerate}\item 
$\epsilon(\kk)$ is even and periodic with respect to $\zz$. 
\item
If,
for some $c>0$, we have $\omega(\kk)
\geq c\dist(\kk,\zz)$,
 then
 \[\liminf_{\kk\to0}\frac{\omega(\kk)}{|\kk|}=c_\ph\ \hbox{ implies }\
\lim_{\kk\to0}\frac{\epsilon(\kk)}{|\kk|}=c_\ph.\]
\end{enumerate}
\end{thm}

\section{Subadditivity of the excitation spectrum of interacting Bose gas}
\label{a00}

In this appendix we describe a heuristic argument in favor of Conjecture 
 \ref{conj1} (5). Recall that this conjecture says that the IES
 of interacting Bose gas in thermodynamic limit
 should be subadditive. Clearly, this would be true if the Bose gas was
 described by a quadratic Hamiltonian of a form \ref{quad}. We will see,
 however, that this conjecture
 follows as well from an assumption saying 
 that one can
 describe excitations by approximately localized operators.

Consider Bose gas in a  box of side length $L$ where $L$ is very large.
Let
$\Phi_0$ be the ground state of the Hamiltonian and $E_0$ its ground state
energy, so that $H\Phi_0=E_0\Phi_0$ and $P\Phi_0=0$.
Let $(E_0+e_i,\kk_i)\in\sp(H,P)$, $i=1,2$. We can find eigenvectors with these
eigenvalues, that is, vectors $\Phi_i$ satisfying $H\Phi_i=(E_0+e_i)\Phi_i$,
$P\Phi_i=\kk_i\Phi_i$.  Let us make the assumption that  it is possible to find
operators $A_i$, which are polynomials in creation and annihilation operator
smeared with functions well localized in  configuration space such that
$PA_i\approx A_i(P+\kk_i)$,  and which approximately 
create the vectors $\Phi_i$ from the
ground state, that is
 $\Phi_i\approx A_i\Phi_0$.
(Note that here a large size of $L$ plays a role).
 By replacing $\Phi_2$ with $\e^{\i \y P}\Phi_2$ for
some $\y$ and $A_2$ with $\e^{\i\y P}A_2 \e^{-\i\y P}$, we can make sure that the
regions of localization of $A_1$ and $A_2$  are separated by a large distance.

 Now consider the vector $\Phi_{12}:=
A_1A_2\Phi_0$. Clearly, \[P\Phi_{12}\approx (\kk_1+\kk_2)\Phi_{12}.\]
$\Phi_{12}$ looks like the vector $\Phi_i$ in
 the region of localization of $A_i$, elsewhere it looks like $\Phi_0$.
 The
Hamiltonian $H$  involves only
 expressions of short range (the potential decays in space). Therefore, 
we expect that
\[H\Phi_{12}\approx (E_0+e_1+e_2)\Phi_{12}.\]
If this were the case, it
 would imply that $(E_0+e_1+e_2,\kk_1+\kk_2)\in\sp(H,P)$, and hence would
show that the IES is subadditive.

Clearly, the argument we presented has its weak points -- it
 is based on approximate locality, which can be violated because of
 correlations due to the Bose-Einstein condensation. 
Nevertheless, we have the impression that many  physicists believe that 
even in the interacting case, in the thermodynamic limit, 
 one can often
 ``compose excitations'' in a sense similar to the one described above.
  (See
 the  discussion of the concept of elementary excitations in interacting
 systems by Lieb \cite{L2}).

\section{Speed of sound at zero temperature}
\label{Speed of sound}

It is well known (e.g. \cite{Hu}) that at any temperature the speed of sound is given by
\[c_\s=\sqrt{\frac{\partial p}{\partial \rho}\Big|_S},\]
where $p$ is the pressure, $\rho$ is the density  and $S$ is the
entropy. (Recall that we assume that the mass of an individual particle is
$1$).

Let $E$ denote the ground state energy (which corresponds to the total energy
at zero temperature), $V$ the volume and $n$ the number of particles. Note
that $n=V\rho$ and $E=Ve(\rho)$, where $e(\rho)$ denotes the energy density.
At zero temperature the    pressure is given by
\[p=-\frac{\partial E}{\partial V}\Big|_n=-e(\rho)+\rho e'(\rho).\]

Clearly, at zero temperature the entropy is zero.
Therefore, 
\begin{eqnarray*}
c_\s^2&=&\frac{\partial p}{\partial \rho}\Big|_{S=0}
=\frac{\partial p}{\partial \rho}\Big|_{T=0}\\
&=&\frac{\partial }{\partial \rho}
\left(-e(\rho)+\rho e'(\rho)\right)=\rho e''(\rho).
\end{eqnarray*}

\section{Wick and anti-Wick symbol}
\label{Wick and anti-Wick symbol}

Let  $a^*_1,a_2^*,\dots,a_n^*$
  and $a_1,a_2,\dots, a_n$ be creation/annihilation
operators.
Let $H$ be an operator given as a polynomial in these
operators. We can write $H$
 in two ways:
\begin{eqnarray*}
H=&
\sum\limits_{\gamma,\delta} h_{\gamma,\delta}(a^*)^\gamma a^\delta=&
\sum_{\gamma,\delta} \tilde  h_{\gamma,\delta} a^\delta(a^*)^\gamma .
\end{eqnarray*}
(We use here the multiindex notation, e.g.
$a^\alpha=a_1^{\alpha_1}\cdots a_n^{\alpha_n}$).
Then the function 
\[\cc^n\ni\alpha\mapsto   H(\alpha)=\sum_{\gamma,\delta}
 h_{\gamma,\delta}\bar\alpha^\gamma\alpha^\delta\]
is called the {\em Wick symbol} of the operator $H$. (Synonyms: lower symbol,
normal symbol,
covariant symbol, $a^*,a$-symbol, $Q$-representation).
The function
\[\cc^n\ni\alpha\mapsto \tilde H(\alpha)=\sum_{\gamma,\delta}
\tilde h_{\gamma,\delta}\bar\alpha^\gamma\alpha^\delta\]
is called the {\em anti-Wick symbol} of the operator $H$. (Synonyms: upper
 symbol,
anti-normal symbol,
contravariant symbol, $a,a^*$-symbol, $P$-representation).

Introduce the standard coherent states: \[W(\alpha):=\exp \left(\sum_{i=1}^n (-\alpha_i a_i^*+\bar\alpha_i a_i) \right),\ \ \ 
\Omega_\alpha:=W(\alpha)\Omega.\]
Note the identities (that can be used as alternative definitions of the Wick
and anti-Wick symbols):
\begin{eqnarray}
H(\alpha)&=&(\Omega_\alpha|H\Omega_\alpha),\label{pss1}\\
H&=&\int_\cc\tilde H(\alpha)|\Omega_\alpha)(\Omega_\alpha|\frac{\d^2\alpha}{\pi}
.\label{pss2}\end{eqnarray}

Let $H$ be a bounded from below self-adjoint operator.
We have the following lower and upper bound for the ground state energy of
$H$, which follow immediately from (\ref{pss1}) and (\ref{pss2}):
\[\inf\{\tilde H(\alpha)\ :\ \alpha\in\cc^n\}
\leq\inf\sp H
\leq
\inf\{H(\alpha)\ :\ \alpha\in\cc^n\}.\]

\section{Bogoliubov transformations}
\label{a1}

In this appendix we recall the  
well-known properties of Bogoliubov transformations
and squeezed vectors. For simplicity we restrict ourselves to one degree of
freedom. 

Let $a^*$, $a$ are creation and annihilation operators and $\Omega$ the vacuum
vector. Recall that $[a,a^*]=1$ and $a\Omega=0$.

Here are the basic identities for Bogoliubov translations and coherent
vectors. Let
\[W_\alpha:=\e^{-\alpha a^*+\bar\alpha a}.\]
Then
\[\begin{array}{rl}
W_\alpha aW_\alpha^*&=a+\alpha,\\[3mm]
W_\alpha a^*W_\alpha^*&=a^*+\bar\alpha,\\[3mm]
W_\alpha^*\Omega&=\e^{-\frac{|\alpha|^2}{2}}\e^{\alpha a^*}\Omega.
\end{array}\]

Here are
 the basic identities for Bogoliubov rotations and squeezed vectors.
Let 
\[U_\theta:=\e^{-\frac\theta2a^*a^*+\frac{\bar\theta}{2}aa}.\]
Then
\[\begin{array}{rl}
U_\theta aU_\theta^*&=\cosh|\theta| a+\frac{\theta}{|\theta|}\sinh|\theta| a^*,\\[3mm]
U_\theta a^*U_\theta^*&=\cosh|\theta| a^*+\frac{\bar\theta}{|\theta|}\sinh|\theta| a,\\[3mm]
U_\theta^*\Omega&=(1+\tanh^2|\theta|)^{\frac14}
\e^{-\frac{\theta}{2|\theta|}\tanh|\theta| a^*a^*}\Omega.
\end{array}\]

Vectors obtained by acting with both Bogoliubov translation and rotation
will be  also called squeezed vectors.

\section{Computations of the Bogoliubov rotation}
\label{a2}

In this appendix we give the computations of the rotated terms in the
Hamiltonian used in Section \ref{s5}.

\begin{eqnarray}
U_\theta a_{\kk}^*a_{\kk}U_\theta^*&=&|s_{\kk}|^2\nonumber\\
&&+c_{\kk}^2{a}_{\kk}^*{a}_{\kk}-c_{\kk}s_{\kk}{a}_{\kk}^*{a}_{-{\kk}}^*
- c_{\kk}\bar s_{\kk} {a}_{\kk}{a}_{-{\kk}}+|s_{{\kk}}|^2{a}_{-{\kk}}^*{a}_{-{\kk}},\nonumber\\
U_\theta a_{\kk}^*a_{-{\kk}}^*U_\theta^*
&=&-\bar s_{\kk}c_{\kk}\nonumber\\
&&+c_{\kk}^2{a}_{\kk}^*{a}_{-{\kk}}^*-c_{\kk}\bar s_{\kk}{a}_{\kk}^*{a}_{\kk} -c_{\kk}\bar s_{\kk} {a}_{-{\kk}}^*{a}_{-{\kk}}+\bar
s_{\kk}^2{a}_{-{\kk}}{a}_{\kk} ;\nonumber\\
U_\theta a_{\kk} a_{-{\kk}}U_\theta^*&=&-s_{\kk} c_{\kk}\nonumber\\
&&+c_{\kk}^2{a}_{\kk}{a}_{-{\kk}}- c_{\kk} s_{\kk}{a}_{\kk}^*{a}_{\kk} -c_{\kk}
s_{\kk} {a}_{-{\kk}}^*{a}_{-{\kk}}+
s_{\kk}^2{a}_{-{\kk}}^*{a}_{\kk}^*;\nonumber
\end{eqnarray}

\begin{eqnarray}
U_\theta a_{{\kk}+{\kk}'}^*a_{\kk} a_{{\kk}'}U_\theta^*&=&
\left(c_0(|s_{\kk}|^2\delta({\kk}')+|s_{{\kk}'}|^2\delta({\kk}))
+\bar s_0 c_{\kk}s_{\kk}\delta({\kk}+{\kk}')\right){a}_0\nonumber
\\
&&
-\left(s_0(|s_{\kk}|^2\delta({\kk}')+|s_{{\kk}'}|^2\delta({\kk}))
+c_0\bar c_{\kk}s_{\kk}\delta({\kk}+{\kk}')\right){a}_0^*\nonumber\\
&&+\hbox{higher order terms};\nonumber\\
U_\theta a_{{\kk}+{\kk}'}a_{\kk}^*a_{{\kk}'}^*U_\theta^*&=&
\left(c_0(|s_{\kk}|^2\delta({\kk}')+|s_{{\kk}'}|^2\delta({\kk}))+s_0c_{\kk}\bar
s_{\kk}\delta({\kk}+{\kk}')\right){a}_0^* \nonumber\\
&&-
\left(\bar s_0(|s_{\kk}|^2\delta({\kk}')+|s_{{\kk}'}|^2\delta({\kk}))
+ c_0 c_{\kk}\bar s_{\kk}\delta({\kk}+{\kk}')\right){a}_0\nonumber\\
&&+\hbox{higher order terms};\nonumber
\end{eqnarray}
\begin{eqnarray}
&&\delta({\kk}_1+{\kk}_2-{\kk}_3-{\kk}_4)U_\theta
a_{{\kk}_1}^*a_{{\kk}_2}^*a_{{\kk}_3}a_{{\kk}_4}U_\theta^*\nonumber\\
=&&
c_{{\kk}_1}\bar s_{{\kk}_1}
c_{{\kk}_3}s_{{\kk}_3}\delta({\kk}_1+{\kk}_2)\delta({\kk}_3+{\kk}_4)\nonumber \\
&+&|s_{{\kk}_1}|^2|s_{{\kk}_2}|^2\left(
\delta({\kk}_1-{\kk}_3)\delta({\kk}_2-{\kk}_4)+\delta({\kk}_1-{\kk}_4)\delta({\kk}_2-{\kk}_3)\right)
\nonumber\\
&+&\Bigl(\bar s_{{\kk}_1}c_{{\kk}_1}
(s_{{\kk}_3}c_{{\kk}_3} {a}_{-{\kk}_3}^*{a}_{-{\kk}_3}
-s_{{\kk}_3}^2{a}_{{\kk}_3}^*{a}_{-{\kk}_3}^*
- c_{{\kk}_3}^2 {a}_{{\kk}_3}{a}_{-{\kk}_3}
+s_{{\kk}_3} c_{{\kk}_3}{a}_{{\kk}_3}^*{a}_{{\kk}_3}) \nonumber\\&&+
s_{{\kk}_3} c_{{\kk}_3}
(\bar s_{{\kk}_1} c_{{\kk}_1} {a}_{{\kk}_1}^*{a}_{{\kk}_1}
-\bar s_{{\kk}_1}^2{a}_{{\kk}_1}{a}_{-{\kk}_1}
-c_{{\kk}_1}^2 {a}_{{\kk}_1}^*{a}_{-{\kk}_1}^*
+\bar s_{{\kk}_1}
c_{{\kk}_1}{a}_{-{\kk}_1}^*{a}_{-{\kk}_1})\Bigr)\nonumber\\&&\times
\delta({\kk}_1+{\kk}_2)\delta({\kk}_3+{\kk}_4)
\nonumber\\
&+&\Bigl(|s_{{\kk}_2}|^2(
|c_{{\kk}_1}|^2{a}_{{\kk}_1}^*{a}_{{\kk}_1}
-c_{{\kk}_1}s_{{\kk}_1}{a}_{{\kk}_1}^*{a}_{-{\kk}_1}^*-
 c_{{\kk}_1}\bar s_{{\kk}_1}{a}_{{\kk}_1}{a}_{-{\kk}_1}
+|s_{{\kk}_1}|^2{a}_{-{\kk}_1}^*{a}_{-{\kk}_1})
\nonumber\\
&&+
|s_{{\kk}_1}|^2(
|c_{{\kk}_2}|^2{a}_{{\kk}_2}^*{a}_{{\kk}_2}
-c_{{\kk}_2}s_{{\kk}_2}{a}_{{\kk}_2}^*{a}_{-{\kk}_2}^*-
 c_{{\kk}_2}\bar s_{{\kk}_2}{a}_{{\kk}_2}{a}_{-{\kk}_2}
+|s_{{\kk}_2}|^2{a}_{-{\kk}_2}^*{a}_{-{\kk}_2})\bigr)\nonumber\\ &&\times
\bigl(\delta({\kk}_1-{\kk}_3)\delta({\kk}_2-{\kk}_4)+
\delta({\kk}_1-{\kk}_4)\delta({\kk}_2-{\kk}_3)\Bigr)
\nonumber\\
&&+\hbox{higher order terms}.\nonumber\end{eqnarray}

\section{Operator inequalities}
\label{sec-ine}

Let us fix a vector $\Psi$ and let $\langle A\rangle$ denote $(\Psi|A\Psi)$.
Let $[A,B]_+:=AB+BA$ denote the anticommutator. Occasionally, we will write
$[A,B]_-:=AB-BA$ for the usual commutator.

\begin{thm}
Suppose that $A,B$ are operators. We have the following inequalities:
\begin{enumerate}\item
 {\bf Schwarz inequality for an anticommutator}
\begin{equation}
\left|\langle [A^*,B]_+\rangle\right|^2\leq\langle[A^*,A]_+\rangle
\langle[B^*,B]_+\rangle.\label{pui1}
\end{equation}
\item {\bf Uncertainty relation for a pair of operators}
\begin{equation}
|\langle[A^*,B]\rangle|^2\leq \langle [A,A^*]_+\rangle\langle
	  [B,B^*]_+\rangle.\label{pui}
\end{equation}\end{enumerate}
\end{thm}

\proof We add the inequalities
\[0\leq (A+tB)^*(A+t B), \ \ \ 0\leq(A\pm tB)(A\pm t B)^*, \]
obtaining
\[0\leq[A,A^*]_++\bar t[B^*,A]_\pm+t[A^*,B]_\pm+|t|^2[B^*,B]_+.\]
Then we take the expectation value of both sides and
set $t=-\frac{\langle[B^*,A]_\pm\rangle}{\langle[B^*,B]_+\rangle}.$ \qed

Suppose that $H$ is an operator bounded from below and $\Psi$ is its ground
state vector:
\[H\Psi=E\Psi,\ \ \ H-E\geq0.\]
 Assume that $\langle A\rangle=\langle B\rangle=0$. Then we will write
\[\langle\langle A,B\rangle\rangle:=
\langle A(H-E)^{-1}B\rangle+\langle B(H-E)^{-1}A\rangle
.\]

\begin{thm}
\begin{eqnarray}
|\langle\langle A^*,B\rangle\rangle|^2&\leq&
\langle \langle A^*,A\rangle\rangle
\langle \langle B^*,B\rangle\rangle;
\label{epsi3a}\\
|\langle[A^*,B]\rangle|^2&\leq&
\langle \langle A^*,A\rangle\rangle\langle[B^*,[H,B]]\rangle
.\label{epsi3}\end{eqnarray}
\end{thm}

\proof
To see (\ref{epsi3a}) we apply the Schwarz inequality to the positive
definite form $\langle\langle A^*,B\rangle\rangle$.

To obtain (\ref{epsi3})  we first use the identity
\begin{equation}
\langle [A^*,B]\rangle=\langle\langle A^*,[H,B]\rangle\rangle,
\label{toz}\end{equation}
and then  (\ref{epsi3a}).
\qed

\begin{thm}
Let ${\mathcal H}_0$ be the space
\[{\mathcal H}_0:=\{f_1(H)A\Psi+g(H)A^*\Psi\ :\ f,g\}^\cl\]
(the smallest invariant 
 subspace of the operator  $H$ containing $A\Psi$ and $A^*\Psi$).
Let $\epsilon:=\inf \sp H\Big|_{{\mathcal H}_0}-E$. Then
\begin{eqnarray}
\epsilon&\leq&
\frac{\langle [A^*,[H,A]]\rangle}{\langle[ A^*,A]_+\rangle}\label{epsi0}
,\\
\epsilon^2&\leq&\frac{\langle[A^*,[H,A]]\rangle}
{\langle\langle A^*,A\rangle\rangle}
,\label{epsi1}
\\
\epsilon^2&\leq&\frac{\langle[A^*,[H,A]]\rangle\langle[ B^*,[H,B]]\rangle}
{|\langle[ A^*,B]\rangle|^2},
\label{epsi2}\\
\epsilon^2&\leq&
\frac{\langle[[A^*,H][H,[H,A]]]\rangle}
{\langle [A^*,[H,A]]\rangle}.\label{epsi9}
\end{eqnarray}
\end{thm}
\proof
To prove (\ref{epsi0}) we add
\[\epsilon \langle A^*A\rangle\leq \langle A^*(H-E)A\rangle,\
 \ \ \epsilon\langle AA^*\rangle\leq\langle A(H-E)A^*\rangle.\]

 To prove (\ref{epsi1}) we add
\[\epsilon^2\langle A^*(H-E)^{-1}A\rangle\leq \langle A^*(H-E)A\rangle,\ \ 
\epsilon^2 \langle A(H-E)^{-1}A^*\rangle\leq \langle A(H-E)A^*\rangle.\]
(\ref{epsi2}) follows from (\ref{epsi1}) and (\ref{epsi3}).
\qed

(\ref{epsi0}) is called the Feynman bound \cite{St1} and 
(\ref{epsi2}) is due to Wagner \cite{Wa,St1}. (\ref{epsi9}) comes from
\cite{Pu,PW}.

\section{Green's functions}
\label{Green's functions}

 We consider a quantum system
described by a bounded from below Hamiltonian $H$. We assume
 that it has a unique ground state
$(\Psi|\cdot\Psi)=\langle\cdot\rangle$,
 $H\Psi=E\Psi$. If $A$
is an operator, then we will write
\begin{eqnarray*}
A(t):=\e^{\i tH}A\e^{-\i tH}.\end{eqnarray*}

The {\em time-dependent 
Green's} function associated to a pair of operators $A$, $B$ is
defined as the function depending on $t\in\rr$
\begin{eqnarray}
G_{A,B}^\td(t)&=&\theta(-t)\langle A(0) B(t)\rangle+\theta(t)\langle
B(t)A(0) \rangle
\nonumber\\
&=&\theta(-t)\langle A\e^{\i t( H-E)}B\rangle
+\theta(t)\langle B\e^{-\i t( H-E)}A\rangle,\label{green}\end{eqnarray}
where $\theta$ is the Heaviside function.

We also introduce the {\em energy-dependent Green's function}, which
is the Fourier transform of
(\ref{green}) (with one of conventional normalizations). It
 is the
distribution on $\omega\in\rr$ defined as
\begin{eqnarray}G_{A,B}(z)&=&
\lim_{\epsilon\searrow0}\i
\int G_{A,B}(t)\e^{-\i \omega t-\epsilon|t|}\d t\label{green1}\\
&=&\nonumber
\lim_{\epsilon\searrow0}
\i\int_0^\infty\left(\langle A\e^{-\i t (H-E-\omega-\i\epsilon)}B\rangle+
\langle B\e^{\i t(H-E+\omega-\i\epsilon)} A\rangle\right)\d t\\&=&
\langle A(H-E-\omega-\i0)^{-1}B\rangle+
\langle B(H-E+\omega-\i0)^{-1}A\rangle
.\nonumber\end{eqnarray}

Finally, the {\em analytic Green's function} is 
 defined for
 $z\in\C\backslash\left(\sp (H-E)\cup\sp(E-H)\right)$ and  is defined as
\begin{eqnarray}G_{A,B}^\an(z)&=&
\langle A(H-E-z)^{-1}B\rangle+
\langle B(H-E+z)^{-1}A\rangle
.\nonumber\end{eqnarray}
The distribution 
$G_{A,B}(\omega)$ is the boundary value of
 the analytic function
$G_{A,B}^\an(z)$, provided that we approach the 
 the real line from the appropriate side. Besides, in the energy gap
 both functions coincide:
\begin{eqnarray*}G_{A,B}(\omega)&=&
  G_{A,B}^\an(\omega-\i0),\ \ \omega\in\sp(E-H)\subset]-\infty,0];\\
G_{A,B}(\omega)&=&G_{A,B}^\an(\omega),\ \ \ \ \omega\in\rr\backslash
\{\sp(E-H)\cup\sp(E-H)\};\\
G_{A,B}(\omega)&=&
  G_{A,B}^\an(\omega+\i0),\ \ \omega\in\sp(H-E)\subset]0,\infty].
\end{eqnarray*} 
Motivated by the above relations, following the 
usual convention, we can
 treat
$G_{A,B}(\omega)$ and $G_{A,B}^\an(z)$ as restrictions of a single
fuction and  drop the subscript $\an$.
Note that
\[G_{A,B}(z)=\overline{G_{B^*,A^*}(\bar z)}
=G_{B,A}(-z).\]

Green's functions are well motivated physically. 
Let us briefly describe their two separate physical applications.

Following \cite{Bog1},
let us first describe
the physical meaning of the {\em static} Green's function
\[G_{A,B}(0)=\langle\langle A,B\rangle\rangle.\] 
Suppose that $\Psi$ is an eigenvector of $H$ (not necessarily a ground
state). Let $B$ be a perturbation with $1_{\{E\}}(H)B\Psi=0$. Suppose that it
  is possible to apply perturbation theory to the family $H_\tau:=H+\tau B$
  obtaining an analytic familly of eigenvectors $\Psi_\tau$ with eigenvalues
  $E_\tau$ such that $E_0=E$ and $\Psi_0=\Psi$. The Rayleigh-Schr\"odinger
  perturbation theory says that
\begin{equation}
\Psi_\tau=\Psi+\tau(H-E)^{-1}B\Psi+O(\tau^2).\label{pertu}\end{equation}
Let $\langle A\rangle_\tau:=(\Psi_\tau|A\Psi_\tau)$. (\ref{pertu}) implies
that 
 for any operator $A$ we have
\begin{equation}\frac{\d}{\d \tau}\langle A\rangle_\tau\Big|_{\tau=0}
=\langle\langle B,  A\rangle\rangle.\label{pertu1}\end{equation}
Thus $\langle\langle A,B\rangle\rangle$ measures the linear response
of eigenvalues of a quantum system.

Let us describe a typical  illustration of the physical meaning of
 $G_{A,B}(z)$ for a general $z$.
 Suppose 
that at time $0$ the system described by a Hamiltonian $H$
is in its ground state. We perturb the Hamiltonian by a weak
perturbation $\lambda B$ and at time
 $t$ we measure the observable
$A$. The shift of the expectation of the measurement is
\begin{eqnarray*}
\delta_\lambda(t)&:=&
\langle\e^{\i t(H+\lambda B)}A\e^{-\i t(H+\lambda B)}\rangle-
\langle A\rangle\\
&\approx&
\i\lambda\int_0^t\d u\langle [B(u),A(t)]\rangle\\
&=&\i\lambda\int_0^t\d s
\, \left( \langle B \e^{\i s(H-E)} A \rangle   - \langle A \e^{- \i 
s(H-E)} B \rangle \right)
,\label{shi}\end{eqnarray*}
where we took the leading term in $\lambda$.
For some $\Im
z<0$, we compute
 the Laplace transform of $\delta_\lambda(t)$, make the
 linear approximation and change the variable $u=t-s$:
\begin{eqnarray*}
\int_0^\infty\e^{-\i tz}
\delta_\lambda(t)\d t
&\approx&\i\lambda\int_0^\infty\e^{-\i tz}\d t\int_0^t\left(\langle 
B\e^{\i s(H-E)}A\rangle-\langle A\e^{-\i s(H-E)}B\rangle\right)\d s\\
&=&\i\lambda\int_0^\infty\e^{-\i uz}\d u\int_0^\infty\left(\langle 
B\e^{\i s(H-E-z)}A\rangle-\langle A\e^{-\i s(H-E+z)}B\rangle\right)\d s
\\&=&\frac{\i\lambda}{z}G_{B,A}(z).
\end{eqnarray*}
Thus,  $G_{A,B}(z)$  measures the
 {\em linear response} of the dynamics a quantum system.

\subsection{The van Hove formfactor}
\label{ap-vH}
A typical experiment  measuring excitation spectrum involves
scattering with a beam of  particles,
 typically neutrons. Following
van Hove \cite{vH}
let us try to describe such an experiment mathematically.

We can assume that a neutron of mass $m$
interacts with each particle of the Bose
gas through a potential $w$. Its incident  momentum is
$\p_\i$. We measure scattered neutrons of momentum
$\p_\f\neq \p_\i$.
The  space describing the Bose gas and a single 
 neutron is $\Gamma_\s(\Lambda)\otimes L^2(\rr^d)$ and the
Hamiltonian is
\begin{eqnarray*}
\tilde H_\lambda&:=&\tilde H_0+\lambda I\\
&=&H\otimes\one+\one\otimes \frac{1}{2m}D_\y^2
+\lambda\int a_\x^*a_\x w(\x-\y)\d \x,\end{eqnarray*}
where $\y$ denotes the position of the neutron, $D_\y$ its
momentum, $\lambda$ is small, and as usual $H$ is the Hamiltonian
of the Bose gas. Let $\Phi_\p$ denote the plane wave function of
momentum $\p$, 
that is $\Phi_\p(\y)=V^{-\frac12}\e^{\i \p\y}$.

Suppose that the initial state of the composite system is
$\Psi\otimes\Phi_{\p_\i}$, where $\Psi$ is the ground state. Let $E$
be the ground 
state energy and $\sigma_\i=\frac{1}{2m}\p_\i^2$ the energy
of the incident neutron.
After time  $2T$ the  evolved state is given by
\begin{eqnarray}
\nonumber
\Theta(T,\p_\i)
&=&\e^{-\i 2T \tilde H_\lambda}\Psi\otimes\Phi_{\p_\i}\\
&\approx&\e^{-\i 2T \tilde H_0}\Psi\otimes\Phi_{\p_\i}\nonumber\\
&&-\i\lambda\int_0^{2T}
\e^{-\i 2T \tilde H_0+\i t(\tilde H_0-E-\sigma_\i)}I\d t \Psi\otimes\Phi_{\p_\i}
\nonumber\\
\nonumber
&=&\e^{-\i 2T(E+\sigma_\i)}
\Psi\otimes\Phi_{\p_\i}\\
&&-
2\i\lambda\e^{-\i T(\tilde H_0+E+\sigma_\i)}
\frac{\sin T(\tilde H_0-E-\sigma_\i)}{\tilde H_0-E-\sigma_\i}
I\Psi\otimes\Phi_{\p_\i}\nonumber
,
\end{eqnarray} 
where we used the so-called Born approximation.
Let $\sigma_\f:=\frac{1}{2m}\p_\f^2$ be the final
energy of the neutron. We introduce also the momentum and energy
transfer
\[\q=\p_\i-\p_\f,\ \ \ \omega:=\sigma_\i-\sigma_\f.\]
To obtain the amplitude of the measurement of the momentum $\p_\f$ we
take the partial scalar product of $\Theta(T,\p_\i)\in
\Gamma_\s(\Lambda)\otimes L^2(\rr^d)$ with
 with
$\Phi_{\p_\f}\in L^2(\rr^d)$ obtaining
the vector in $\Gamma_\s(L^2(\Lambda))$
equal\begin{eqnarray*}
&&\!\!\!\!\!\!\!\!\!\!\!\Theta(T,\omega,\q)\ :=\ \left(\Phi_{\p_\f}\Big|\Theta(T,\p_\i)\right)
\\
&=&-\frac{2\lambda\i}{V}
\e^{-\i T(H_0+E+\sigma_\f+\sigma_\i)}
\frac{\sin T(H_0-E-\omega)}{H_0-E-\omega}
\int \int w(\x-\y)a_\x^*a_\x\Psi\e^{\i \q\y}
\d\y\d\x
\\
&=&-\frac{2\lambda\i}{V}
\e^{-\i T(H_0+E+\sigma_\f+\sigma_\i)}
\frac{\sin T(H_0-E-\omega)}{H_0-E-\omega}
\hat w(\q) N_\q\Psi.
\end{eqnarray*}
Note that the number of states in a cube $\d \q_1\cdots\d \q_d$ equals
equals $V(2\pi)^{-d}\d \q_1\cdots\d \q_d$. Therefore,
the scattering crosssection per unit time
in the Born approximation is 
\begin{eqnarray}\nonumber
&&\frac{1}{2T}
\|\Theta(T,\omega,\q)\|^2V(2\pi)^{-d}\\&=&
\frac{2\lambda^2}{V^2}|\hat w(\q)|^2\left(
\Psi\Big|N_{\q}^*\frac{\sin^2 T(H_0-E-\omega)}{T(H_0-E-\omega)^2}
N_{\q}\Psi\right) V(2\pi)^{-d}\nonumber\\
&\mathop{\longrightarrow}\limits_{T\to\infty}&
(2\pi)^{1-d}\lambda^2\rho|\hat w(\q)|^2
S(\omega,\q),\label{vH1}\end{eqnarray}
where $\rho=\frac{\langle N\rangle}{V}$ is as usual the density and
\begin{eqnarray}
S(\omega,\q)=\langle N\rangle^{-1}
(\Psi|N_\q^*\delta(H-E-\omega)N_\q\Psi)
\label{formfactor}\end{eqnarray}
is sometimes called the {\em van Hove formfactor.}

It is interesting to note that (\ref{vH1}) depends on the incoming and
outgoing data only through the momentum and energy transfer.


\begin{thebibliography}{}
\bibitem{Be} Beliaev, S.T.: Energy spectrum of a non-ideal Bose gas,
  {\it JETP-USSR} {\bf 7} no.2 (1958) 299-307 
\bibitem{Bi} Bijl, A.: Physica {\bf 8} (1940) 655
\bibitem{Bog} Bogoliubov, N. N.: {\it J. Phys. (USSR) }{\bf 9} (1947) 23;
 {\em J. Phys. USSR } {\bf 11} (1947) 23, reprinted 
in D. Pines {\em The Many-Body Problem } (New York, W.A. Benjamin 1962)
\bibitem{Bog1}  Bogoliubov, N. N.: Quasi-averages in problems of statistical
  mechanics, in colection of papers, Vol. 3, Naukova Dumka, Kiev 1971, 174-243 
\bibitem{BS} Brueckner, K.A., Sawada, K.: Phys. Rev. {\bf 106} (1957) 1117
\bibitem{CS} Critchley, R.H. and Solomon, A. I.: A variational approach to
  superfluidity, Journ. of Stat. Phys. {\bf 14} (1976) 381-293
\bibitem{F} Feynmann, R.: in {\em Progress in Low Temperature Phsyics},
  ed. C.J.Gorter, North Holland, Amsterdam 1955, vol. 1, chap 2.
\bibitem{GN} Gavoret, J., Nozi\`{e}res, P.: Structure of the perturbation
  expansion for the Bose liquid at zero temperature {\it Ann. of Phys.} {\bf 28
}  (1964) 349-399 
\bibitem{Gi} Girardeau, M.D.: J. Math. Phys. 1 (1960) 516
\bibitem{Gira} Girardeau, M., Arnowitt, R.: Theory of Many-Boson
  Systems: Pair Theory, {\it Phys. Rev.} {\bf 113} (1959) 755-761
\bibitem{Hu} Huang, K.: Statistical Mechanics, John Wiley and Sons, Inc. New
  York 1963
\bibitem{Ho}Hohenberg, P.C.: Existence of long-range order in one and two
  dimensions, Phys. Rev., {\bf 158} (1967) 156-158
\bibitem{HP} Hugenholtz, N. M., Pines, D.: Ground state energy and excitation
  spectrum of a system of interacting bosons, Phys. Rev. {\bf 116} 
(1959) 489-506
\bibitem{LaL} Landau, L. D., Lifschitz, E.M.: Statistical Physics, Nauka,
  Moscow 
\bibitem{LHY} Lee, T.D., Huang, K., Yang, C.N., Phys. Rev. {\bf 106} (1957) 1135
\bibitem{L1} Lieb, E.H.: Simplified approach to the ground state energy of an
  imperfect Bose gas, Phys. Rev. {\bf 130} (1963) 2518-2528
\bibitem{L2} Lieb, E.H.: Exact Analysis of an Interacting
  Bose Gas. II. The Excitation Spectrum, {\it
    Phys. Rev.} {\bf 130} no. 4 (1963) 1616-1624

\bibitem{LL} Lieb, E.H., Liniger, W.: Exact Analysis of an 
Interacting
  Bose Gas. I. The General Solution and the Ground State, {\it
    Phys. Rev.} {\bf 130} no. 4 (1963) 1605-1616

\bibitem{LSY} Lieb, E. L., Seiringer, R.,
  Yngvason, J.: Superfluidity of dilute trapped Bose gases,
  Phys. Rev. B 66 (2002) 134529


\bibitem{LSSY1} Lieb, E. L., Seiringer, R.,
Solovej, J. P.,  Yngvason, J.:
Justification of $c$-number substitutions in bosonic hamiltonians,
Phys. Rev. Lett. 94 (2005) 080401


\bibitem{LSSY} Lieb, E. L., Seiringer, R.,
Solovej, J. P.,  Yngvason, J.:
The Mathematics of the Bose Gas and its Condensation
Oberwolfach Seminars,  Birkh\"auser Verlag, 2005 
\bibitem{M} Maris, H.~J.: Phonon-phonon 
interactions in liquid helium,
  Rev. Mod. Physi 49 (1977) 341-359
\bibitem{PN} Pines, D., Nozieres, P.: The theory of quantum liquids,
Vol. 2 Superfluid Bose liquids,
  Addison-Wesley 1990
\bibitem{PW} Pines, D., Woo, C.-W.: Phys. Rev. Lett. 24 (1970) 1044
\bibitem{Pr} Price, P.J.: Phys. Rev. 94 (1954) 257
\bibitem{PSt} Pitaevskii, L. and Stringari, S.: J. Low Temp. Phys. {\bf 5} (1991)
\bibitem{Pop}
 Popov, V.N.:
  "Functional Integrals in Quantum Field Theory and Statistical
Physics",
 D. Reidel Publishing Company, 1983
\bibitem{Pu} Puff, R.D.: Phys. Rev. A {\bf 137} (1965) 406
\bibitem{RKODKHK}
 Raman, C.,  Kohl, M.,  Onofrio, R., Durfee,  D.S., Kuklewicz,  C.E.,
 Hadzibabic, Z.
and Ketterle, W.: Phys. Rev. Lett. {\bf 83}, 2502 (1999); Evidence for a
Critical Velocity in a Bose Einstein Condensed Gas.
\bibitem{Ro} Robinson, D. W.:  On the ground state energy of the Bose gas,
  {\it Comm. Math. Phys.} {\bf 1} (1965) 159-174
\bibitem{SOKD}
Steinhauer, J., Ozeri, R., Katz, N., Davidson, N.: Excitation  
spectrum
of a Bose-Einstein condensate, Phys. Rev. Lett. 88 (2002) 120407
\bibitem{St1}  Stringari, S.: Sum rules for density and particle expectations
  in Bose superfluids, Physical Review B, {\bf 46} (1992) 2974-2984
\bibitem{St2} Stringari, S.: Sum rules and Bose-Einstein condensation,
  International Workshop on Bose-Einstein condensation, Levico, Italy, 1993, cond-math/9311024v1

\bibitem{Ta} Takano, F.: Excitation  spectrum in many-boson systems,
 {\it Phys. Rev.} {\bf 123} (1961) 699-705
\bibitem{Ta3} Takesaki, H.: Theory of Operator Algebras III, Springer 2003
\bibitem{vH} van Hove, L.: Correlations in space and timee and Born
  approximation scattering in systems of interacting particles, Phys.
  Rev. 95 (1954) 249-262
\bibitem{Wa} Wagner, H.: Z. Physik {\bf 195} (1966) 273
\bibitem{WC} Woods, A.D.B., Cowley, R.A.: Structure 
and excitations of
  liquid helium, Rep. Prog. Phys. 36 (1973) 1135-1231
\bibitem{WdS} Wreszinski, W.F., da Silva, M. A. Jr: Onsager's inequality, the
  Landau-Feynmann ansatz and superfluidity, J. Phys. A: Math. Gen. 38 (2005)
  6293-6310 


\bibitem{ZB} Zagrebnov, V.A., Bru, J.B.: The Bogoliubov model of
  weakly imperfect Bose gas, {\it Phys. Rep.} {\bf 350} no. 5-6 (2001) 291-434


\end{thebibliography}
\end{document}